\documentclass[12pt, letter]{article}
\usepackage{amsmath}
\usepackage{graphicx,psfrag,epsf}
\usepackage{enumerate}
\usepackage{natbib}

\newcommand{\blind}{0}

\usepackage[figuresright]{rotating}
\usepackage{amsfonts}
\usepackage{amsmath}
\usepackage[font=scriptsize,labelfont=bf]{caption}
\usepackage{graphicx,multicol}
\usepackage{color}
\usepackage{booktabs}  
\usepackage{float}
\usepackage[section]{placeins}
\usepackage[caption = false]{subfig}
\usepackage{multirow}
\usepackage{pdflscape}

\newcommand\bigCI{\perp\!\!\!\perp}
\DeclareMathOperator*{\argmin}{argmin}

\numberwithin{equation}{section}

\newtheorem{remark}{Remark}[section]

\begin{document}

	
	\def\spacingset#1{\renewcommand{\baselinestretch}%
		{#1}\small\normalsize} \spacingset{1}

	\vspace{-3cm}
	\if0\blind
	{
		\title{\bf Robust graphical lasso  based on multivariate Winsorization}
	\author{
		\small Ginette LAFIT, Francisco J. NOGALES, Marcelo RUIZ and Ruben H. ZAMAR 
	}
		\maketitle
			\makeatletter{\renewcommand*{\@makefnmark}{}
			{\footnotetext{ \scriptsize Ginette Lafit, Postdoctoral research fellow, Research Group of Quantitative Psychology and Individual Differences, KU Leuven–University of Leuven, Leuven, Belgium (E-mail:  ginette.lafit@kuleuven.be), Francisco J. Nogales is  Professor, Department of Statistics and UC3M-BS Institute of Financial Big Data, Universidad Carlos III de Madrid, Espa\~na (E-mail: fcojavier.nogales@uc3m.es),  Ruben H. Zamar is Professor, Department of Statistics, University of British
					Columbia, 3182 Earth Sciences Building, 2207 Main Mall, Vancouver, BC  V6T 1Z4, Canada (Email:
					ruben@stat.ubc.ca) and Marcelo Ruiz is Professor, Departamento de Matem\'atica, FCEFQyNat, Universidad Nacional de R\'io Cuarto, C\'ordoba, Argentina (E-mail: mruiz@exa.unrc.edu.ar). }}\makeatother}
	} \fi
	
	\if1\blind
	  \fi
	\vspace{-1cm}
	
	\begin{abstract}
 We propose the use of a  robust covariance estimator based on multivariate Winsorization in the context of the Tarr--M\"{u}ller--Weber framework for sparse estimation of the precision matrix of a Gaussian graphical model. Likewise Croux--\"{O}llerer's  precision matrix estimator, our proposed estimator  attains the maximum finite sample breakdown point  of 0.5 under cellwise contamination.  We conduct an extensive Monte Carlo simulation study to assess  the performance of ours and the currently existing proposals. We find that ours has a competitive behavior,  regarding the the estimation of the precision matrix and the recovery of the graph.  We demonstrate the usefulness  of the proposed methodology in a real  application to breast cancer data.  
	\end{abstract}

\noindent%
{\it Key words:}  Gaussian graphical model; Precision matrix; Sparse robust estimation; Cellwise contamination; Winsorization.

 \newpage

\section{Introduction}
\label{intro}

Let  $\boldsymbol{X}=(X_1,\ldots,X_p)'$ be  a $p$-variate random vector with Gaussian distribution with mean vector $\boldsymbol{\mu}$ and  covariance matrix $\Sigma$. 
We  assume  that $\Sigma$ is positive definite ($\Sigma \succ 0$) and its inverse, the precision matrix,  will be denoted by $\Omega=(\omega_{ij})_{i,j=1 \ldots,p}$; furthermore, we assume that $\boldsymbol{\mu}=\boldsymbol{0}$. 
Abbreviated, the model is 
\begin{eqnarray}\label{modelo}
	\boldsymbol{X}\sim \text{N}(\boldsymbol{0}, \Sigma).
\end{eqnarray}

Given  $V=\{1,\ldots, p \}$, let $V^2=V\times V$ and $V^2_{-d}=\{(i,l)\in V^2: i\neq l\}$.  For a given pair  $(i,l)\in V^2_{-d}$, let 
$V\backslash\left\{  i,l\right\}= \left\{j\in V:  i\neq j\neq l\right\}$ and $X_{V\backslash\left\{  i,l\right\} }=\{ X_j: j\in V\backslash\left\{  i,l\right\} \}$.

For a random vector   $\boldsymbol{X}$ satisfying \eqref{modelo}, a {\it Gaussian graphical model} (GGM) is the undirected graph $(V,E)$ where $V$ is the set of nodes and $E$ is the set of edges,  which is defined by
\begin{equation} \label{pmp0}
(i,l) \notin E \text{ if and only if } \text{ corr}  \left( X_i,X_l | X_{V\backslash\left\{  i,l\right\}  } \right)=0,
\end{equation}
where $\text{ corr}  \left( X_i,X_l | X_{V\backslash\left\{  i,l\right\}  } \right)$ is the conditional correlation coefficient of  $X_i$ and $X_l$ given $X_{V \setminus \{i,l\}}$. For brevity, \eqref{pmp0} can be 
rewritten as 
\begin{equation} \label{pmp}
	(i,l) \notin E \text{ if and only if }  X_i \bigCI X_l \mid X_{V \setminus \{i,l\}}.
\end{equation}	
 	
So,  the set of edges $E$ can be   expressed  as 
\begin{eqnarray}\label{erepresent2}
	E=	\left\{(i,l) \in  V^2_{-d}:\, \text{ corr}  \left( X_i,X_l | X_{V\backslash\left\{  i,l\right\}  } \right) \neq 0 \right\}.
\end{eqnarray}
It is well known that there exists a characterization of the conditional correlation in terms of the elements of  the precision matrix. More specifically
 \begin{equation} \label{param}
\forall (i,l)\in V^2_{-d}:\,  	 \text{ corr}  \left( X_i,X_l | X_{V\backslash\left\{  i,l\right\}  } \right)= - \frac{\omega_{il}}{ \sqrt{\omega_{ii}\omega_{ll} }}.
\end{equation}
Hence, we have the following parametrization for $E$: 
\begin{eqnarray}\label{paramprec}
E= \{(i,l) \in  V^2_{-d}:\, \omega_{i,l} \neq 0 \}.
\end{eqnarray}

Given a sample of $\boldsymbol{X}$,  the goal of covariance selection  is to estimate the conditional dependence structure by determining the set of  nonzero entries of the precision matrix  $\Omega$      \cite[see][]{dempster1972covariance, lauritzen1996graphical,edwards2000introduction}. Generally, in high-dimensional statistics,  it is assumed that there are just a few entries of $\Omega$ which are different from zero, that is, that $\Omega$ is sparse.

Until a few decades ago,  statistical procedures assumed that datasets included many observations of a few and carefully chosen variables. Nowadays, datasets contain a large number of variables relative to the sample size, producing blessings but also curses of dimensionality \citep{donoho2000, donoho2017}. Therefore, in a high-dimensional setting, the estimation of precision matrices faces  significant challenges.

Let $\mathbb{X}=\left(\boldsymbol{x}_1',\ldots  \boldsymbol{x}_n'\right)'$ be a $n\times p$ data matrix where $\boldsymbol{x}_1,\ldots,\boldsymbol{x}_n$ is a sample. 
If $n>p$ then the sample covariance matrix $\displaystyle S=\frac{1}{n}\sum_{i=1}^n (\boldsymbol{x}_i-\bar{\boldsymbol{x}})(\boldsymbol{x}_i-\bar{\boldsymbol{x}})^{\prime}$ is well conditioned and a well known  optimal estimate of   $ \Sigma $. In this case  $  [(n-p-1)/(n-1)]S^{-1}$ is an unbiased estimate of $\Omega$. On the other hand, when $p>n$  the sample covariance matrix is not invertible. 
 
To deal with this problem   several covariance selection procedures based on regularization  have been developed under the assumption that $\Omega$ is sparse.  For instance, if  $\widehat{\Sigma}$ is an estimator of $ \Sigma$, the Graphical lasso (Glasso) proposed by  \cite{friedman2008sparse} is defined by 
\begin{equation}\label{eq2-4}
\widehat{\Omega} = \argmin_{ \{ U: U^{'}=U, U \succ 0\}} \left\{\text{tr}(U \widehat{\Sigma}) -\text{log}\text{det} (U) + \lambda \parallel U \parallel_{1}\right\}
\end{equation}
 where the optimization is over the set of symmetric positive definite matrices,
\begin{equation}\label{eq2-5}
\parallel U \parallel_{1  } \; =: \sum_{i, j} |u_{ij}| \quad \text{for} \; i,j=1,\ldots,p, 
\end{equation}
is the $\ell_1$ norm of the matrix $U=(u_{ij})_{i,j=1 \ldots,p}$ and $\lambda \geq 0$ is a  regularization or penalty parameter usually determined by crossvalidation. Note that the larger the value of $\lambda$ is, the more sparse the precision matrix estimate becomes. 

For $\lambda= 0$, if $\widehat{\Sigma}=S\succ 0$ then the solution of \eqref{eq2-4}  is the classical maximum likelihood estimate of $\Omega$.   On the other hand \cite{banerjee2008model} proved that, for any symmetric and  positive semidefinite   matrix  $\widehat{\Sigma}$  and   $\lambda>0$, the equation   \eqref{eq2-4} has a strictly positive definite solution $\widehat{\Omega}$ even if $p>n$.

In contrast to univariate data sets, in multivariate settings, outliers can appear in  complex ways. In this regard, two types of contamination mechanisms have been introduced in the robustness literature: the Tukey-Huber contamination model (THCM) and the independent contamination model (ICM). In the THCM it is assumed that a a relative large proportion $\epsilon$ ($\epsilon>0\text{.}5$) of the rows in the data table are contaminated.   In the ICM,   introduced by \cite{alqallaf2009propagation},  each cell of the data matrix has a probability to be independently contaminated. This second mechanism is a better fit for the high-dimensional setting where the variables are likely to be obtained from different sources and measured separately \citep{agostinelli2015}.

The vast majority of the work in the area of robust statistics has concentrated on the estimation of the covariance matrix under these  two types of contamination models. Robust conditional correlation coefficient estimation has been  studied when $p$ is small.  \cite{raosievers} introduced a measure that uses residuals based on rank estimates of regression parameters when $p=3$. Only recently a few papers have focused on estimation of  the precision matrix in the context of ICM. 

\cite{tarr2016robust} and \cite{ollecroux} showed that Glasso is not robust in the presence of cellwise outliers.  Therefore, in order to obtain a robust estimate of the precision matrix, they proposed a  plug-in approach, using a robust covariance matrix estimator  $\widehat{\Sigma}$ in equation \eqref{eq2-4}. 
There are several robust  estimators of  $ \Sigma$ but, unfortunately, their computation is very  time-consuming and  may not be possibly well defined  when the dimension $p$ is high \citep{khan2007robust}. To overcome this problem, resistant pairwise procedures can be used  to avoid sensitivity to two-dimensional outliers, like in  \cite{tarr2016robust} and   \cite{ollecroux} proposals.
 \cite{tarr2016robust} proposed to use pairwise robust covariances estimates, whereas \cite{ollecroux} use pairwise robust correlation estimates.

\cite{huber2011robust} proposed a robust estimator of the correlation coefficient by using one-dimensional Winsorization. \cite{alqallaf2002} proposed the use of Huberized pairwise correlation coefficients based on one dimensional Winsorization.   A limitation of this approach is that the pairwise Huberized estimates and covariance estimates     do not take into account the orientation of the (pairwise) bivariate data. To overcome this limitation, \cite{khan2007robust} developed an adjusted bivariate Winsorization estimation, obtaining a robust estimator of the correlation matrix under cellwise contamination. Here, we use this estimator to introduce a new robust Graphical lasso procedure, RGlassoWinsor.  We compare the performance of our method with  other existing approaches under cellwise and casewise contamination.
 
 Section \ref{ohd} discusses  the main differences between the THCM and ICM.  Section \ref{RGlassoWinsor} introduces our proposal. Section \ref{simul} presents the results of an extensive simulation experiment comparing the currently existing estimators of the robust precision matrix  with our new robust Graphical lasso procedure.   Section \ref{realdata} contains an application to breast cancer data. Section \ref{conclrema} concludes with some remarks. The Appendix gives some additional simulation results.

 \section{Outliers  in high-dimensional data}  \label{ohd}

In this section, we briefly outline the main differences between   THCM and ICM.

Consider   a  set of $n$ independent observations  of the  multivariate Gaussian vector $\boldsymbol{X}=(X_1,\ldots,X_p)^{\prime}$ satisfying \eqref{modelo}, let $\epsilon\in (0,1)$ be the  fraction of contamination and define  the random vector
\begin{equation}\label{vbjs}
	\boldsymbol{B} = (B_1,\ldots, B_p)^{\prime} \text{ with } B_j \sim \text{Bernoulli}(\epsilon), \, j=1,\ldots, p.
\end{equation}

Suppose that instead of $\boldsymbol{X}$ we observe 
\begin{equation}\label{contam}
\boldsymbol{Y} = (I-D) \boldsymbol{X} +D \boldsymbol{Z}
\end{equation}
\noindent
where $I$ is the $p \times p$ identity matrix, $\boldsymbol{Z}$ is a $p$-variate  random vector with an arbitrary and unspecified outlier generating distribution and  $D$ is a  diagonal matrix with diagonal elements $B_1, \ldots, B_p$. 
 Moreover, we assume  that $\boldsymbol{X}$, $\boldsymbol{B}$ and $\boldsymbol{Z}$ are independent.

The classical  THCM assumes that   the  random vector $\boldsymbol{B} = (B_1,\ldots, B_p)^{\prime}$   satisfy  $P(B_1=B_2=\ldots=B_p)=1$. So,   we either see a perfect realization of the random vector $\boldsymbol{X}$,  with probability $1-\epsilon$, or a realization of the random vector  $\boldsymbol{Z}$, with probability $\epsilon$.

Motivated by the THCM,  robust procedures identify and downweight possibly contaminated cases. However, in a high-dimensional setting, this strategy is inconvenient for two reasons. The most obvious is that in high-dimension, when $n$ is relatively small compared with   $p$, discarding a single observation may result in a substantial loss of information. A perhaps less obvious reason was highlighted by \cite{alqallaf2009propagation}, where they argued that there are situations where the contaminating mechanism may be independent for different variables. Consequently, they proposed the  ICM that assumes that $B_1,\ldots,B_p$ are   independent random variables and satisfy  
\begin{equation}\label{eq2-3}
P(B_1=1)=\ldots = P(B_p=1)=\epsilon.
\end{equation}
Hence, a case is uncontaminated,  $\boldsymbol{Y}=\boldsymbol{X}$, with probability $P(  \boldsymbol{B}=\boldsymbol{0})=(1-\epsilon)^p$, which quickly decreases 
 below $1/2$ as $p$ increases.  Equivalently, the probability that at least one component of $\boldsymbol{Y}$ is contaminated is $1-(1-\epsilon)^p$.   For example if $p=60$ and $\epsilon=$0.05 this probability  equals to  0.95. If $p\geq 200$ (not an uncommon case these days) this probability becomes nearly 1 .

\begin{figure} [H]	
	\begin{multicols}{2}
		\includegraphics[width=\linewidth]{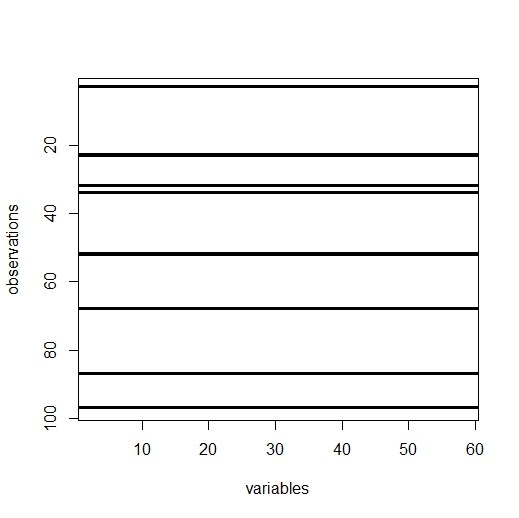}\par
		\caption*{ (a)}
		\includegraphics[width=\linewidth]{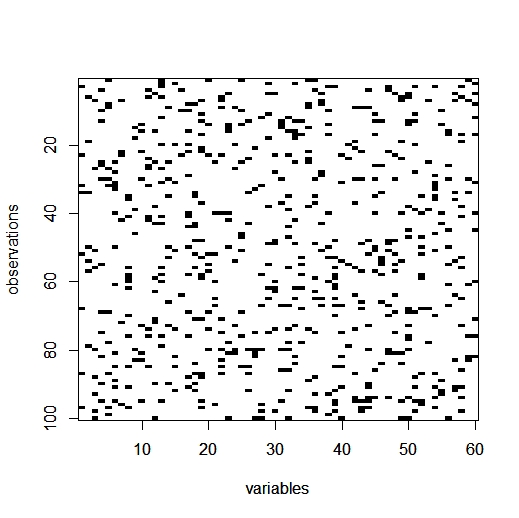}\par
		\caption*{(b)}
	\end{multicols}
	\caption{   Panels (a) and (b) represent the data matrix of dimension $100\times 60$ corresponding to the random vector $\boldsymbol{B}$ of dimension $60$ given in \eqref{vbjs} generated under  THCM  and ICM respectively. Uncontaminated cells are in color white and contaminated cells are in color  black.  }
	\label{simB}
\end{figure}

The indicator matrix $D$, whose diagonal is a sequence of Bernoulli random variables, determines the structure of the contamination model. Figure \ref{simB} shows a representation of  a sample of size $n=100$ of 	$\boldsymbol{B}$ with dimension $p=60$, contamination fraction  $\epsilon=0.10$, under both contamination models: THCM in panel (a) and ICM in panel (b). On each panel, uncontaminated cells are in color white and contaminated cells are in color black.    For  THCM the actual proportion of contaminated cells is 0.08, coinciding  with the percentage of contaminated observations (rows). But, for  ICM the  proportion of contaminated cells is 0.10 but   all  the observations have at least one contaminated cell ($\approx 1-(0\text{.}9)^{60}$), hence the totality of the cases or rows are contaminated. This phenomenon is called ``propagation of outliers'' in \cite{alqallaf2009propagation}

THCM is also  called casewise contamination model, where a minority  of observations or cases (rows) of the data matrix  contains outliers and the size of this minority does not depend on the number  $p$ of variables. ICM is also denominated cellwise contamination model because the contamination is produced randomly affecting the cells of the data table.

The classical robustness theory based on the affine equivariant  Tukey-Huber contamination model relays and enforces the concept of equivariance.                                                                                                               
\cite{alqallaf2009propagation} showed that under the cellwise contamination model, a  standard high-breakdown affine equivariant estimators propagate outliers, and this causes their very poor performance when $p$ is large. The reason is that affine equivariant robust estimators depend on linear combinations of the observations which have a very high probability of being contaminated under ICM for moderate and large $p$. Notice that under ICM,  the majority of cases will have at least some contaminated component.    \cite{agostinelli2015}  addressed the problem of robust estimation of location and scatter under the two contamination models.

\section {Robust lasso  for covariance selection   }
\label{RGlassoWinsor}

\subsection { Plug-in strategy }
\label{plugst}

Hereafter  $\mathbf{y}_i=(y_{i1},\ldots, y_{ip})^{'}, i=1,\ldots, n$ denotes a sample of observations of a $p$-multivariate random vector $\boldsymbol{Y}=(Y_1,\ldots, Y_p)'$ satisfying  \eqref{contam}  and let  $\mathbb{Y}=(\mathbf{y}_1,\ldots, \mathbf{y}_n)\in \mathbb{R}^{n\times p}$ be the corresponding  data table.   Let $R$ denote the correlation matrix; i.e. if $  \Sigma=(\Sigma_{ij})$ then $R=(R_{ij})$ with  $R_{ij}= \Sigma_{ij}/\sqrt{ \Sigma_{ii}\Sigma_{jj}}$.

Following  \cite{tarr2016robust}  and using \eqref{eq2-4}   we will construct a robust estimation procedure of the precision matrix  as follows:
\begin{eqnarray}\label{rglasso}
	\widehat{\Omega} = \argmin_{ \{ U: U^{'}=U, U \succ 0\}} \text{tr}( U \widehat{\Sigma}) -\text{log}\text{det}U + \lambda \parallel U \parallel_{1} 
\end{eqnarray}
where $\widehat{\Sigma}$ is a robust estimator of the covariance matrix.



\subsection {Adjusted multivariate Winsorization  }
\label{plug}

To control the effect of bivariate outliers on the pairwise estimation of $ \Sigma$, we apply the procedure proposed by \cite{khan2006robust}. In this procedure, the  robust estimator $\widehat{R}^W$ of the correlation matrix $R$ is defined in two steps by first computing the pairwise correlation matrix, $\widehat{R}^0$, using an adjusted Winsorization scheme, which takes into consideration the orientation of bidimensional data. Later, based on $\widehat{R}^0$, a robust estimator of the covariance matrix  $\Sigma$  is defined. 

The two steps to compute $\widehat{R}^W$ are given below:

\begin{itemize}
	
	\item[1)] Initial estimate $\widehat{R}^0$.
	
Given  $j, k \in \{1, \ldots, p\}$, with $j \neq k$,     let consider the bivariate sample  $\{(y_{ij}, y_{ik})^{'}$, $i=1,\ldots n\}$ and compute   for every $l=j,k$
\begin{eqnarray} \label{fstep0}
m_l=\text{median}(y_{1l}, \ldots,y_{nl} ) , s_l= \text{mad} (y_{1l}, \ldots,y_{nl} ),\ 
\end{eqnarray}
 where   ``$\text{mad}$'' denotes the median absolute deviation.  Define now  the the standardized samples
\begin{eqnarray} \label{sstep0}
 \tilde{y}_{il}=\frac{y_{il}-m_l}{s_l}, \, i=1,\ldots, n
\end{eqnarray}
for every $l=j,k$. 

As \cite{khan2007robust} noted, one dimensional Winsorization does not account for the orientation of the  bidimensional data and does not address the effect of bivariate outliers. Therefore, they propose a bivariate adjusted Winsorization 
that uses two tuning constants denoted $c_1$ and $c_2$. The constant $c_1$ is used on  the two quadrants that contain the majority  of the standardized data and the constant $c_2$, smaller than $c_1$, is used on the other two quadrants. Typically $c_1=2$ or $2\text{.}5$  and $c_2=\sqrt{h}c_1$ with $h=n_2/n_1$, where $n_1$ is the number of observations in the two major  quadrants and $n_2=n-n_1$.

The bivariate Winsorized data $(v_{ij},v_{ik})^{'}$, $i=1,\ldots, n$ are computed as follows. If $(\tilde{y}_{ij}, \tilde{y}_{ik})$ lies in one of the  major (more populated) quadrants, let 
\begin{eqnarray} \label{psima}
v_{il}=\psi_{c_1} \left( \tilde{y}_{il}   \right), \; i=1,\ldots, n;\, l=j, k,
\end{eqnarray}
where  $\psi_{c_1}$ is the Huber function $\psi_c(x)=\min \left\{ \max\left\{-c,x\right\}, c  \right\}$ with tunning constant $c=c_1$. On the other hand, if $(\tilde{y}_{ij}, \tilde{y}_{ik})$ lies in one of the  minor (less populated) quadrants   then 
\begin{eqnarray} \label{psimi}
	v_{il}=\psi_{c_2} \left( \tilde{y}_{il}   \right), \; i=1,\ldots, n;\, l=j, k.
\end{eqnarray}

The elements $\widehat{R}^0_{jk}$ of the matrix $\widehat{R}^0$ are now defined as follows. For $j=k$ we set  $\widehat{R}^0_{jj}=1$, and for $j\neq k$ we set 
 \begin{eqnarray*} \label{entriesr0}
\widehat{R}^0_{jk}=\text{corr}(\mathbf{v}_j, \mathbf{v}_k)
\end{eqnarray*}
where $\mathbf{v}_j=(v_{1j},\ldots, v_{nj} )^{'}$  and $\mathbf{v}_k=(v_{1k},\ldots, v_{nj} )^{'}$.

\item[2)] Final estimate $\widehat{R}^W$.   

As before, consider $\{(y_{ij}, y_{ik})^{'}$, $i=1,\ldots n\}$  a  bivariate sample of the two variables $Y_j$ and $Y_k$, with  $j \neq k$ (columns $j$ and $k$ of the data table). Let 
\begin{equation*}
	A_{jk} = 
	\begin{pmatrix}
	1 & \widehat{R}^0_{jk} \\
	\widehat{R}^0_{kj} & 1\\
	\end{pmatrix}.
\end{equation*}
be the $2\times2$ submatrix of $\widehat{R}^0$. Perform  now, for every $l=j,k$, the following bivariate transformation 
\begin{eqnarray} \label{bivtrans}
	u_{il}=y_{il}\min \left( \sqrt{c/D_{jk} (y_{ij},   y_{ik})} , 1\right), i=1,\ldots, n; l=j,k, 
\end{eqnarray}
where   $D_{jk}$  is  the Mahalanobis distance based on the correlation matrix	$A_{jk}$ and evaluated in $(y_{ij},   y_{ik})$. The  tunning constante $c=5\text{.}99$ corresponds to  the $95\%$ quantile of a $\chi^2_2$ distribution. By this transformation the outliers
  are shrunken to the border of an ellipse, including the majority of the data.

We now define the Winsorized correlation estimate $\widehat{R}^W=	(\widehat{R}^W_{jk})$  as follows. For  $j\neq k$, we set   
 \begin{eqnarray*} \label{entriesr0w}
 	\widehat{R}^W_{jk}=\text{corr}(\mathbf{u}_j, \mathbf{u}_k),
 \end{eqnarray*}
 where $\mathbf{u}_j=(u_{1j},\ldots, u_{nj} )^{'}$ and $\mathbf{u}_k=(u_{1k},\ldots, u_{nj} )^{'}$  and, for $j=k$, we set $\widehat{R}^W_{jj}=1$.

\end{itemize}

Finally, based on $\widehat{R}^W$, a robust estimator of $\Sigma$ is defined as
\begin{eqnarray} \label{covwins}
\widehat{\Sigma}^W =\text{diag}(s_1,\dots, s_p) \widehat{ R }^W \text{diag}(s_1,\dots, s_p) 
\end{eqnarray}
where $s_j$ is the robust estimator of the dispersion introduced in \eqref{fstep0}.   In order to guarantee positive definiteness of  $\widehat{\Sigma}^W$ we compute the nearest positive   definite matrix \citep{higham}. Finally, the robust Glasso estimator of the precision matrix based on bivariate adjusted Winsorization, called RGlassoWinsor and denoted by $	\widehat{\Omega}^W$,  is defined by \eqref{rglasso} with  $\widehat{\Sigma}=\widehat{\Sigma}^W$.

\begin{remark}
By Theorem 19.1 and Proposition 19.1 in \cite{ollecroux} the finite sample breakdown point under ICM of $\widehat{\Omega}^W$ satisfies 
$$
\epsilon_{n}\left(\widehat{\Omega}^W\right) \geq \epsilon_{n}^{+}(\widehat{\Sigma}^W) \geq \max_{j=1,\ldots,p} \epsilon_{n}^{+}(s_j)=1/2
$$
where $\epsilon_{n}^{+}(\widehat{\Sigma}^W)$ is the explosion finite-sample  breakdown point (EBP) under ICM  contamination of $\widehat{\Sigma}^W$ and $\epsilon_{n}^{+}(s_j)$ is the EBP of the univariate scale estimator scale $s_j$, $j=1,\ldots,p$.
\end{remark}

\section{Simulation experiment and numerical results}
\label{simul}

We conducted a Monte Carlo simulation experiment to investigate the performance of RGlassoWinsor compared with other procedures.

\subsection{Simulation settings}
In the following, we describe the precision matrix models, the contamination scenarios and the precision matrix estimation procedures considered in our simulation study.

\medskip

\subsubsection*{Precision matrix models}
\noindent \textbf{}

We consider two dimension values ($p = 60, 200$) and five $\Omega$ models.

\begin{itemize}
	\item[]  \textbf {Model 1.}  Autoregressive model  of order $1$, denoted $\text{AR}(1)$. In this case we set $\Sigma_{ij}=0.4^{|i-j|}$ for $i,j=1,\ldots p$ and $\Omega=\Sigma^{-1}$.
		\item[]  \textbf {Model 2.}  Block diagonal matrix model, denoted BG. In this case the precision matrix $\Omega$ has $q$ blocks of  size $p/q$. Each block   has diagonal elements equal to $1$ and  off-diagonal elements equal to $0.5$. For $p=60,200$    we use $q=10$ and $40$  blocks,  respectively.    
	\item[]  \textbf{Model 3.}  Random model, denoted $\text{Rand}$. \cite{huge}, in the   R package \texttt{huge}, compute the  $\Omega$ matrix of this model as follows. First they consider $\Theta=(\theta_{ij})$ an adjacency matrix of dimension $p$ such that every diagonal entry $\theta_{ii}=0$, each pair of off-diagonal elements  is randomly set $ \theta_{ij}= \theta_{ji}=1$ with probability $\text{prob}=3/p$ (the default value) and defined as $0$ otherwise. Then  they define the set of edges of the graph, establishing that two different nodes, $i$ and $j$, are connected if and only if $\theta_{ij}=1$. 
	Finally, given $\Theta$, is possibly to choose real constants $v$  and $s$ such that $\Omega= v\Theta + s I_p$   is positive definite, with $I_p$ the identity matrix.



	\item[]  \textbf { Model 4.}  Nearest neighbors model of order 2, denoted
	$\text{NN}(2)$. For each node  we randomly select two neighbors and choose a pair of symmetric entries of  $\Omega$ using the ``NeighborOmega'' function of the R package \texttt{Tlasso} \citep{Tlasso}.  
		\item[]  \textbf { Model 5.}  Hub model,  denoted  
	$\text{Hub}$. As in Model 3,  consider    $\Theta=(\theta_{ij})$  an adjacency matrix defined as follows. The row/columns are evenly partitioned into $3$  ($10$) disjoint groups if $p=60$ (if $p=200$). Each group is associated with a ``center'' row $i$ in that group. Each pair of off-diagonal elements, $ i\neq j$,  are set $ \theta_{ij}=\theta_{ij}=1$ 	if   $j$ also belongs to the same group as $i$ and $0$ otherwise. It results in $57$  ($190$) edges in $E$ if $p=60$ (if $p=200$). The precision matrix $\Omega$ is defined as in Rand Model and computed using the same R package \texttt{huge} \citep{huge}.
\end{itemize}

\begin{figure} [H]
	
\begin{multicols}{3}
\includegraphics[width=1\linewidth]{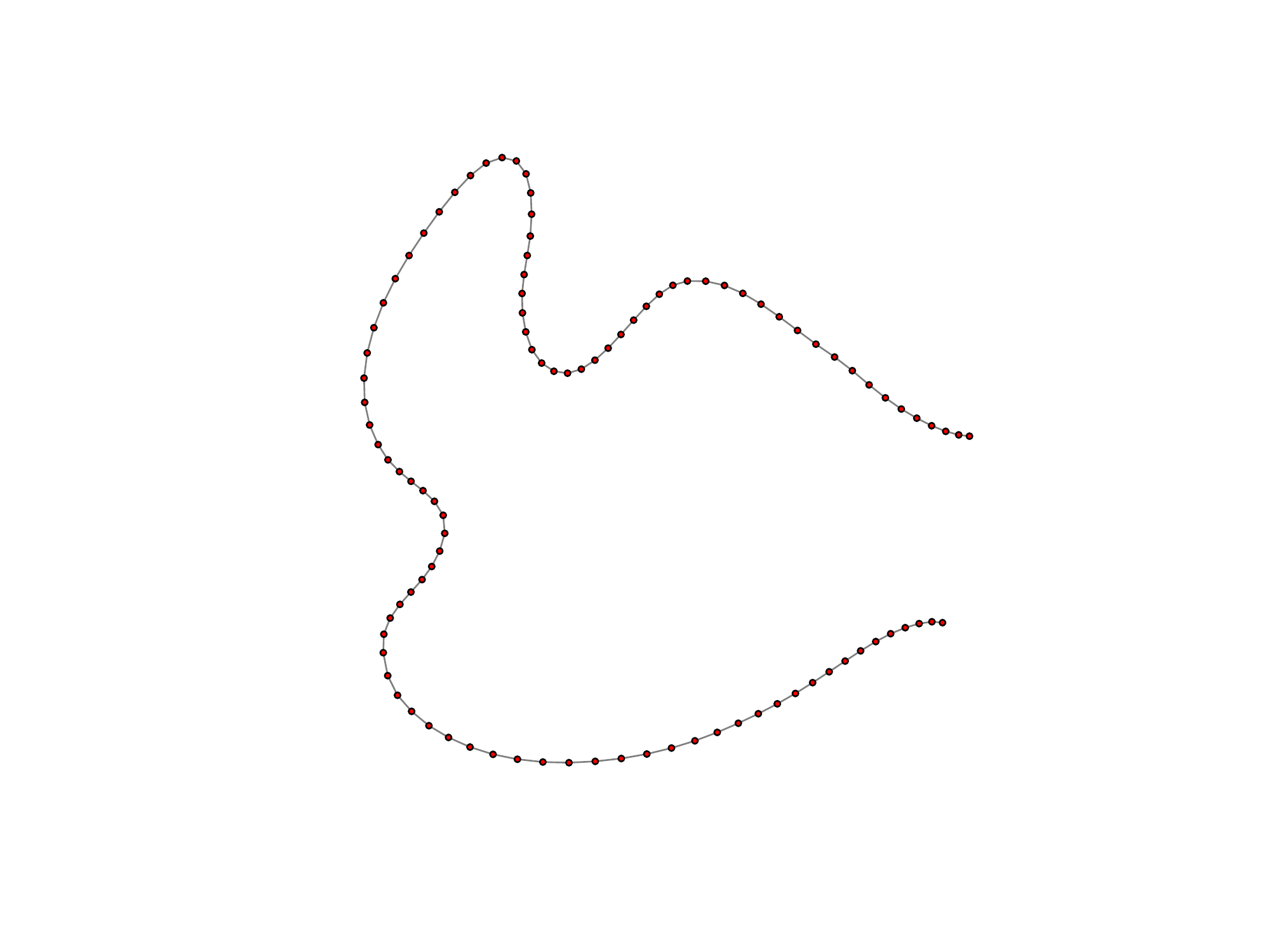}\par
\caption*{ Model AR(1)}
\includegraphics[width=1\linewidth]{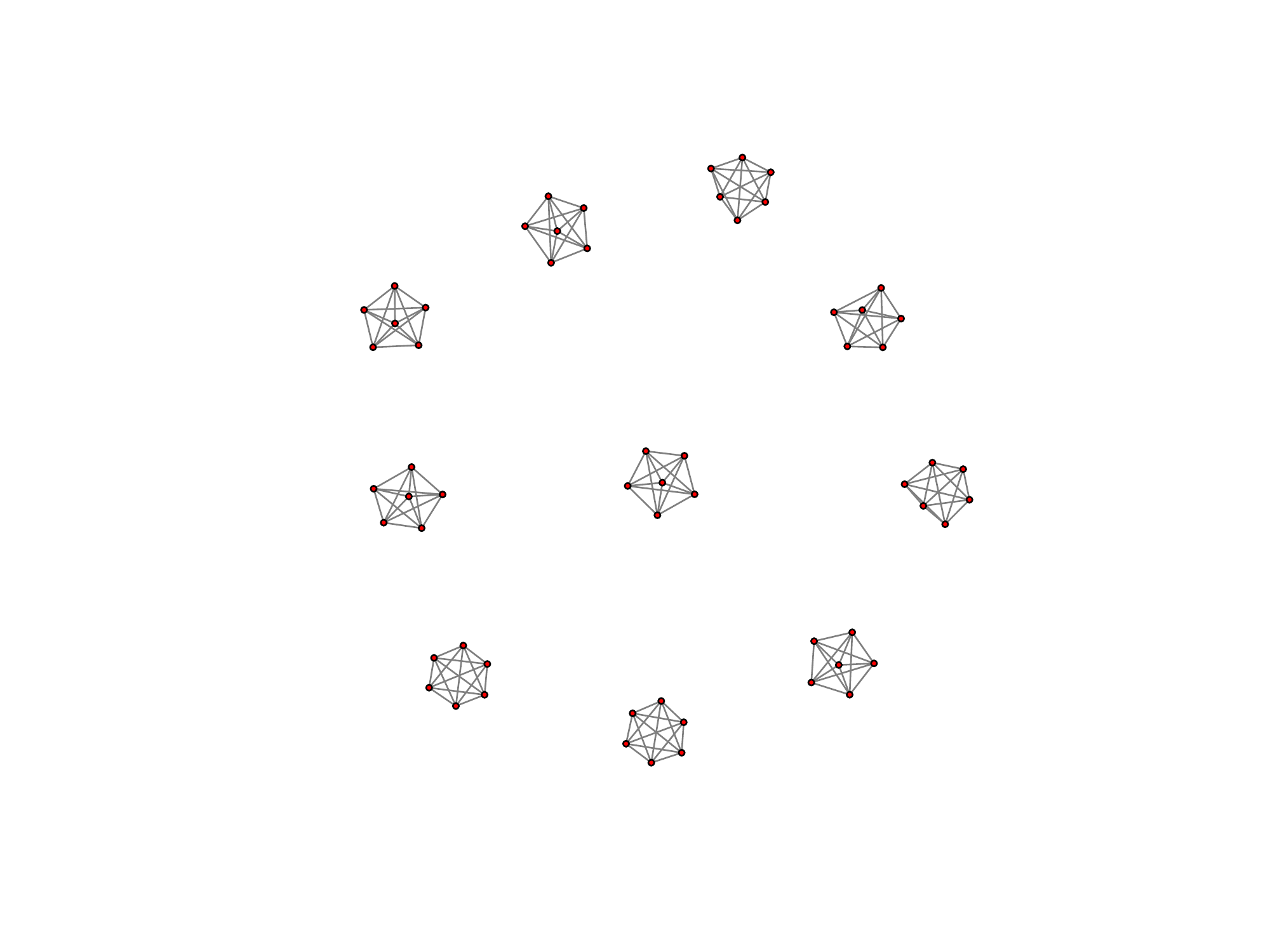}\par
\caption*{Model  $\text{BG}$}
\includegraphics[width=1\linewidth]{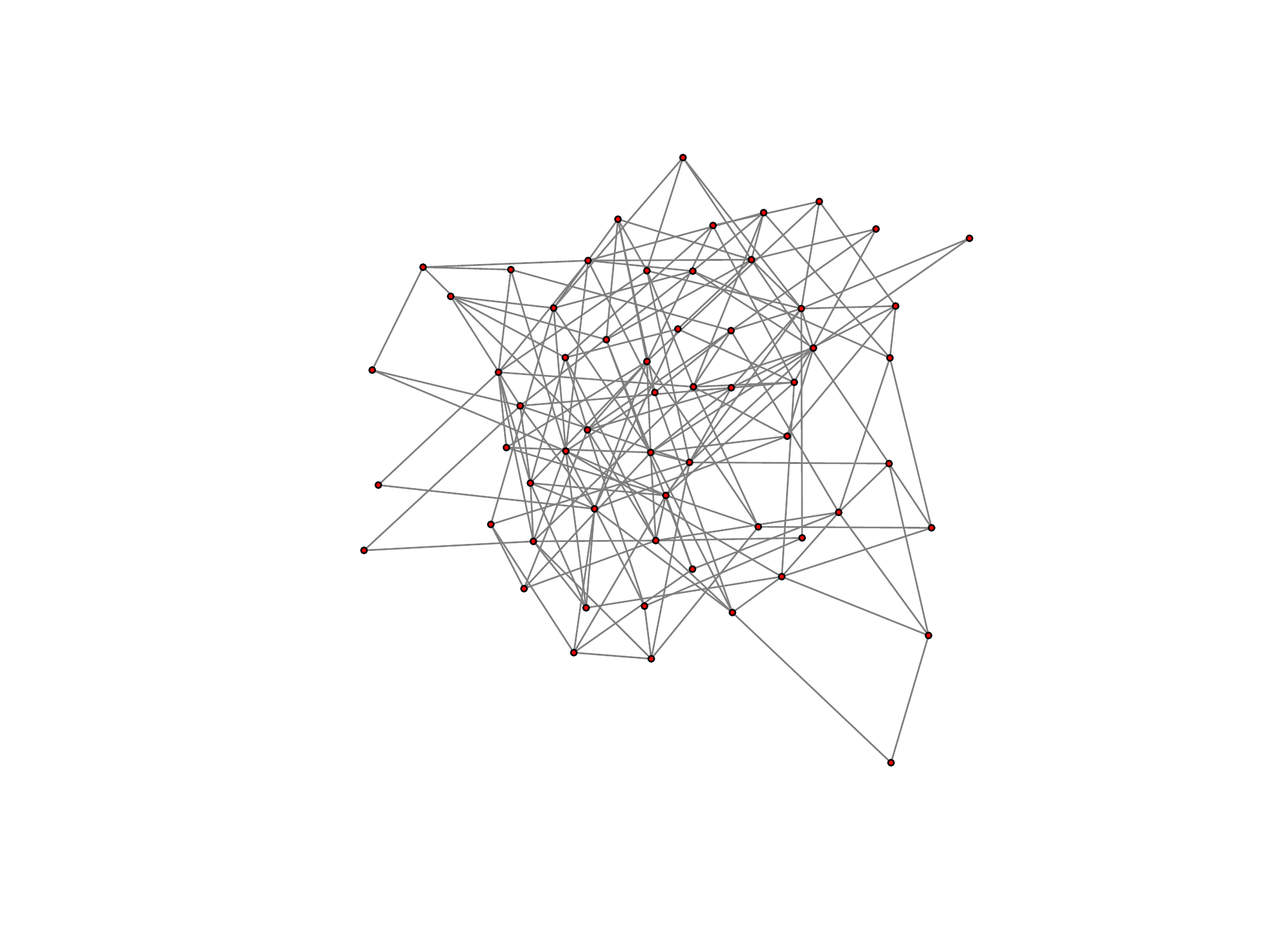}\par
\caption*{ Model  $\text{Rand}$}
\end{multicols}	
\begin{multicols}{2}
	\includegraphics[width=0.9\linewidth]{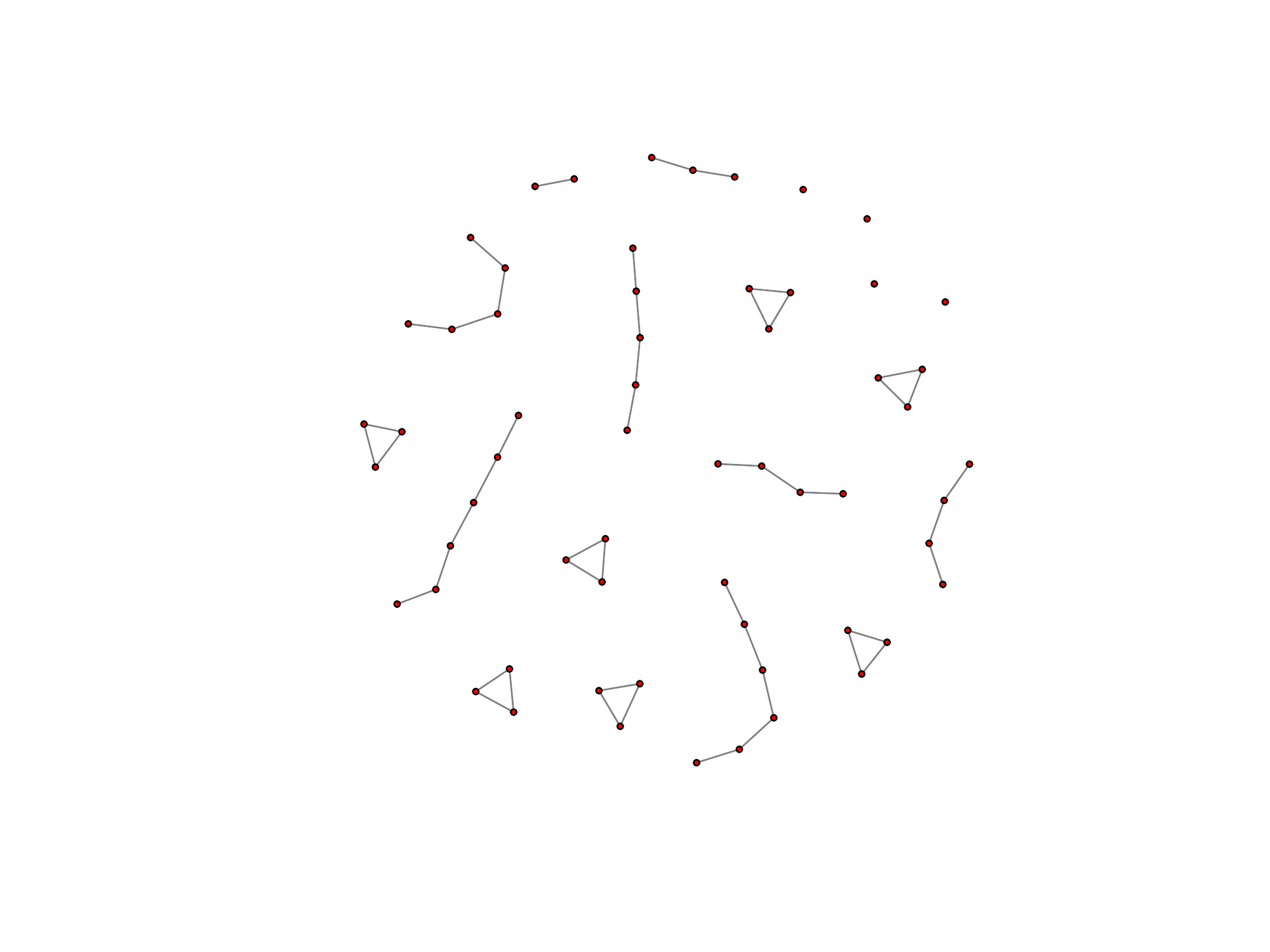}\par
		\caption*{Model  $\text{NN(2)}$}
		\includegraphics[width=0.9\linewidth]{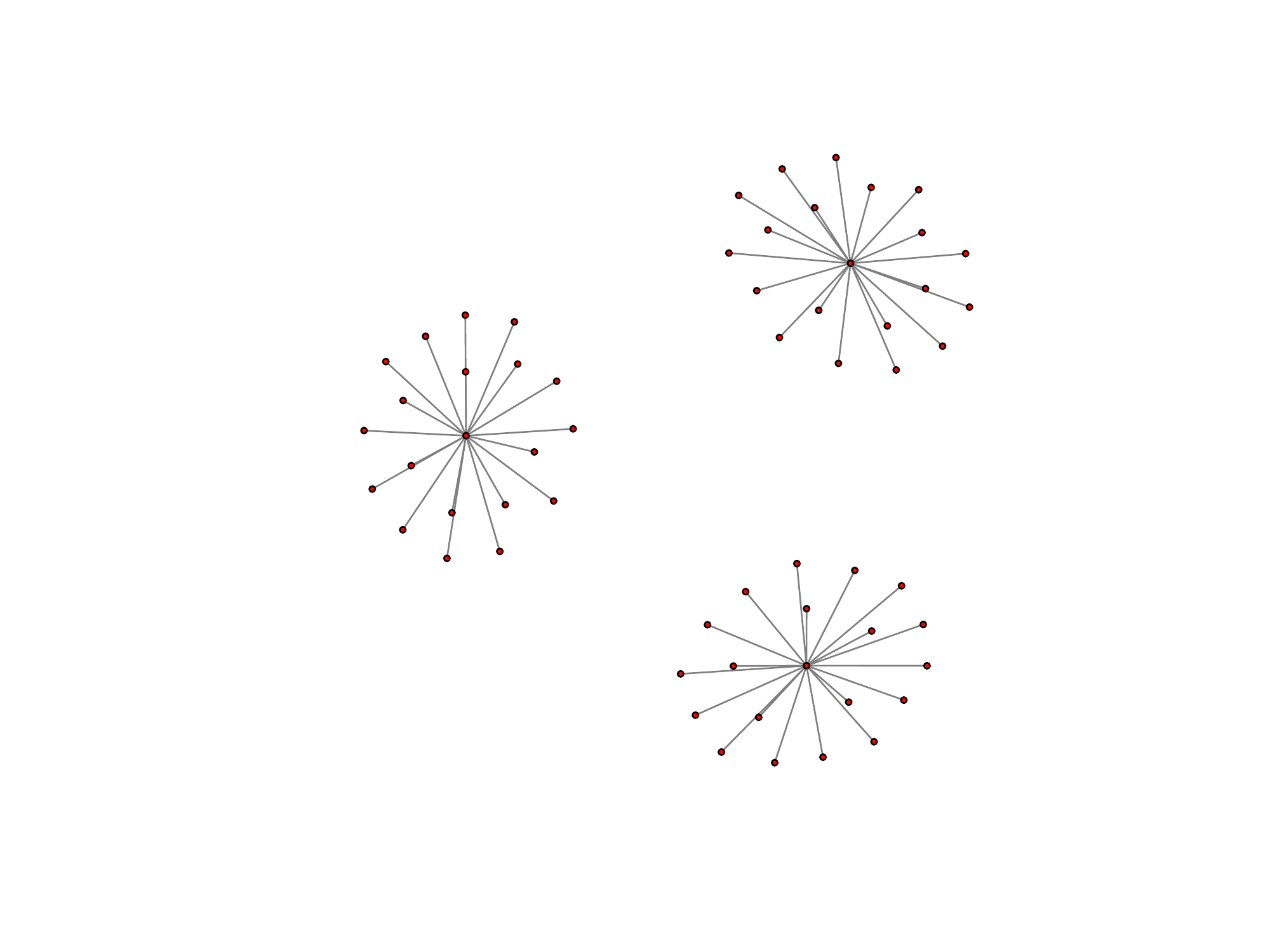}\par
		\caption*{Model  $\text{Hub}$}	
			\end{multicols}	
	\caption{ Graphs of  $\text{AR(1)}$, $\text{BG}$, $\text{Rand}$, $\text{NN}(2)$ and $\text{Hub}$,  graphical models for $p=60$ nodes}
	\label{graphs}
\end{figure}

Figure \ref{graphs} displays graphs  from Models 1-5 with $p=60$.

\medskip

\subsubsection*{Contamination scenarios}
As in  \eqref{contam}, let   $	\boldsymbol{Y} = (I-B) \boldsymbol{X} +B \boldsymbol{Z}$ and consider the following scenarios. 

\begin{itemize}
	\item[i)] \textit{Clean data}.  $\boldsymbol{Y} =\boldsymbol{X}\sim \text{N}(\boldsymbol{0}, \Sigma)$ correspondind to $\epsilon=0$. 
	\item[ii)] \textit{Cellwise or ICM}.   	Here $\boldsymbol{Z}\sim \text{N}(\boldsymbol{\mu}_1, \sigma^2 \Sigma)$  where  $\boldsymbol{\mu}_1=(10,\ldots, 10)^{\prime}$, $\sigma=0\text{.}2 $ and contamination fractions $\epsilon:  0.01, 0.05$, $0.10$. 
\item[iii)] \textit{Casewise or THCM}.  Let	  $\boldsymbol{Z}=\boldsymbol{z}$   with $\boldsymbol{z}=k \boldsymbol{v}$, $\boldsymbol{v}$ is the eigenvector corresponding to the smallest eigenvalue of $\Sigma$ satisfying $ \boldsymbol{v} ^{\prime} \Sigma_{0}^{-1}\boldsymbol{v}=1 $ and $k=100$. 	We consider $\epsilon:  0.05$ and $0.10$.

\end{itemize}

For every  $p$, $\epsilon$ and $\Omega$ model  we generate $N=100$ random samples $\mathbb{Y}_1, \ldots,  \mathbb{Y}_N$, of size $n=100$ of $\boldsymbol{Y}$.

\medskip

\subsubsection*{Precision matrix estimators}

We will compare the performance of the following  estimators of $  \Omega $.

\begin{itemize}
	
	\item[1.] The classical Glasso estimator defined by \eqref{eq2-4}.
		
	\item[2.]  RGlassoQn and RGlassotau. 
\cite{tarr2016robust} estimates a robust initial covariance matrix based on the approach proposed by \cite{gnanadesikan1972robust}. Noting that the covariance of two random variables $X$ and $Y$ can be written as
	\begin{equation}\label{Robust_Scale}
	\text{Cov}(X,Y)= \frac{1}{4 \alpha \beta} \left[ \text{Var}(\alpha X + \beta Y ) -  \text{Var}(\alpha X - Y ) \right],
	\end{equation}  
	where $\alpha = 1/\sqrt{\text{Var}(X)}$ and  $\beta = 1/\sqrt{\text{Var}(Y))}$, a robust estimate of the bivariate covariance $\widehat{\Sigma}_{lj}$ can be obtained by replacing  $ \text{Var}$ in \eqref{Robust_Scale} with a robust variance estimator like $Q_n$ or $\tau$-scale estimators defined by \cite{maronnazamar} and  \cite{rouscroux}.  Based on these robust estimators of the covariance matrix, using \eqref{rglasso},  \cite{tarr2016robust}  derived a robust estimator of $\Omega$, denoted by RGlassoQn  and RGlassotau.   We use the R package \texttt{robustbase} to compute the robust variance estimators $Q_n$ and $\tau$-scale \citep{todorov2009object}.
	\item[3.] RGlassoGauss, RGlassoSpearman  and RGlassoQuadrant. 	\cite{ollecroux} propose a robust estimator $\widehat{\Sigma}_{lj}^{R}$ of the bivariate correlations
	\begin{eqnarray}\label{ollecov}
	\widehat{\Sigma}_{lj}^{R}=\text{scale}(\mathbf{y}_l) \text{scale}(\mathbf{y}_j) r(\mathbf{y}_l, \mathbf{y}_k)
	\end{eqnarray}
	where $r(\cdot)$ and  $\text{scale}(\cdot)$ are  robust  correlation and scale estimators, respectively. For instance $\text{scale}(\cdot)$ is  $\text{Q}_n$ (or the mad) and for $r(\cdot)$  there are different possibilities, like Gaussian rank correlation, Spearman correlation and Quadrant correlation. This proposal leads, using \eqref{rglasso}, to three robust  estimators called RGlassoGauss, RGlassoSpearman and RGlassoQuadrant. 
	
	\item[4.]  Our proposal,  RGlassoWinsor estimator. 	To compute the robust bivariate adjusted correlation estimator defined in steps 1 and 2 of Section \ref{RGlassoWinsor} we use the function ``corhuber'' of the R package \texttt{robustHD}. 

\end{itemize}

In proposals 1) and 3), to make the pairwise correlation matrices positive-definite, we compute the nearest positive definite matrix using the function ``nearPD'' of the R package \texttt{Matrix} \citep{matrix}.
To solve the regularized equation \eqref{rglasso} we use the R-package \texttt{huge}. There are different alternatives to select the optimal regularization parameter and we use 5-fold cross-validation as it is indicated by  \cite{zhao} and \cite{ollecroux}.

\medskip

\subsubsection*{Estimation performance evaluation}

 We wish to evaluate two  different features of the  procedures: (i) their performance as estimates of $\Omega$; and (ii) how well they recover the true graphical model graph.

The { \it numerical performance} of
$\widehat{\Omega}$   is measured by the mean squared error (MSE) defined by the Frobenius norm of the difference between $\Omega$ and the predicted precision matrix  $\widehat{\Omega}$
\begin{eqnarray*}
	m_F=|| \widehat{\Omega}-\Omega ||_{F}=\sqrt{\sum_{ij}|\omega_{ij}-\hat{\omega}_{ij}|^2}
\end{eqnarray*}
and also quantified by  the Kullback-Leibler divergence
\begin{eqnarray*}
	D_{KL} =\frac12 \left(\text{tr} \left\{\widehat{\Omega}\Omega^{-1}\right\} - \text{log}\left\{\text{det}\left[\widehat{\Omega}\Omega^{-1}\right]\right\}-p\right).
\end{eqnarray*}

To evaluate the \textit{graph recovery  or classification performance} we compute the true positive and true negative rates- also called sensitivity and specificity, respectively- defined by 
\begin{eqnarray*}
	\text{TPR}  = \frac{\mathrm{TP}}{\#E}   \text{ and }  
\text{TNR} = \frac{\mathrm{TN}}{\#NE}  
\end{eqnarray*} 
where $E=\left\{(i, j)\in V^2_{-d}:  \omega_{i j}\neq 0\right\}$ is the set of edges,  $NE=\left\{(i, j)\in V^2_{-d}:  \omega_{i j}=0\right\}$ is the set of non connected nodes and
$$
\begin{aligned}
	&\mathrm{TP} =\#\left\{(i, j)\in V^2_{-d}:  \hat{\omega}_{i j} \neq 0 \wedge \omega_{i j}\neq 0\right\},\, \mathrm{TN} =\#\left\{(i, j)\in V^2_{-d}:   \hat{\omega}_{i j} = 0 \wedge \omega_{i j}=0\right\}
\end{aligned}
$$
denotes the size of the sets of true positives and true negatives, respectively.

A related measure is the Matthews  correlation coefficient (MCC) given by
\begin{equation}
	\text{MCC}= \frac{\text{TP} \times \text{TN}   -    \text{FP} \times \text{FN}}{\sqrt{(\text{TP} + \text{FP})(\text{TP} + \text{FN})(\text{TN} + \text{FP})(\text{TN} + \text{FN})}},
\end{equation}
where
$$
\begin{aligned}
	&\mathrm{FP} =\#\left\{(i, j)\in V^2_{-d}:  \hat{\omega}_{i j} \neq 0 \wedge \omega_{i j}=0\right\},\,    \mathrm{FN} =\#\left\{(i, j)\in V^2_{-d}: \hat{\omega}_{i j} = 0 \wedge \omega_{i j}\neq0\right\}  
\end{aligned}
$$
denotes  the number of false positives and false negatives sets, respectively.

 Note that larger values of 	TPR , TNR and  MCC  indicate better performances \citep{fan2009network, baldi2000}.

\begin{figure}[H]
	\captionsetup[subfloat]{farskip=0.5pt,captionskip=0.5pt}
	\centering
		\subfloat[Estimated Model $\text{Rand}$ ]{\includegraphics[width = 1.8in]{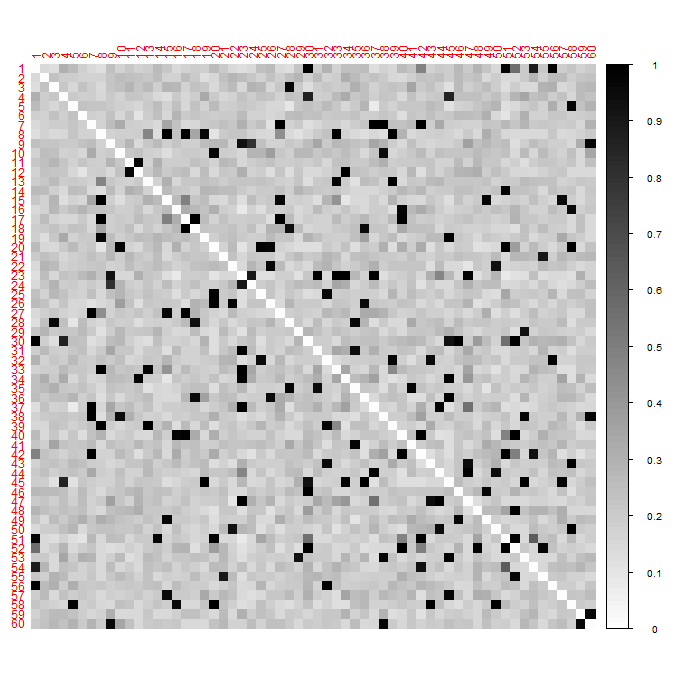}} 
\subfloat[True graph of model $\text{Rand}$ ]{\includegraphics[width = 1.8in]{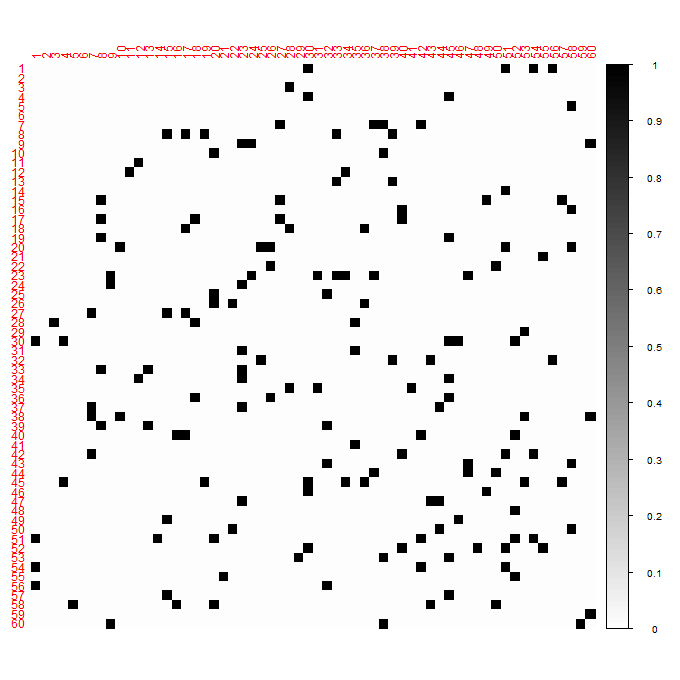}} 	

	\caption{Heatmap for the frequency of  adjacency for each pair of nodes  with $p=60$ and true graph of model  $\text{Rand}$. The axes  display  the graph  $p$-nodes in a given order.}
	\label{heats}
\end{figure}

Heatmaps are  useful   to visualize the graph recovery performance of a given procedure.   As an example, for $p=60$, the axes in the panels of Figure \ref{heats}   display the graph  nodes in a given order.
Panel (a) shows $N=100$ estimated $\text{Rand}$ models by Glasso from simulation replicates where each cell displays a gray level proportional to how frequently the corresponding pair of nodes appear in the estimated graph in the $N$ simulation replicates. So, a white color in a given cell $(i,j)$ means that nodes $i$ and $j$ are never adjacent in the simulated graphs and, a pair of nodes that are always adjacent in the simulated graphs  are represented by a  black coloured  cell.    The heatmap of Panel (a) is compared with the figure of Panel (b) that represents the  graph of true model $\text{Rand}$ where a black or white cell correspond to a pair of connected or non connected nodes, respectively.

\begin{figure}[H]
	\captionsetup[subfloat]{farskip=0.5pt,captionskip=0.5pt}
	\centering
	\subfloat[$\text{AR}(1)$]{\includegraphics[width = 1in]{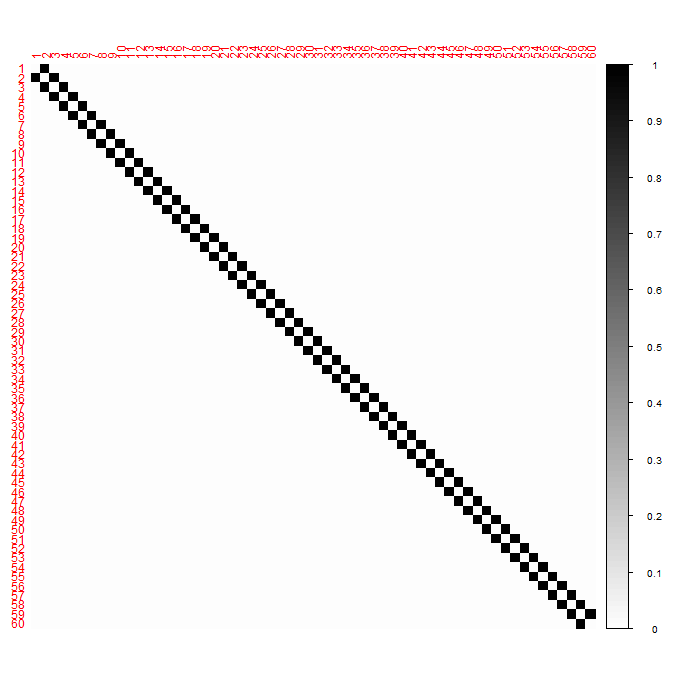}}\hfill
	\subfloat[$\text{BG}$]{\includegraphics[width = 1in]{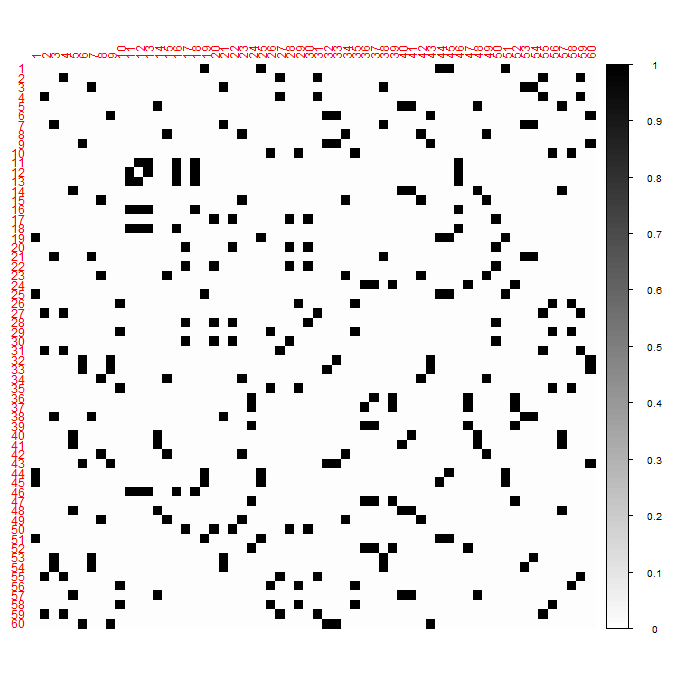}}\hfill
		\subfloat[$\text{Rand}$ ]{\includegraphics[width = 1in]{Model_Random_p_60_n_100_epsilon_0_Heat}}\hfill
	\subfloat[$\text{NN}(2)$ ]{\includegraphics[width = 1in]{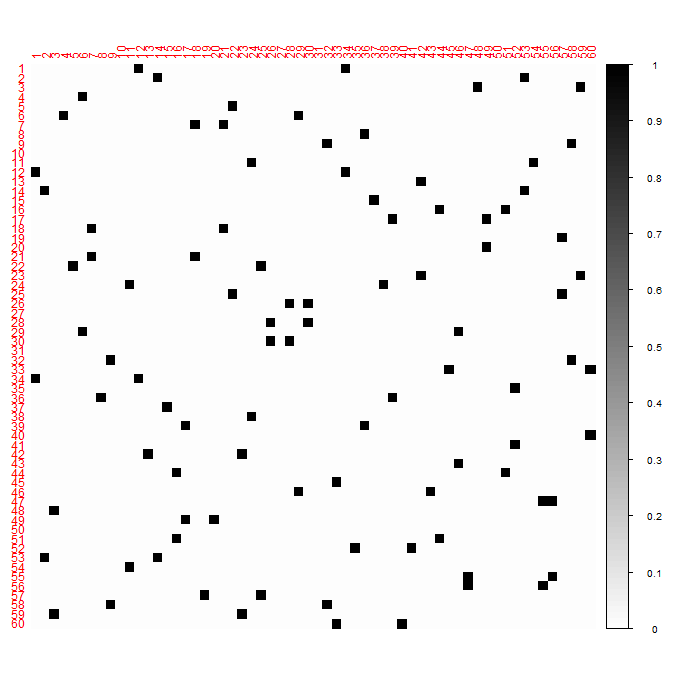}}\hfill
	\subfloat[$\text{Hub}$ ]{\includegraphics[width = 1in]{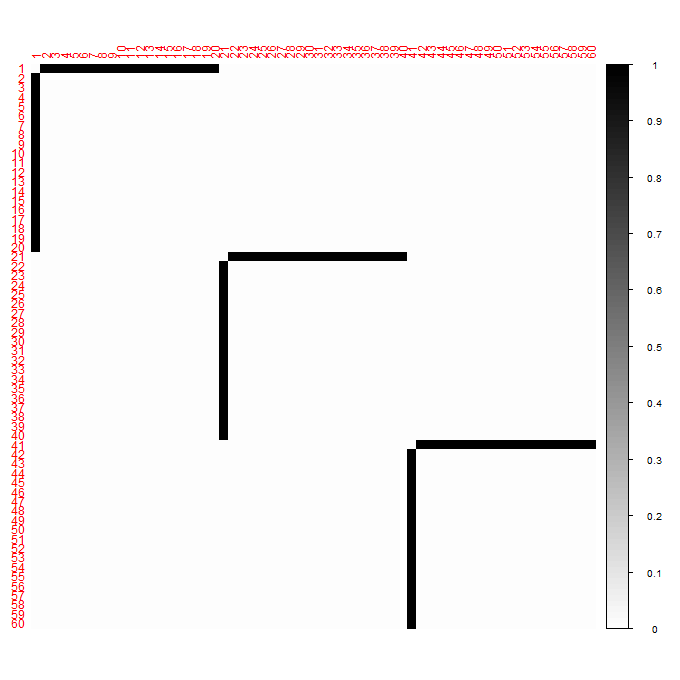}}\hfill
	\caption{True  models graphs  with $p=60$. The axes  display  the graph  $p$-nodes in a given order.}
	\label{Fig_True_Models_p_60}
\end{figure}

Finally,  Figure \ref{Fig_True_Models_p_60}  represent the   five true models graphs.

\subsection{Estimation and graph recovery performances}

In this section we analyze the numerical and graph recovery performances of the estimation of the different GGM, represented by   its precision matrix $\Omega$, for clean data and under both contamination scenarios. To abreviate we will group  RGlassoQn and RGlassotau  under the name of Group 2, and RGlassoGauss, RGlassoSpearman  and RGlassoQuadrant  will be named  Group 3.

To fix some ideas we first focus on  the estimation results for the  $\text{AR}(1)$ model. Tables \ref{num_performance_se_ar1_p_60} to \ref{MCC_se_AR1_p_200}   show the estimation performance under ICM and Tables \ref{num_performance_se_ar1_p_60_THCM} to \ref{MCC_se_ar1_p_200_THCM}  in the Appendix B, under THCM.




	\begin{table}[H]
		\mbox{}\hfill
		\begin{minipage}[t]{.48\linewidth}
			\centering{ 
				
				\scalebox{0.46}{
					\begin{tabular}{l|rr|rr|rr|rr}
						\hline
						& \multicolumn{2}{c|}{$\epsilon=0$} & \multicolumn{2}{c|}{$\epsilon=0.01$} & \multicolumn{2}{c|}{$\epsilon=0.05$} & \multicolumn{2}{c}{$\epsilon=0.10$} \\
						
						\hline
						& $D_{KL}$ & $m_{F}$ & $D_{KL}$ & $m_{F}$ & $D_{KL}$ & $m_{F}$ & $D_{KL}$ & $m_{F}$  \\
						\hline
						RGlassoWinsor & 5.223 & 4.365 & 5.739 & 4.689 & 8.336 & 5.798 & 13.089 & 7.053\\
						& (0.039) & (0.024) & (0.046) & (0.024) & (0.069) & (0.024) & (0.102) & (0.022)\\
						Glasso & 4.232 & 4.063 & 30.007 & 8.960 & 76.465 & 10.828 & 103.705 & 11.241\\
						& (0.028)  & (0.021)  & (0.339)  & (0.030)  & (0.240)  & (0.005)  & (0.187)  & (0.002)\\
						RGlasso$Q_n$ & 8.118 & 5.830 & 10.314 & 6.477 & 29.604 & 9.100 & 57.220 & 10.406\\
						& (0.080)  & (0.027)  & (0.131)  & (0.034)  & (0.450)  & (0.040)  & (0.428)  & (0.013)\\
						RGlassoTau & 5.687 & 4.737 & 7.071 & 5.373 & 24.044 & 8.548 & 71.010 & 10.742\\
						& (0.044)  & (0.023)  & (0.070)  & (0.030)  & (0.501)  & (0.054)  & (0.593)  & (0.013)\\
						RGlassoGauss & 4.595 & 4.278 & 5.732 & 4.854 & 10.540 & 6.516 & 16.375 & 7.697\\
						& (0.033)  & (0.021)  & (0.048)  & (0.025)  & (0.080)  & (0.022)  & (0.095)  & (0.016)\\
						RGlassoSpearman & 4.968 & 4.478 & 5.889 & 4.936 & 10.303 & 6.455 & 16.274 & 7.670\\
						& (0.042)  & (0.025)  & (0.049)  & (0.025)  & (0.076)  & (0.021)  & (0.096)  & (0.016)\\
						RGlassoQuad & 10.545 & 6.560 & 11.682 & 6.843 & 16.151 & 7.693 & 22.521 & 8.515\\
						& (0.073)  & (0.020)  & (0.093)  & (0.023)  & (0.109)  & (0.019)  & (0.130)  & (0.015)\\
						\hline
				\end{tabular}}
			}
			
			\caption{Model $\text{AR}(1)$ under ICM.    Comparison  of  means and standard deviations (in brackets) of  $\text{m}_F$  and $D_{KL}$ over $N=100$ replicates. $p=60$, $n=100$.} 
			
			\label{num_performance_se_ar1_p_60}%

		\end{minipage}\hfill
		\begin{minipage}[t]{.48\linewidth}
			\centering
			\scalebox{0.46}{
				\begin{tabular}{l|rr|rr|rr|rr}
					\hline
					& \multicolumn{2}{c|}{$\epsilon=0$} & \multicolumn{2}{c|}{$\epsilon=0.01$} & \multicolumn{2}{c|}{$\epsilon=0.05$} & \multicolumn{2}{c}{$\epsilon=0.10$}  \\
					
					\hline
					& $D_{KL}$ & $m_{F}$ & $D_{KL}$ & $m_{F}$ & $D_{KL}$ & $m_{F}$ & $D_{KL}$ & $m_{F}$   \\
					\hline
					RGlassoWinsor & 23.481 & 9.541 & 25.191 & 9.986 & 33.635 & 11.564 & 49.845 & 13.611\\
					& (0.125) &(0.038) & (0.126) &(0.035) & (0.193) & (0.034) & (0.233) & (0.025)\\
					Glasso & 19.469 & 8.784 & 94.502 & 16.112 & 257.189 & 19.867 & 350.501 & 20.628\\
					& (0.085) & (0.044) & (0.467) & (0.025) & (0.576) & (0.007) & (0.409) & (0.003)\\
					RGlasso$Q_n$ & 63.930 & 14.998 & 78.345 & 15.856 & 149.160 & 18.274 & 255.526 & 19.914\\
					& (0.335) & (0.024) & (0.465) & (0.025) & (0.746) & (0.017) & (1.105) & (0.011)\\
					RGlassoTau & 29.859 & 11.082 & 38.890 & 12.440 & 135.343 & 17.916 & 306.799 & 20.354\\
					&(0.169) & (0.031) & (0.250) &(0.034) & (0.834) & (0.022) & (1.245) & (0.009)\\
					RGlassoGauss & 21.102 & 9.295 & 25.163 & 10.216 & 41.158 & 12.711 & 60.306 & 14.603\\
					& (0.107) & (0.039) & (0.106) & (0.028) & (0.180) & (0.026) & (0.217) & (0.019)\\
					RGlassoSpearman & 23.131 & 9.794 & 26.254 & 10.438 & 41.564 & 12.756 & 61.643 & 14.686\\
					& (0.116) & (0.035) & (0.110) & (0.029) & (0.189) & (0.027) & (0.222) & (0.019)\\
					RGlassoQuad & 48.458 & 13.612 & 53.087 & 14.044 & 70.442 & 15.322 & 91.330 & 16.416\\
					& (0.181) & (0.021) & (0.214) & (0.022) & (0.254) & (0.019) & (0.336) & (0.017)\\
					\hline
			\end{tabular}}

		\caption{Model $\text{AR}(1)$ under ICM. Comparison  of  means and standard deviations (in brackets) of  $\text{m}_F$  and $D_{KL}$ over $N=100$ replicates. $p=200, n=100$. } 
		\label{num_performance_se_ar1_p_200}%
		
	\end{minipage}\hfill
	\mbox{}
	
\end{table}

In terms of numerical performance, Glasso is slightly better than other methods for clean data, but it is clearly non robust under both  contamination models for all positive contamination fractions.  In both contamination models, our proposal, RGlassoWinsor, has the best numerical performance.  Note that the mean squared error, $m_F$, and the Kullback-Leibler divergence,  $D_{KL}$, grow when the dimension $p$ increases, for  both, clean  and contaminated data. $D_{KL}$ and  $m_F$ are higher for  cellwise contamination model than  the casewise contamination model.

\begin{table}[H]
	\mbox{}\hfill
	\begin{minipage}[t]{.48\linewidth}
		\centering{ 
			
			\scalebox{0.5}{
				\begin{tabular}{l|rr|rr|rr|rr}
					\hline
					& \multicolumn{2}{c|}{$\epsilon=0$} & \multicolumn{2}{c|}{$\epsilon=0.01$} & \multicolumn{2}{c|}{$\epsilon=0.05$} & \multicolumn{2}{c}{$\epsilon=0.10$}  \\
					
					\hline
					& TPR & TNR   & TPR & TNR & TPR & TNR & TPR & TNR     \\
					\hline
					RGlassoWinsor & 0.991 & 0.842 & 0.989 & 0.853 & 0.962 & 0.887 & 0.799 & 0.934\\
					& (0.001) & (0.003) & (0.001) & (0.003) & (0.003) & (0.003) & (0.009) & (0.003)\\
					Glasso & 0.997 & 0.816 & 0.140 & 0.986 & 0.033 & 0.985 & 0.045 & 0.968\\
					& (0.001) & (0.003) & (0.015) & (0.001) & (0.003) & (0.001) & (0.003) & (0.001)\\
					RGlasso$Q_n$ & 0.865 & 0.952 & 0.786 & 0.968 & 0.131 & 0.998 & 0.003 & 1.000\\
					& (0.006) & (0.002) & (0.011) & (0.002) & (0.018) & (0.000) & (0.001) & (0.000)\\
					RGlassoTau & 0.960 & 0.875 & 0.928 & 0.902 & 0.397 & 0.988 & 0.012 & 1.000\\
					& (0.003) & (0.002) & (0.004) & (0.002) & (0.025) & (0.001) & (0.001) & (0.000)\\
					RGlassoGauss & 0.996 & 0.834 & 0.987 & 0.835 & 0.888 & 0.882 & 0.655 & 0.924\\
					& (0.001) & (0.003) & (0.002) & (0.003) & (0.005) & (0.003) & (0.008) & (0.003)\\
					RGlassoSpearman & 0.990 & 0.850 & 0.983 & 0.851 & 0.914 & 0.890 & 0.718 & 0.927\\
					& (0.001) & (0.003) & (0.002) & (0.003) & (0.004) & (0.002) & (0.008) & (0.002)\\
					RGlassoQuad & 0.729 & 0.934 & 0.688 & 0.941 & 0.553 & 0.957 & 0.339 & 0.978\\
					& (0.008) & (0.002) & (0.010) & (0.002) & (0.012) & (0.002) & (0.012) & (0.001)\\
					\hline
			\end{tabular}}
		}
		
		\caption{ Model $\text{AR}(1)$ under ICM.  Comparison  of  means and standard deviations (in brackets) of  TPR and TNR  over $N=100$ replicates. $p=60, n=100$. } 
		\label{TPTN_ar1_p_60}%
	\end{minipage}\hfill
		\begin{minipage}[t]{.48\linewidth}
	\centering
	\scalebox{0.5}{		
		\begin{tabular}{l|r|r|r|r}
			\hline
			& \multicolumn{1}{c|}{$\epsilon=0$} & \multicolumn{1}{c|}{$\epsilon=0.01$} & \multicolumn{1}{c|}{$\epsilon=0.05$} & \multicolumn{1}{c}{$\epsilon=0.10$}   \\
			
			\hline
			RGlassoWinsor & 0.389 & 0.402 & 0.444 & 0.471\\
			& (0.004) &(0.004) & (0.005) & (0.005)\\
			Glasso & 0.360 & 0.147 & 0.026 & 0.013\\
			& (0.003) & (0.013) & (0.003) & (0.003)\\
			RGlasso$Q_n$ & 0.567 & 0.594 & 0.261 & 0.023\\
			& (0.005) & (0.005) & (0.017) & (0.005)\\
			RGlassoTau & 0.420 & 0.455 & 0.441 & 0.056\\
			& (0.004) & (0.004) & (0.014) & (0.006\\
			RGlassoGauss & 0.380 & 0.378 & 0.399 & 0.360\\
			& (0.003) & (0.003) & (0.004) & (0.004)\\
			RGlassoSpearman & 0.398 & 0.397 & 0.424 & 0.401\\
			& (0.004) & (0.004) & (0.003) & (0.004)\\
			RGlassoQuad & 0.428 & 0.425 & 0.391 & 0.331\\
			& (0.005) & (0.004) & (0.005) & (0.006)\\
			\hline
	\end{tabular}}
	\vspace{0.15cm}
	\centering 
	\caption{    Model $\text{AR}(1)$ under ICM.    Comparison  of  means and standard deviations (in brackets) of  MCC over $N=100$ replicates. $p=60$, $n=100$.} 
	\label{MCC_se_ar1_p_60}%
	
\end{minipage}\hfill
\mbox{}

\end{table}

\begin{table}[H]
	\mbox{}\hfill
	\begin{minipage}[t]{.48\linewidth}
		\centering{ 
			
			\scalebox{0.5}{
				\begin{tabular}{l|rr|rr|rr|rr}
					\hline
					& \multicolumn{2}{c|}{$\epsilon=0$} & \multicolumn{2}{c|}{$\epsilon=0.01$} & \multicolumn{2}{c|}{$\epsilon=0.05$} & \multicolumn{2}{c}{$\epsilon=0.10$}  \\
					
					\hline
					& TPR & TNR & TPR & TNR & TPR & TNR & TPR & TNR   \\
					
					\hline
					RGlassoWinsor & 0.971 & 0.932 & 0.961 & 0.941 & 0.904 & 0.958 & 0.620 & 0.982\\
					& (0.002) & (0.002) & (0.002) & (0.002) & (0.003) & (0.001) & (0.009) & (0.001)\\
					Glasso & 0.986 & 0.914 & 0.207 & 0.984 & 0.024 & 0.989 & 0.028 & 0.983\\
					& (0.001) & (0.002) & (0.010) & (0.000) & (0.001) & (0.000) & (0.001) & (0.001)\\
					RGlasso$Q_n$ & 0.041 & 1.000 & 0.021 & 1.000 & 0.001 & 1.000 & 0.000 & 1.000\\
					& (0.003) & (0.000) & (0.002) & (0.000) & (0.000) & (0.000) & (0.000) & (0.000)\\
					RGlassoTau & 0.867 & 0.963 & 0.796 & 0.976 & 0.017 & 1.000 & 0.002 & 1.000\\
					& (0.004) & (0.001) & (0.006) & (0.001) & (0.002) & (0.000) & (0.000) & (0.000)\\
					RGlassoGauss & 0.978 & 0.931 & 0.959 & 0.937 & 0.794 & 0.952 & 0.487 & 0.973\\
					& (0.001) & (0.002) & (0.002) & (0.001) & (0.005) & (0.001) & (0.008) & (0.001)\\
					RGlassoSpearman & 0.967 & 0.935 & 0.957 & 0.936 & 0.842 & 0.952 & 0.565 & 0.973\\
					& (0.002) & (0.002) & (0.002) & (0.001) & (0.004) & (0.001) & (0.008) & (0.001)\\
					RGlassoQuad & 0.648 & 0.965 & 0.607 & 0.970 & 0.455 & 0.979 & 0.304 & 0.985\\
					& (0.005) & (0.001) & (0.006) & (0.001) & (0.009) & (0.001) & (0.008) & (0.001)\\
					\hline
			\end{tabular}}
		}
		
		\caption{ Model $\text{AR}(1)$ under ICM.  Comparison  of  means and standard deviations (in brackets) of  TPR and TNR  over $N=100$ replicates. $p=200, n=100$. } 
		\label{TPTN_p_200}%
	\end{minipage}\hfill
\begin{minipage}[t]{.48\linewidth}
	\centering
	\scalebox{0.5}{
		\begin{tabular}{l|r|r|r|r}
			\hline
			& \multicolumn{1}{c|}{$\epsilon=0$} & \multicolumn{1}{c|}{$\epsilon=0.01$} & \multicolumn{1}{c|}{$\epsilon=0.05$} & \multicolumn{1}{c}{$\epsilon=0.10$}  \\
			
			\hline
			RGlassoWinsor & 0.342 & 0.365 & 0.405 & 0.405\\
			& (0.004) &  (0.004) &  (0.005) &  (0.003)\\
			Glasso & 0.312 & 0.137 & 0.013 & 0.008\\
			& (0.004) &  (0.005) &  (0.001) &  (0.001)\\
			RGlasso$Q_n$ & 0.185 & 0.121 & 0.016 & 0.001\\
			& (0.008) &  (0.008) &  (0.003) &  (0.001) \\
			RGlassoTau & 0.404 & 0.451 & 0.108 & 0.017\\
			& (0.004) &  (0.004) &  (0.006) &  (0.003)\\
			RGlassoGauss & 0.345 & 0.350 & 0.332 & 0.269\\
			& (0.004) &  (0.003) &  (0.003) &  (0.002)\\
			RGlassoSpearman & 0.348 & 0.347 & 0.352 & 0.307\\
			& (0.003) &  (0.003) &  (0.004) &  (0.002)\\
			RGlassoQuad & 0.311 & 0.317 & 0.284 & 0.227\\
			& (0.002) &  (0.003) &  (0.003) &  (0.003)\\
			\hline
	\end{tabular}}
	\vspace{0.15cm}
	\centering 
	\caption{  Model $\text{AR}(1)$ under ICM.   Comparison  of  means and standard deviations (in brackets) of  MCC over $N=100$ replicates. $p=200$, $n=100$.} 
	\label{MCC_se_AR1_p_200}%
	
\end{minipage}\hfill
\mbox{}
\end{table}

Even when there is no contamination and considering the MCC as graph recovery measure, the performance of Glasso is poor.   Under cellwise contamination model,   MCC means produced by RGlassoWinsor and those produced by Group 3  estimators remain almost constant and  even slightly increase  when the contamination fractions  increase as it shown in Tables 	\ref{MCC_se_ar1_p_60} and 	\ref{MCC_se_AR1_p_200}.  Conversely,  MCC means of the  Group 2 estimators dramatically decrease when the contamination fraction $\epsilon$ increases. A better explanation  can be found by looking at the Tables 	\ref{TPTN_ar1_p_60} and \ref{TPTN_p_200},  while the mean of TPR remains relatively high for RGlassoWinsor and for the estimators of Group 3,   the mean of TPR goes to zero for the estimators of Group 2. Note that, although not so extreme,  a similar phenomena occurs under   THCM, as  shown in Tables \ref{TPTN_ar1_p_200_THCM} and 
\ref{MCC_se_ar1_p_200_THCM}.

  \begin{figure}[H]
 	\captionsetup[subfloat]{farskip=0.pt,captionskip=0.5pt}
 	\centering
 	\subfloat[True $\text{AR}(1)$]{\includegraphics[width = 1.5in]{Model_AR1_p_60_n_100_epsilon_0_Heat}} \hspace{0.5cm}
 	\subfloat[Glasso ]{\includegraphics[width = 1.5in]{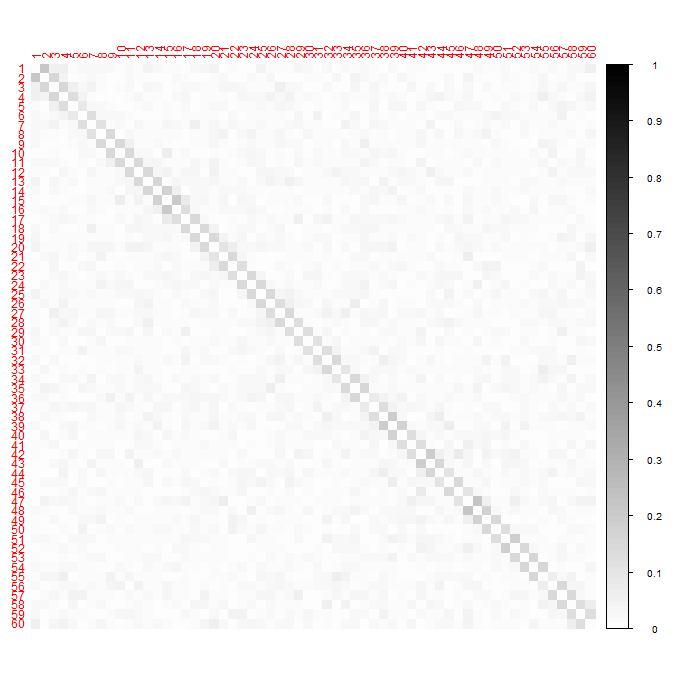}}  \hspace{0.5cm}
 	\subfloat[RGlassoWinsor]{\includegraphics[width = 1.5in]{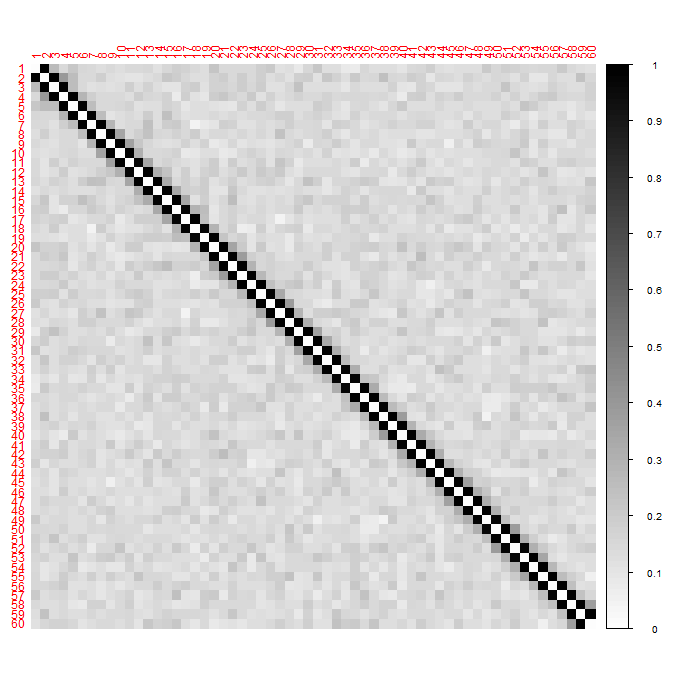}} \hspace{0.5cm}
  	\caption{Graph of true model $\text{AR}(1)$ and heatmaps for the frequency of adjancency for each pair of nodes over $ N=100$ replicates.  $p=60$ and $n=100$. ICM with $\epsilon=0.01$. The axes  display  the graph  $p$-nodes in a given order.}
 	\label{Fig_Glasso_Winsor_AR1_p_60_epsilon_1}
 \end{figure}  


 \begin{figure}[H]
 	\captionsetup[subfloat]{farskip=0.pt,captionskip=0.5pt}
 	\centering
 	\subfloat[True $\text{AR}(1)$]{\includegraphics[width = 1.5in]{Model_AR1_p_60_n_100_epsilon_0_Heat}} \hspace{0.5cm}
 	\subfloat[Glasso  ]{\includegraphics[width = 1.5in]{Glasso_Cor_AR1_p_60_n_100_epsilon_1_Heat}}  \hspace{0.5cm}
  	\subfloat[RGlassoWinsor ]{\includegraphics[width = 1.5in]{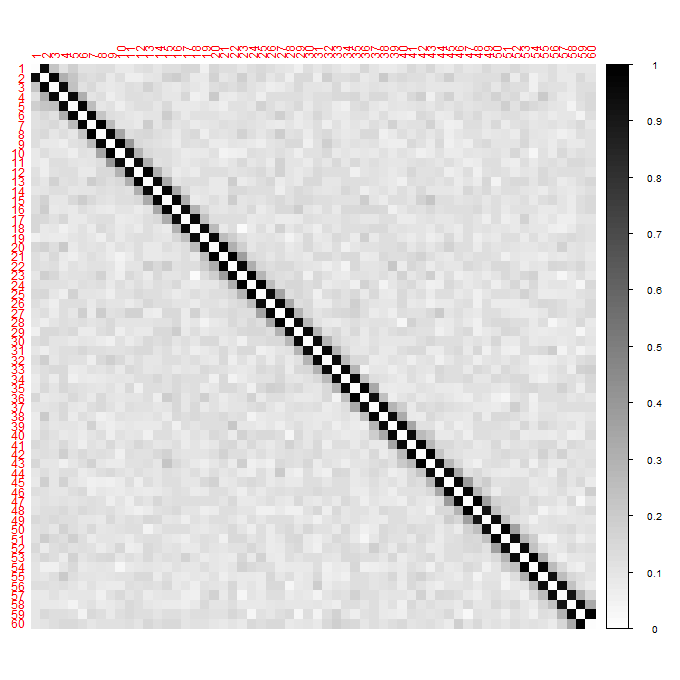}} \hspace{0.5cm}
 		\caption{Graph of true model $\text{AR}(1)$ and heatmaps for the frequency of adjancency for each pair of nodes over $ N=100$ replicates.  $p=60$ and $n=100$. ICM with $\epsilon=0.05$. The axes  display  the graph  $p$-nodes in a given order.}
 	
 	\label{Fig_Glasso_Winsor_AR1_p_60_epsilon_5}
 \end{figure}

  \begin{figure}[H]
 	\captionsetup[subfloat]{farskip=0.pt,captionskip=0.5pt}
 	\centering
 	\subfloat[True $\text{AR}(1)$]{\includegraphics[width = 1.5in]{Model_AR1_p_60_n_100_epsilon_0_Heat}} \hspace{0.5cm}
 	\subfloat[Glasso ]{\includegraphics[width = 1.5in]{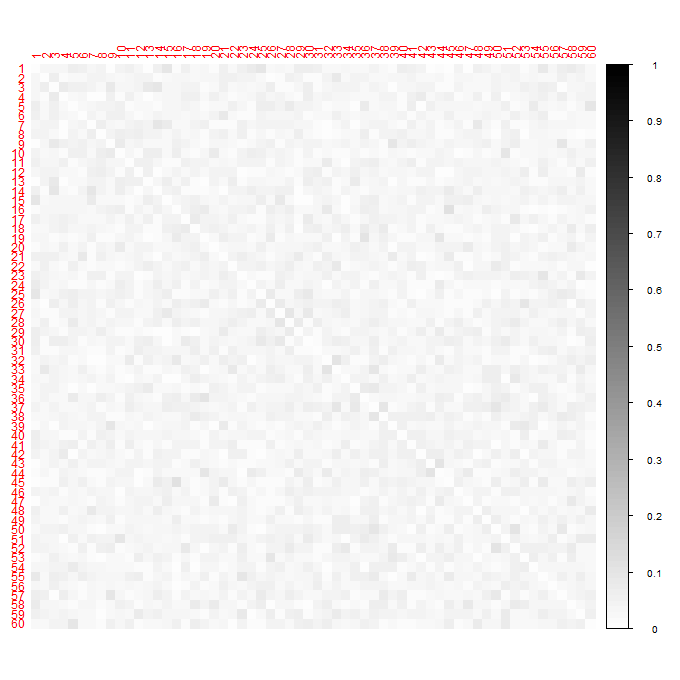}}  \hspace{0.5cm}
  	\subfloat[RGlassoWinsor]{\includegraphics[width = 1.5in]{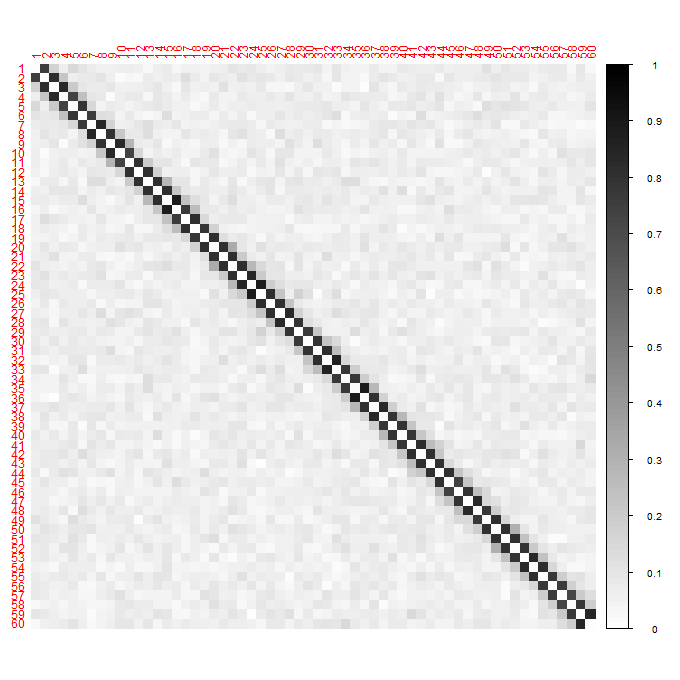}} \hspace{0.5cm}
  	\caption{Graph of true model $\text{AR}(1)$ and heatmaps for the frequency of adjancency for each pair of nodes over $ N=100$   replicates.  $p=60$ and $n=100$. ICM with $\epsilon=0.10$. The axes  display  the graph  $p$-nodes in a given order.}
 	\label{Fig_Glasso_Winsor_AR1_p_60_epsilon_10}
 \end{figure}  

 Figures \ref{Fig_Glasso_Winsor_AR1_p_60_epsilon_1}, \ref{Fig_Glasso_Winsor_AR1_p_60_epsilon_5} and \ref{Fig_Glasso_Winsor_AR1_p_60_epsilon_10} show the performance of  Glasso and RGlassoWinsor for contaminated data under ICM. Notice that for contaminated data, Glasso cannot recover the true set of edges, introducing  a large number of false negatives. Although RGlassoWinsor introduces false positives, it better recovers the true set of edges.

In the following paragraphs we set general conclusions  about the behavior of the estimators for all analyzed $\Omega$ models,  based on
Tables  \ref{num_performance_se_ar1_p_60} to \ref{MCC_se_AR1_p_200} of this section, Tables   \ref{num_performance_se_Block_p_60} to \ref{MCC_se_hub_p_200}  in Appendix A (ICM),  Tables \ref{num_performance_se_ar1_p_60_THCM} 
 to  \ref{MCC_se__p_200_THCM}  in Appendix B (THCM).
 
 Tables  \ref{num_performance_ranking_ICM_p_60}
to
\ref{graph_recovery_performance_ranking_THCM_p_200} below report the average ranks for all the compared estimation methods, evaluated across all the considered  precision matrix models. Rank 1 and rank 7 corresponds to the best and worst performing method, respectively.   The   average ranks of the best two performing methods are shown in bold face.


For $\epsilon=0$, Glasso performs slightly better than the other estimators  and shows  a non-robust performance, being the worst ranked for contaminated data.  For almost all contamination scenarios RGlassoWinsor is the best ranked and the estimators of Group 2, specially RGlassoSpearman and RGlassoGaus, have the closest rankings. Note that for dimension  $p=200$  RGlassoSpearman has a slightly better average rank than RGlassoWinsor, under THCM.

\begin{table}[H]
\mbox{}\hfill
\begin{minipage}[t]{.48\linewidth}
\centering{ 
			
\scalebox{0.6}{
			\begin{tabular}{l|r|r|r|r|r|r|r|r}
				\hline
		\multirow{2}{*}{}	&	\multicolumn{2}{c|}{$\epsilon=0$} & \multicolumn{2}{c|}{$\epsilon=0.01$} & \multicolumn{2}{c|}{$\epsilon=0.05$} & \multicolumn{2}{c}{$\epsilon=0.10$}  
		\\
		\cline{2-9}
		&$m_F$  & $D_{KL}$ & $m_F$  & $D_{KL}$ & $m_F$  & $D_{KL}$ & $m_F$  & $D_{KL}$  \\
				
				\hline
RGlassoWinsor   & 3.4 & 4 & {\bf 1.6} & {\bf 2 }& {\bf 1} & {\bf 1}& {\bf 1}& {\bf 1}\\
Glasso          & {\bf 1}  & {\bf 1} & 6.4 & 6.4 & 7 & 7 & 6.6& 6.6 \\
RGlasso$Q_n$    & 6.4 & 6.4 & 5.8 & 6 & 6 & 6 & 5.2 & 5.2 \\
RGlassoTau      &5.4 &  5.4 & 4.2 & 4.8 & 5 & 5 & 6.2 &6.2 \\
RGlassoGauss    & {\bf 2.2}    &{\bf  2} & {\bf 2.2} & {\bf 1.2} & {\bf 2.4} & {\bf 2.6} &  2.6 & {\bf 2.4} \\
RGlassoSpearman & 3.4 & 3  & 2.6 & 2.8  & 2.6 & 2.4& {\bf 2.4} &   2.6  \\
RGlassoQuad     & 6.2 & 6.2 & 5.2 & 5 & 4 & 4 & 4& 4\\			 
\hline 
		\end{tabular}}
	
		}
		
		\caption{   Average rank of the estimation methods based on $m_F$ and $	D_{KL}$   under ICM. $p=60$,  $n=100$.} 
		
		\label{num_performance_ranking_ICM_p_60}%

	\end{minipage}\hfill
	\begin{minipage}[t]{.48\linewidth}
		\centering
		\scalebox{0.6}{
				\begin{tabular}{l|r|r|r|r}
					\hline
					&	\multicolumn{1}{c|}{$\epsilon=0$} & \multicolumn{1}{c|}{$\epsilon=0.01$} & \multicolumn{1}{c|}{$\epsilon=0.05$} & \multicolumn{1}{c}{$\epsilon=0.10$}   \\
					
					\hline
					RGlassoWinsor   & \bf{ 3.2} &  \bf{ 2.2} & \bf{ 1.2} & \bf{ 1.4 }\\
					Glasso          & 4.4       &  6.4  &  7  &  6.8 \\
					RGlasso$Q_n$    & \bf{ 3.4}   &  3.4  & 5.8   & 5.4 \\
					RGlassoTau      & 3.8  & \bf{3.2}   &  4.6  & 6.2 \\
					RGlassoGauss    & 3.8  & 4.2  &   3.5 &  3 \\
					RGlassoSpearman  & 4.2  & 3.6  &   \bf{2 } & \bf{ 1.6 }\\
					RGlassoQuad     & 5.6  & 5  &  4.2  & 4 \\			 
					\hline
			\end{tabular}}
		
		\caption{  Average rank of the estimation methods based on MCC under ICM. $p=60$,     $n=100$.} 
		
		\label{graph_recovery_performance_ranking_ICM_p_60}%
		
	\end{minipage}\hfill
	\mbox{}
	
\end{table}

\begin{table}[H]
	\mbox{}\hfill
	\begin{minipage}[t]{.48\linewidth}
		\centering{ 
			
			\scalebox{0.6}{
				\begin{tabular}{l|r|r|r|r|r|r|r|r}
				\hline
				\multirow{2}{*}{}	&	\multicolumn{2}{c|}{$\epsilon=0$} & \multicolumn{2}{c|}{$\epsilon=0.01$} & \multicolumn{2}{c|}{$\epsilon=0.05$} & \multicolumn{2}{c}{$\epsilon=0.10$}  
				\\
				\cline{2-9}
				&$m_F$  & $D_{KL}$ & $m_F$  & $D_{KL}$ & $m_F$  & $D_{KL}$ & $m_F$  & $D_{KL}$  \\
				
				\hline
					RGlassoWinsor   & 3.6  &3.4  & {\bf  1.6}  & {\bf 1.8} & {\bf  1} & {\bf  1} &  {\bf  1.2} & {\bf  1}\\
					Glasso          & {\bf  1} & {\bf 1.4}    &  6.2 & 6.2 & 6.6 &6.6 &6.6& 6.8 \\
					RGlasso$Q_n$    &  6.8  &6.8&  6 & 6.2   & 5.6   & 5.8& 5.4 & 5\\
					RGlassoTau      &  5.6  &5.2 &  5  & 5&  5.4  & 5.2& 6 &6.2\\
					RGlassoGauss    &  {\bf  2}   &{\bf 2}  &  {\bf  1.6} & {\bf 1.2} & {\bf 2 }& {\bf  2} & {\bf 2} & {\bf  2.2}\\
					RGlassoSpearman &  3.4  &3.4  & 2.8 &3  & 3   & 3& 2.8 &2.8 \\
					RGlassoQuad    & 5.6  &5.8   & 4.8  & 4.6&4.4 &4.4 &  4&4\\		 
					\hline
			\end{tabular}}
			
		}
		
		\caption{  Average rank of the estimation methods based on $m_F$ and $D_{KL}$ under ICM. $p=200$,  $n=100$.} 
		
		\label{num_performance_ranking_ICM_p_200}%

	\end{minipage}\hfill
	\begin{minipage}[t]{.48\linewidth}
		\centering
		\scalebox{0.6}{
			\begin{tabular}{l|r|r|r|r}
				\hline
				&	\multicolumn{1}{c|}{$\epsilon=0$} & \multicolumn{1}{c|}{$\epsilon=0.01$} & \multicolumn{1}{c|}{$\epsilon=0.05$} & \multicolumn{1}{c}{$\epsilon=0.10$}   \\
				
				\hline
				RGlassoWinsor    &  3.2     & {\bf 1.4}     & {\bf 1}   & \bf{ 1.6}  \\
				Glasso           & 3.6    & 6.2    &  6.6  &  6.4  \\
				RGlasso$Q_n$     &  4.2   & 5.4    & 4.4   & 4.6   \\
				RGlassoTau       &  {\bf 3 }    & 2.6  &  4.2  &   6 \\
				RGlassoGauss     &  {\bf 2.4}   & {\bf 3}    &   3.2 &   3 \\
				RGlassoSpearman  &  3.4  & 3.4    & {\bf  2.2}  & {\bf  1.8}  \\
				RGlassoQuad      &  6.2  &   5.2  &  4.6  &   4.4 \\			 
				\hline
		\end{tabular}}
		
		\caption{ Average rank of the estimation methods based on MCC under ICM. $p=200$,  $n=100$.} 
		
		\label{graph_recovery_performance_ranking_ICM_p_200}%
		
	\end{minipage}\hfill
	\mbox{}
	
\end{table}

\begin{table}[H]
	\mbox{}\hfill
	\begin{minipage}[t]{.48\linewidth}
		\centering{ 
			
			\scalebox{0.6}{	\begin{tabular}{l|r|r|r|r}
					\hline
					\multirow{2}{*}{}	&	\multicolumn{2}{c|}{$\epsilon=0.05$} & \multicolumn{2}{c}{$\epsilon=0.10$}  
					\\
					\cline{2-5}
					&$m_F$  & $D_{KL}$ & $m_F$  & $D_{KL}$  \\
					
					\hline
					
					RGlassoWinsor   &{\bf 1.2}       & {\bf 1.8} &   {\bf 1.4} &  {\bf 2}     \\
					Glasso           &5.2       &7&       7  & 7 \\
					RGlasso$Q_n$    &  6   &5.8   & 5.8      &  5.8  \\
					RGlassoTau      &  4.6   &4.4  &  4.6  &     4.4 \\
					RGlassoGauss     &  {\bf 2.2}  &2.4   &   {\bf  2.2}     & 2.4 \\
					RGlassoSpearman   & 2.6  &{\bf 1.8 }   &   2.4    &{\bf 1.8}   \\
					RGlassoQuad     & 4.8  &4.8    &    4.6     & 4.6 \\
					\hline
				\end{tabular}
			}
			
		}
		
		\caption{  Average rank of the estimation methods based on $m_F$ and $D_{KL}$ under THCM. $p=60$,  $n=100$.} 
		
		\label{num_performance_ranking_THCM_p_60}%

	\end{minipage}\hfill
	\begin{minipage}[t]{.48\linewidth}
		\centering
		\scalebox{0.6}{	\begin{tabular}{l|r|r }
				\hline
				&	\multicolumn{1}{c|}{$\epsilon=0.05$} & \multicolumn{1}{c}{$\epsilon=0.10$}     \\
				
				\hline
				RGlassoWinsor   &  {\bf 2.6}    & {\bf 2 }        \\
				Glasso           & 6.8    &  7        \\
				RGlasso$Q_n$    &   3.6   &  4.6       \\
				RGlassoTau       &  3    & 3.4         \\
				RGlassoGauss     &  4.6    & 4.8          \\
				RGlassoSpearman   & {\bf 2.6}     & {\bf 2.6}       \\
				RGlassoQuad     &  3.8    & 3.6         \\	 
				\hline
			\end{tabular}
			}
		
		\caption{ Average rank of the estimation methods based on MCC under THCM. $p=60$,  $n=100$.} 
		
		\label{graph_recovery_performance_ranking_THCM_p_60}%
		
	\end{minipage}\hfill
	\mbox{}
	
\end{table}

\begin{table}[H]
	\mbox{}\hfill
	\begin{minipage}[t]{.48\linewidth}
		\centering{ 
			
			\scalebox{0.6}{	\begin{tabular}{l|r|r|r|r}
					\hline
					\multirow{2}{*}{}	&	\multicolumn{2}{c|}{$\epsilon=0.05$} & \multicolumn{2}{c}{$\epsilon=0.10$}  
					\\
					\cline{2-5}
					&$m_F$  & $D_{KL}$ & $m_F$  & $D_{KL}$  \\
					
					\hline
					RGlassoWinsor   &  {\bf 2}    &{\bf 1}&     {\bf 1.4}      & {\bf 2} \\
					Glasso           &  4.8    & 7&  5.2       &6.2   \\
					RGlasso$Q_n$    &   7   &  6&    6.6  &6.4    \\
					RGlassoTau      &  4.8    &  5&  4.8   &5.2     \\
					RGlassoGauss    &   {\bf 1} &{\bf 2}  &      {\bf 1.4  }   & {\bf  1.4} \\
					RGlassoSpearman   & 3     &   3&     2.8  &2.4   \\
					RGlassoQuad     &    5.2  & 4&     4.6    &4.2  \\
					\hline
				\end{tabular}
			}
			
		}
		
		\caption{  Average rank of the estimation methods based on $m_F$ and $D_{KL}$ under THCM. $p=200$,  $n=100$.} 
		
		\label{num_performance_ranking_THCM_p_200}%

	\end{minipage}\hfill
	\begin{minipage}[t]{.48\linewidth}
		\centering
		\scalebox{0.6}{	\begin{tabular}{l|r|r }
				\hline
				&	\multicolumn{1}{c|}{$\epsilon=0.05$} & \multicolumn{1}{c}{$\epsilon=0.10$}     \\
				
				\hline
				RGlassoWinsor   & {\bf 2.4 }    & {\bf 2.2 }         \\
			Glasso           &  5.4    &     5.6      \\
			RGlasso$Q_n$    & 6.4     &     5.6      \\
			RGlassoTau      &  3    &     3.8      \\
			RGlassoGauss    &  3.2    &   2.6       \\
			RGlassoSpearman   &  {\bf 2}    &     {\bf 2.2 }     \\
			RGlassoQuad     &    5  &   5.2        \\
				\hline
			\end{tabular}
		}
		
		\caption{ Average rank of the estimation methods based on MCC under THCM. $p=200$,  $n=100$.} 
		
		\label{graph_recovery_performance_ranking_THCM_p_200}%
		
	\end{minipage}\hfill
	\mbox{}
	
\end{table}

\section{Real data example}
\label{realdata}

In preoperative  chemotherapy when all  invasive cancer cells are eradicated, the patient is said to have reached the state of {\it pathological complete response}, abbreviated as pCR.  This pCR  is associated with  the long-term cancer-free survival of a person. On the contrary,  residual disease (RD) indicates that the disease has not been eradicated. Measurements of the  expression level (activity)  of genes  may be able to predict  if a patient can reach a pCR. 

\cite{hess2006pharmacogenomic} use normalized gene expression data of patients in stages I-III of breast cancer,    to identify patients that may achieve pCR  under  preoperative chemotherapy. Their data base has 22283 gene expression levels for 133 patients, with 34 pCR and 99 RD.  \cite{hess2006pharmacogenomic} and \cite{natowicz2008prediction} identify 26 important genes for predicting survival and response to adjuvant chemotherapy. Following \cite{ambroise2009inferring} and \cite{tang2021robust}, we estimate the precision matrix for the 26 key genes on the two classes pCR   and RD.

\cite{raymaekers2020handling}  proposed a  method that detect cellwise outliers, implemented in the R package  \texttt{cellWise} \citep{cellWise}. The function ``cellHandler'' of \texttt{cellWise}  flags cellwise outliers in the data matrix, based on  robust estimates  of the mean $\boldsymbol{\mu}$ and   covariance matrix $\Sigma$ with 0.95\% as cutoff used in the detection of cellwise outliers.  We compare the performance of RGlassoWinsor and RGlassoGauss because  both have shown similar rankings. 
Using the sample median and a  robust estimates of $\Sigma$ provided by Winsorization, $\widehat{\Sigma}^{W}$,  and   Gaussian rank correlations, $\widehat{\Sigma}^G$, (see \eqref{covwins} and \eqref{ollecov}) we first detect outliers in the data set.

\begin{figure} [H]
	\centering 
	\begin{multicols}{2}
		\includegraphics[width=\linewidth]{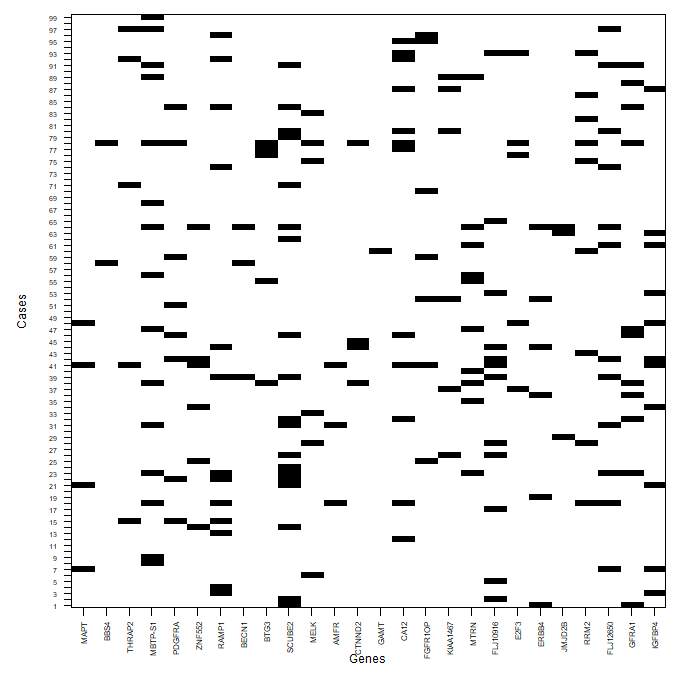}\par
		\caption*{ a)}
		\includegraphics[width=\linewidth]{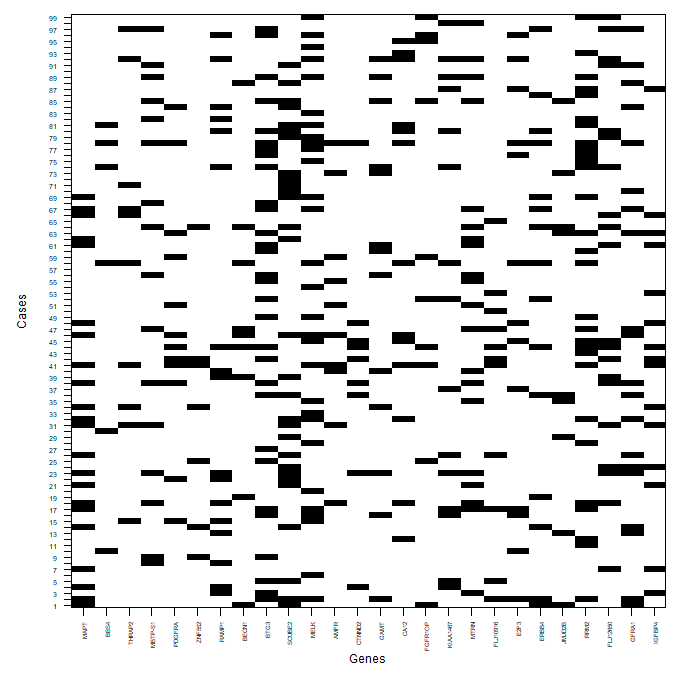}\par
		\caption*{b)}		
	\end{multicols} 
	\caption{ Cellwise outliers detected by ``cellHandler'' for the  RD class based on a) $\widehat{\Sigma}^W$ and b) $\widehat{\Sigma}^G$. A black colored cell indicates an outlier.}
\label{Fig_outl_detct_RD}
\end{figure}

Figure \ref{Fig_outl_detct_RD} illustrates cellwise outliers flagged by ``cellHandler'' based on both robust covariance estimates for the RD class. The rows represent the patients or cases and the columns represent the variables or genes expressions. A black colored cell indicates that its value is an outlier.

 Of the total of 2574 ($99\times 26$) cells of the data matrix of the RD class, 384 (15\%) are contaminated according to ``cellHandler'' based on $\widehat{\Sigma}^W$. The first five most contaminated variables correspond to genes RRM2,  SCUBE2, MELK, CA12 and BTG3.  Using $\widehat{\Sigma}^G$, 409 (almost 16 \%) are flagged as contaminated and the first five most contaminated variables corresponds to genes BTG3, RRM2, MELK, SCUBE2, MAPT.

 A similar procedure shows that for the pCR group, using $\widehat{\Sigma}^W$, of the total of 884 (34x26) cells of the data matrix, ``cellHandler'' flags 166  (19\%)   cells as contaminated and the first five most contaminated variables correspond to genes PDGFRA, CA12, SCUBE2, BBS4 and IGFBP4.   Using $\widehat{\Sigma}^G$, 108  (almost 12 \%) cells are flagged as contaminated and the first five most contaminated variables corresponds to genes CA12, SCUBE2, IGFBP4, KIAA1467 and MTRN.

\begin{figure}[h]
	\center
	\centering
	\subfloat[Glasso]{\includegraphics[width = 2.5in]{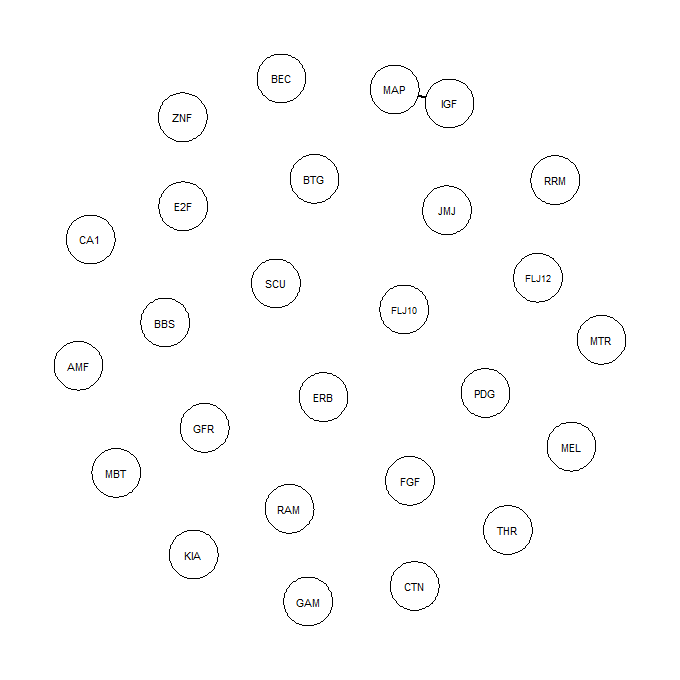}}\hfill 
	\subfloat[RGlassoWinsor]{\includegraphics[width = 2.5in]{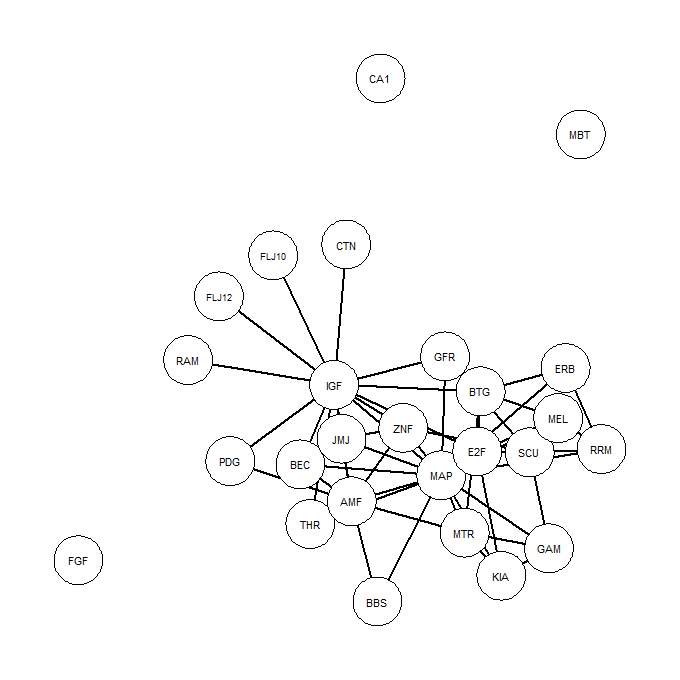}}\hfill\\
	\subfloat[RGlassoTau]{\includegraphics[width = 2.5in]{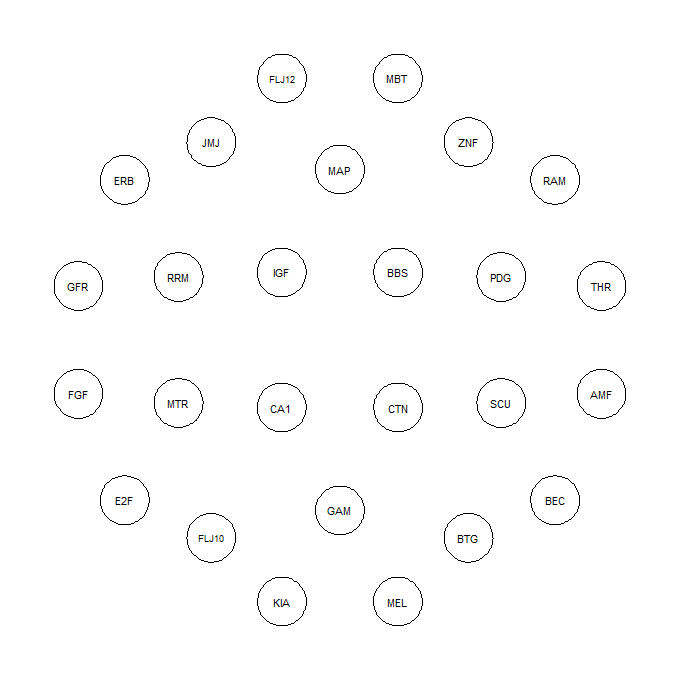}}\hfill
	\subfloat[RGlassoGauss]{\includegraphics[width = 2.5in]{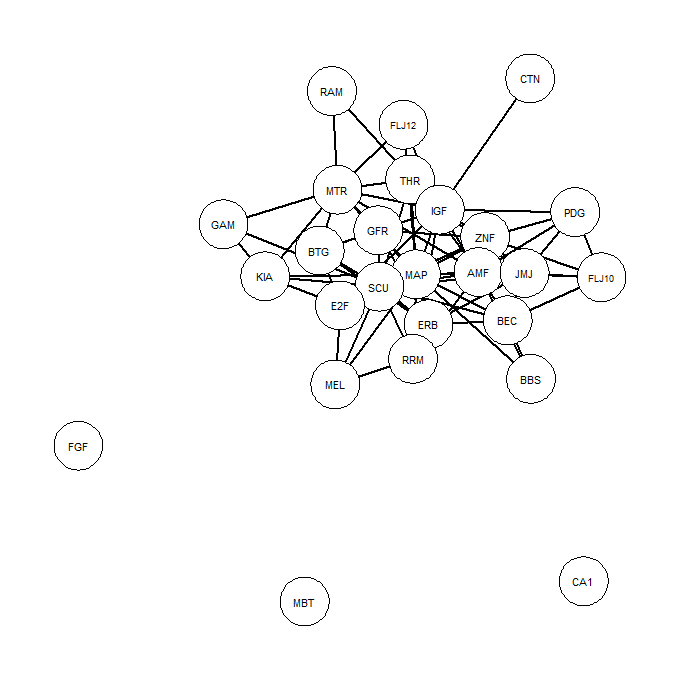}}\hfill
	\caption{Estimated graph of the GGM for the 26 genes corresponding to RD class. } 
	\label{Fig_Breast_Cancer_RD}
\end{figure}

\begin{figure}[h]
	\center
	\centering
	\subfloat[Glasso]{\includegraphics[width = 2.5in]{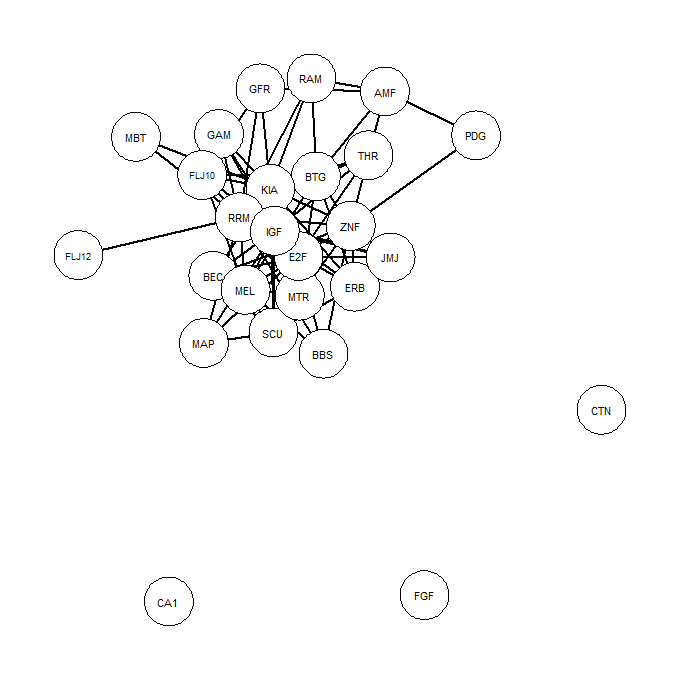}}\hfill 
	\subfloat[RGlassoWinsor]{\includegraphics[width = 2.5in]{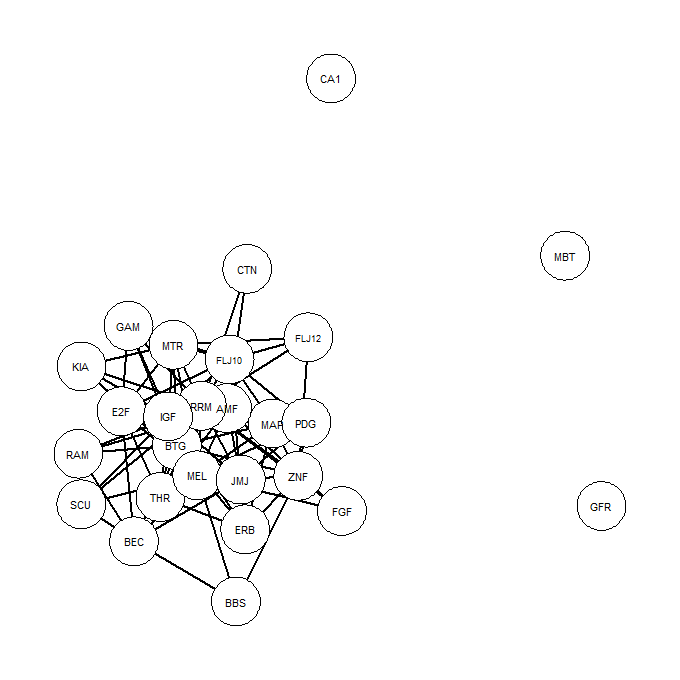}}\hfill\\
	\subfloat[RGlassoTau]{\includegraphics[width = 2.5in]{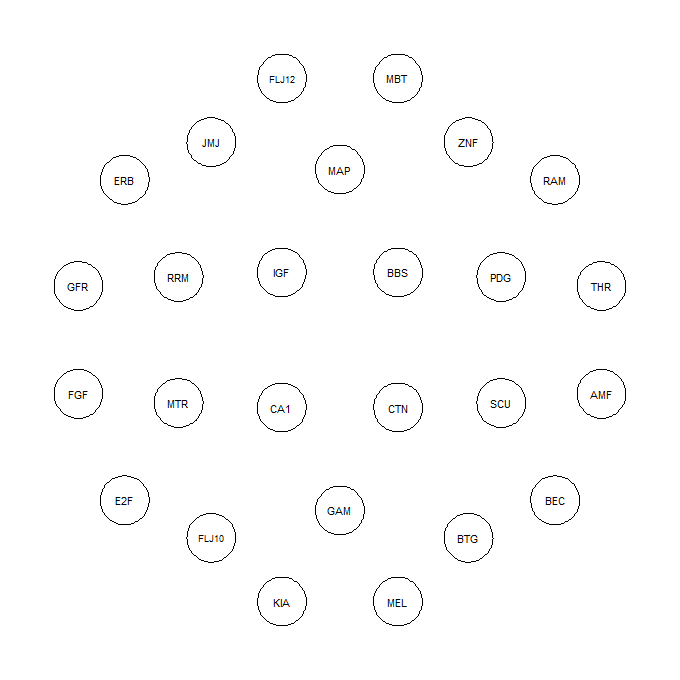}}\hfill
	\subfloat[RGlassoGauss]{\includegraphics[width = 2.5in]{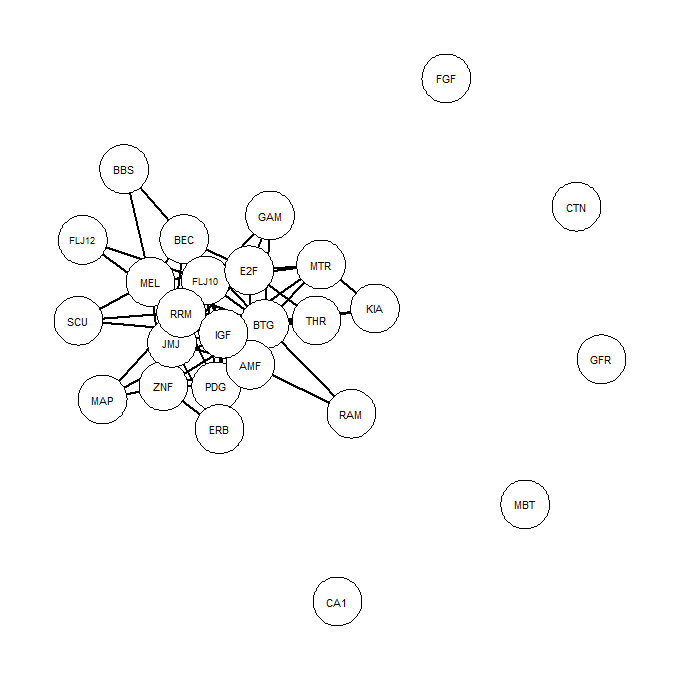}}\hfill
	\caption{Estimated graph of the GGM for the 26 genes corresponding to  PCR class. } 
	\label{Fig_Breast_Cancer_pCR}
\end{figure}

  Figures \ref{Fig_Breast_Cancer_RD} and \ref{Fig_Breast_Cancer_pCR}   display the resulting network obtained using Glasso, RGlassoWinsor, RGlassoTau and RGlassoGauss, the latter two representing Groups 2 and 3 of procedures. Table \ref{Table_Breast_Cancer_Networks} exhibits the estimated network density for the 26 genes for each class, for all procedures, using a regularization parameter  chosen by 5-fold cross-validation.  

Excluding the estimated networks by RGlasso$Q_n$ and RGlassoTau, the undirected graphs differ according to the class membership which may suggest that  genes regulation differs according the participants response to the treatment \citep{ambroise2009inferring}.

In the pCR class, RGlassoWinsor produce a less sparse network than Glasso and RGlassoSpearman, but a similar structure. But, in the RD class  while Glasso and RGlassoTau does not detect any conditional relationship between nodes (genes), RGlassoWinsor  and the procedures of Group 3 detect several edges between genes. 



\begin{table}[h]
	\center

	\begin{tabular}{lrr}
		\hline
		& \multicolumn{1}{c}{pCR class} & \multicolumn{1}{c}{RD class} \\ \hline
		RGlassoWinsor & 0.280  & 0.169 \\
		Glasso & 0.243 & 0.003 \\
		RGlasso$Q_n$ & 0.000 & 0.000 \\
		RGlassoTau & 0.000 & 0.000 \\
		RGlassoGauss & 0.203 & 0.249 \\
		RGlassoSpearman & 0.197 & 0.237 \\
		RGlassoQuad & 0.117 & 0.234 \\ \hline
	\end{tabular}
	\caption{Estimated network density for the 26 genes from breast cancer gene expressions data.} 
\label{Table_Breast_Cancer_Networks}
\end{table}

\section{\textcolor{blue}{Concluding remarks}}
\label{conclrema}

This paper introduces a  new robust graphical lasso  procedure called  RGlassoWinsor based on  adjusted bivariate Winsorization estimation of the covariance matrix for high-dimension covariance selection or precision matrix estimation.

RGlassoWinsor is compared with the currently existing  robust estimators of the precision matrix, introduced by \cite{tarr2016robust} and  \cite{ollecroux}, by using different performance measures regarding graph recovery and sparse estimation of the precision matrix. 

Our proposal shows a good performance for all the precision models, dimensions and contamination scenarios considered in this research. For clean data Glasso is slightly better than other methods, but it is clearly non robust.  Under contamination and for almost all performance measures, our proposal, RGlassoWinsor,  has the best overall performance.

Moreover our  procedure  attains the maximum finite sample breakdown point of 0.5 under cellwise contamination.

Finally, we demonstrate  the usefulness of   RGlassoWinsor   in an application to the analysis of breast cancer data.

 \section*{Acknowledgements}
 
The authors thanks the generous support of NSERC, Canada, Universidad Carlos III de Madrid,  Espa\~na and Universidad Nacional de R\'io Cuarto, Argentina.


\newpage

\section*{Appendix A}

This section contains numerical and  performance classification performance  results for the seven precision matrix estimators applied to   $\text{BG}$,  $\text{NN}(2)$, $\text{Rand}$ and $\text{Hub}$ models. We compare two scenarios $p=60$ and $p=200$ with $n=100$ under ICM.  

\begin{figure}[H]
	\captionsetup[subfloat]{farskip=0.5pt,captionskip=0.5pt}
	\centering
	\subfloat[Glasso-$\epsilon=0$]{\includegraphics[width = 1in]{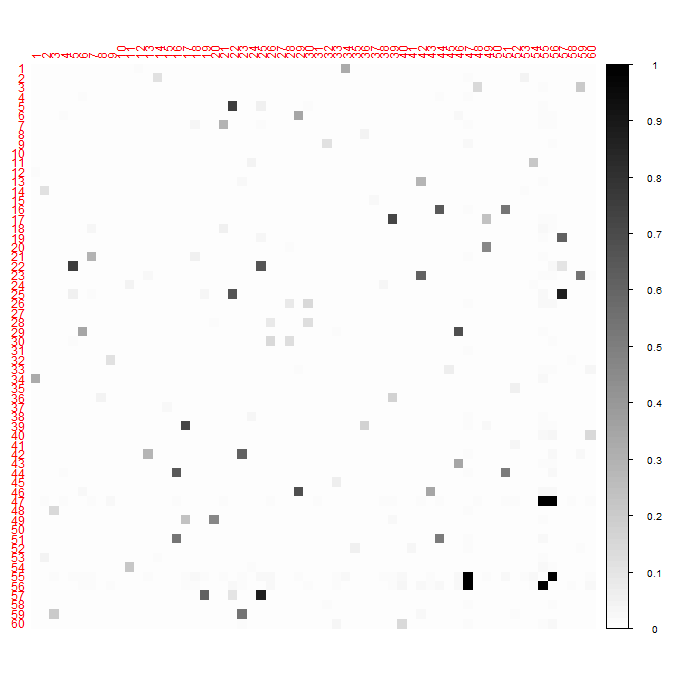}}\hfill
		\subfloat[RGlassoWinsor-$\epsilon=0$]{\includegraphics[width = 1in]{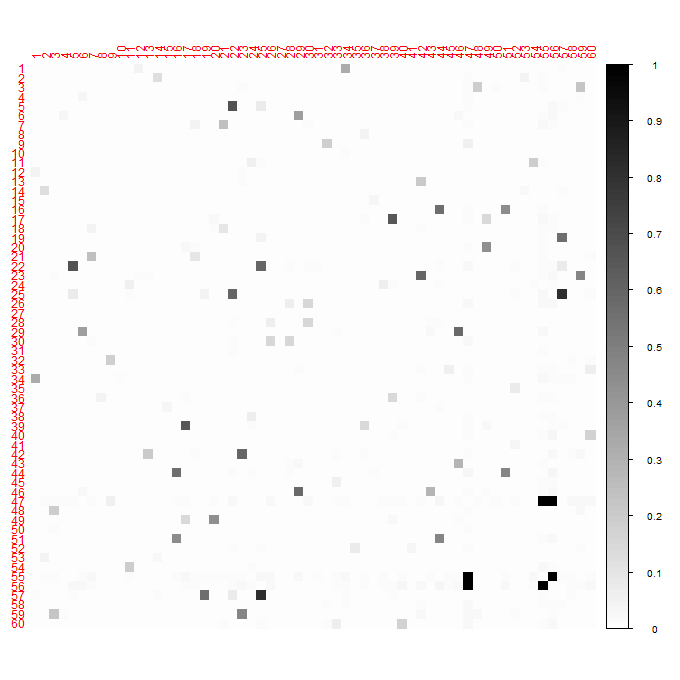}}\hfill 
	\subfloat[Glasso-$\epsilon=0.01$]{\includegraphics[width = 1in]{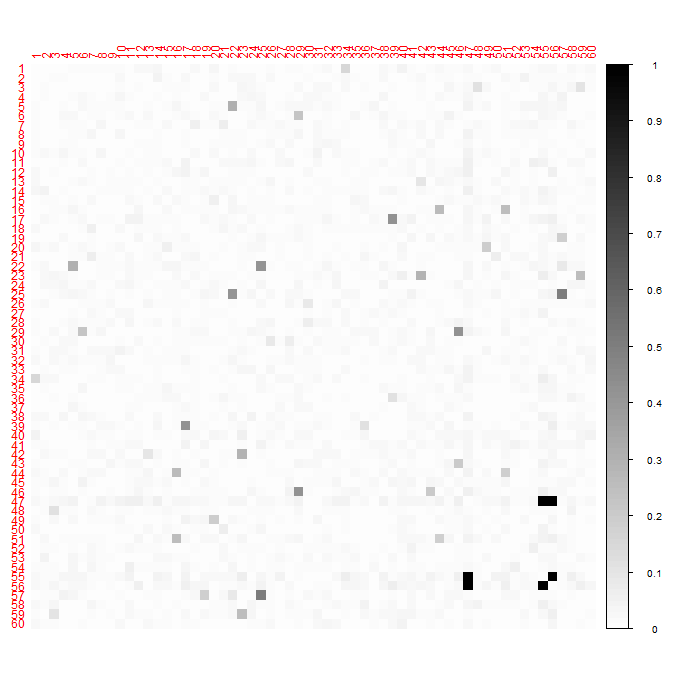}}\hfill 
 	\subfloat[RGlassoWinsor-$\epsilon=0.01$]{\includegraphics[width = 1in]{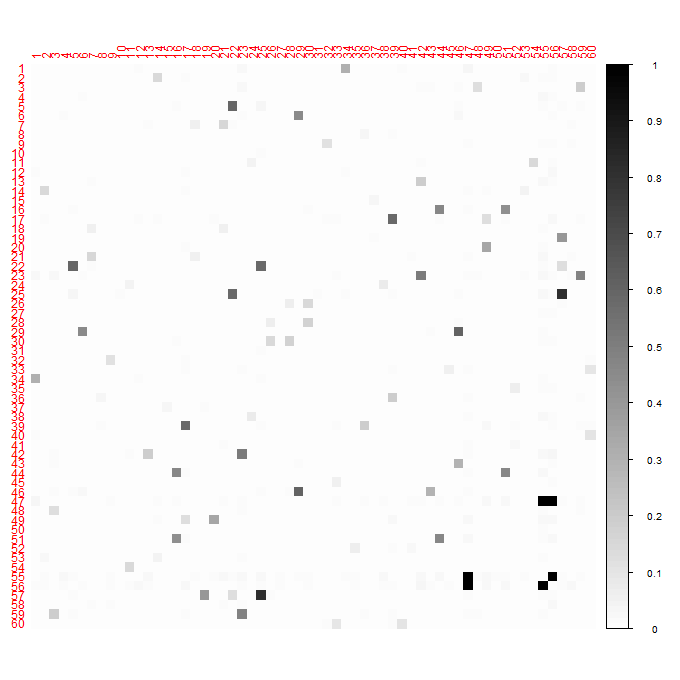}}\hfill \\
\subfloat[Glasso-$\epsilon=0.05$]{\includegraphics[width = 1in]{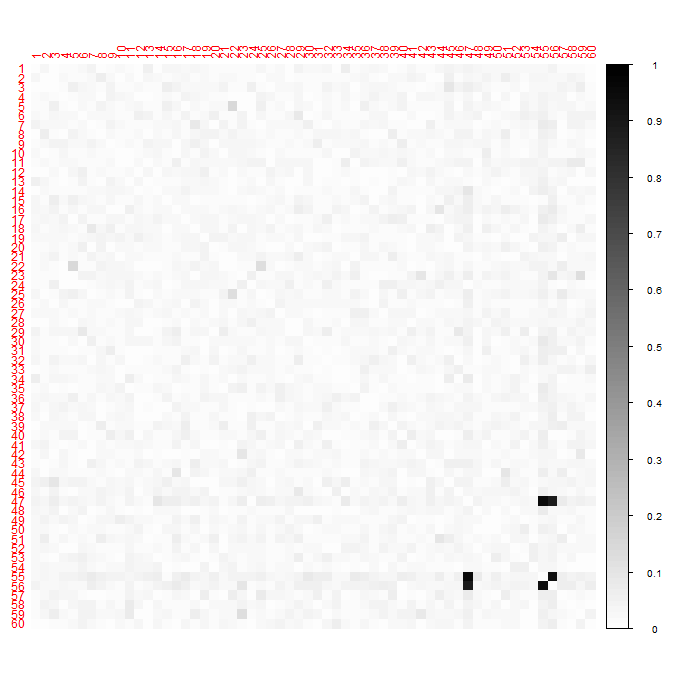}}\hfill 
	\subfloat[RGlassoWinsor-$\epsilon=0.05$]{\includegraphics[width = 1in]{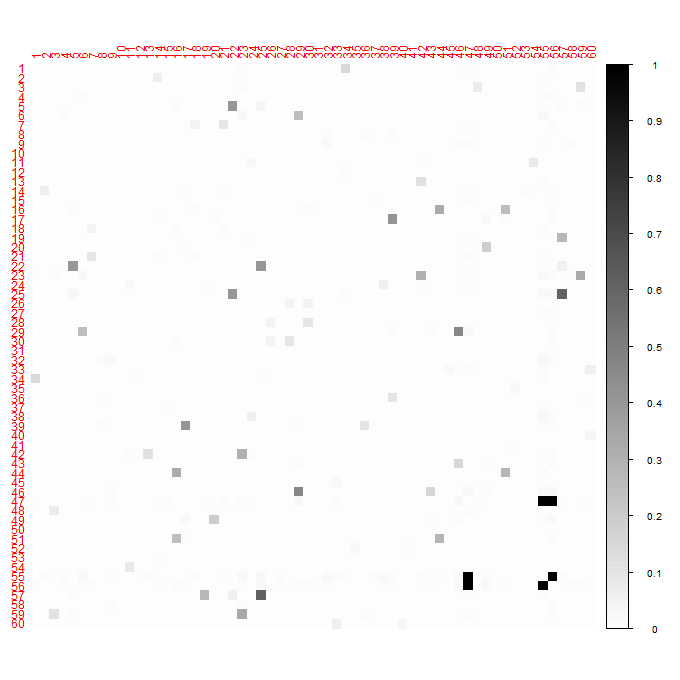}}\hfill 
\subfloat[Glasso-$\epsilon=0.10$]{\includegraphics[width = 1in]{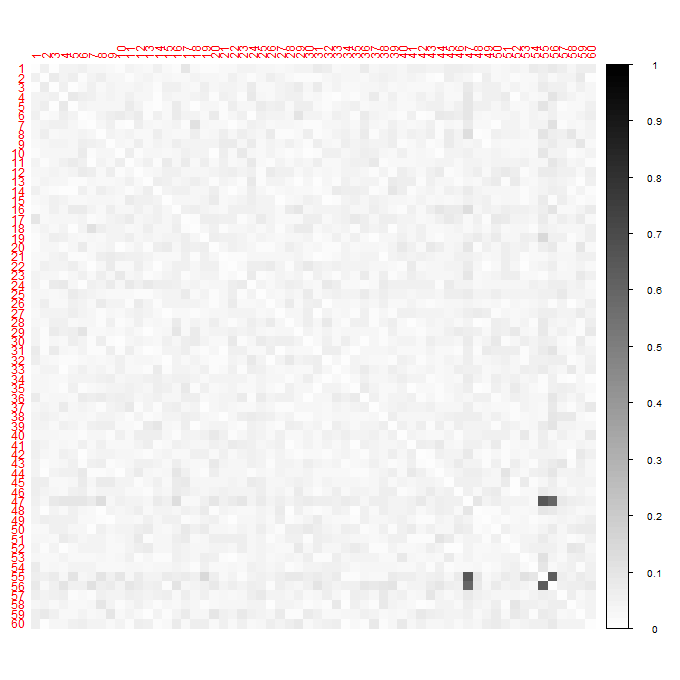}}\hfill
\subfloat[RGlassoWinsor-$\epsilon=0.10$]{\includegraphics[width = 1in]{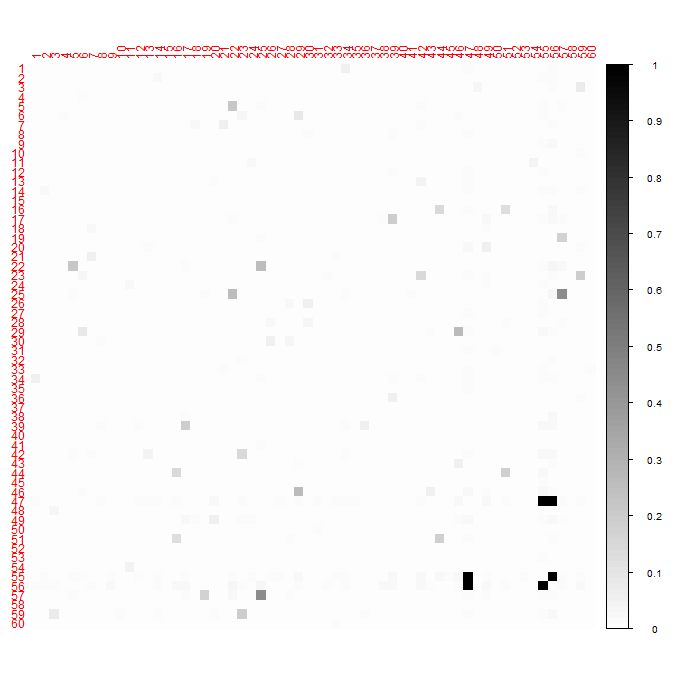}}\hfill 
 	\caption{Heatmaps for the frequency of adjancency for each pair of nodes over $ N=100$ rreplicates under ICM.  $p=60$ and $n=100$.  The axes  display  the graph  $p$-nodes in a given order. }
	\label{Fig_Glasso_Winsor_NN_p_60}
\end{figure}

\begin{figure}[H]
	\subfloat[True Model $\text{Rand}$ ]{\includegraphics[width = 1in]{Model_Random_p_60_n_100_epsilon_0_Heat}}\hfill\centering
	\subfloat[Glasso-$\epsilon=0$]{\includegraphics[width = 1in]{Glasso_Cor_Random_p_60_n_100_epsilon_0_Heat}}\hfill
		\subfloat[RGlassoWinsor-$\epsilon=0$]{\includegraphics[width = 1in]{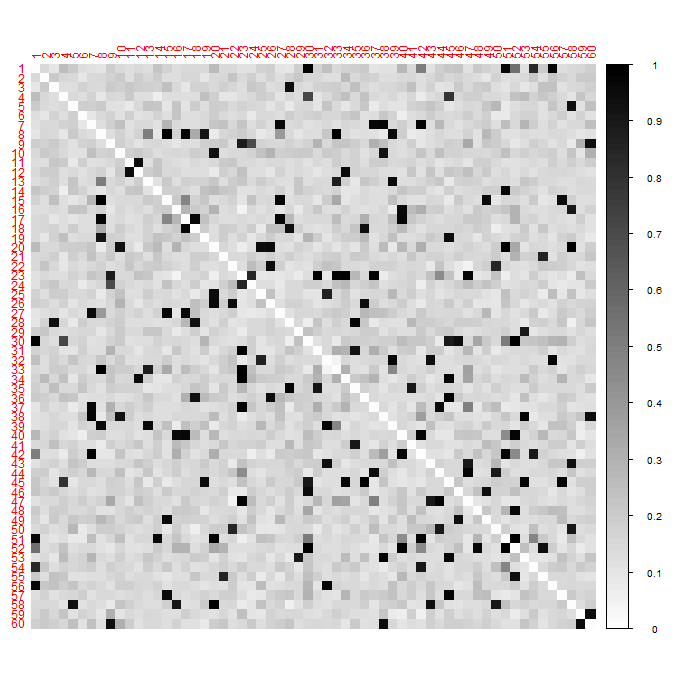}}\hfill
	\subfloat[Glasso-$\epsilon=0.01$]{\includegraphics[width = 1in]{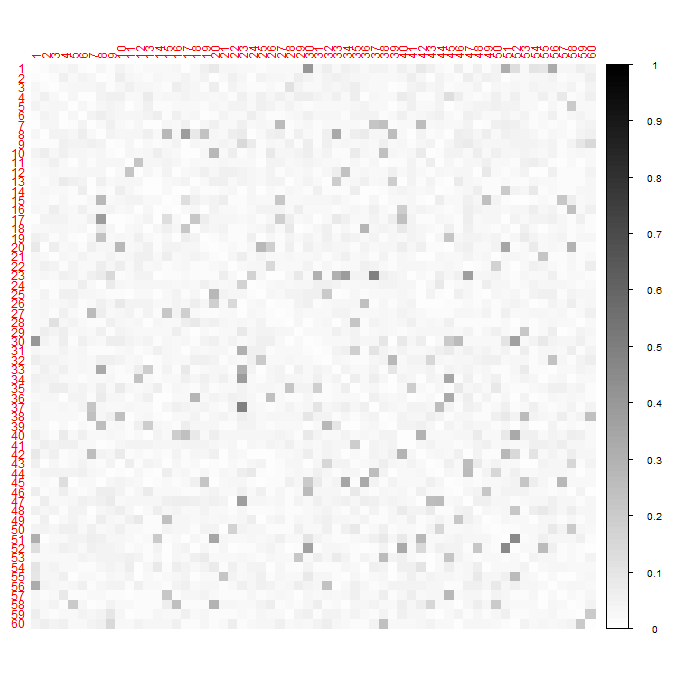}}\hfill\\
\subfloat[RGlassoWinsor-$\epsilon=0.01$]{\includegraphics[width = 1in]{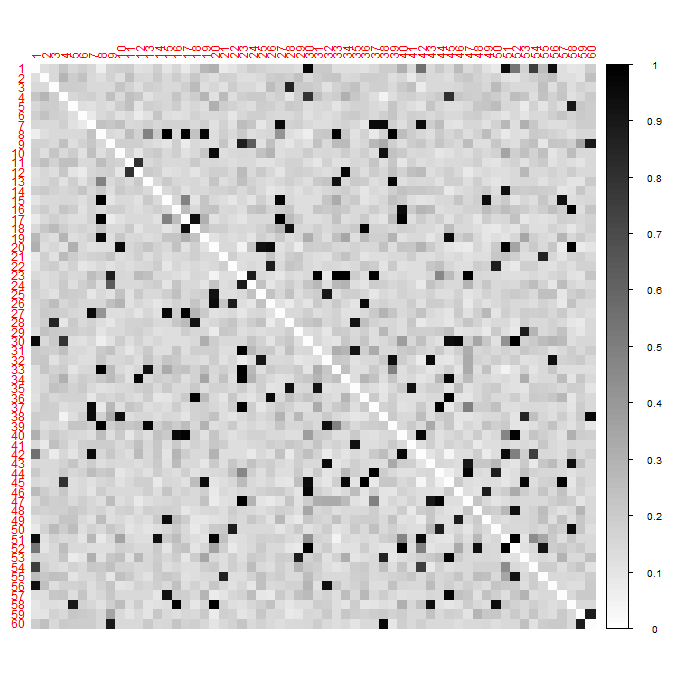}}\hfill
		\subfloat[Glasso-$\epsilon=0.05$]{\includegraphics[width = 1in]{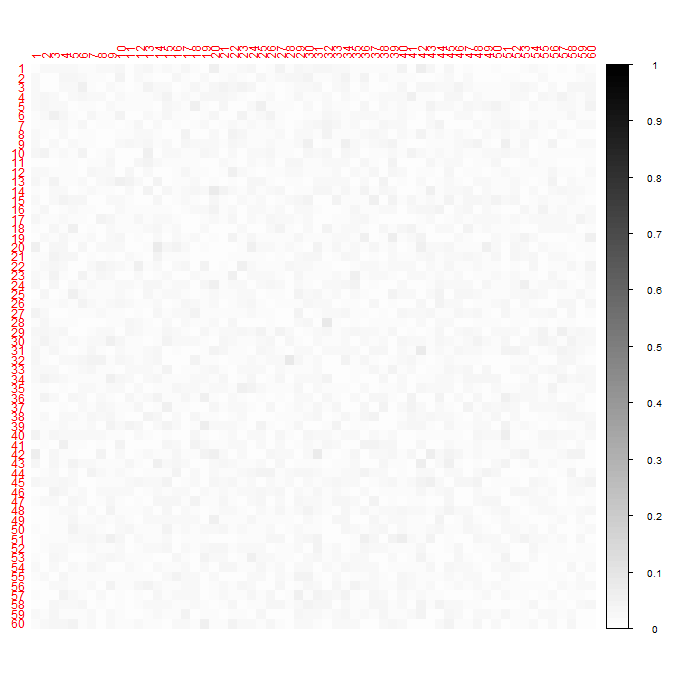}}\hfill
\subfloat[RGlassoWinsor-$\epsilon=0.05$]{\includegraphics[width = 1in]{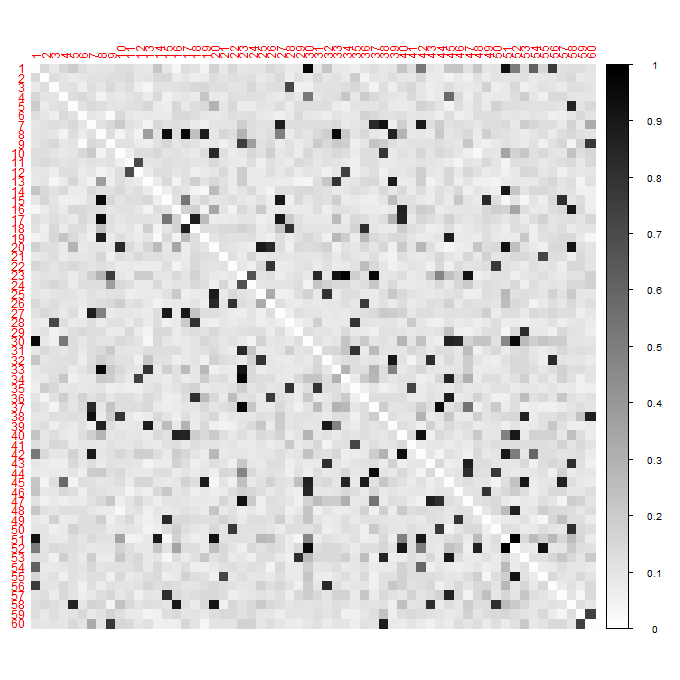}}\hfill
		\subfloat[Glasso-$\epsilon=0.10$]{\includegraphics[width = 1in]{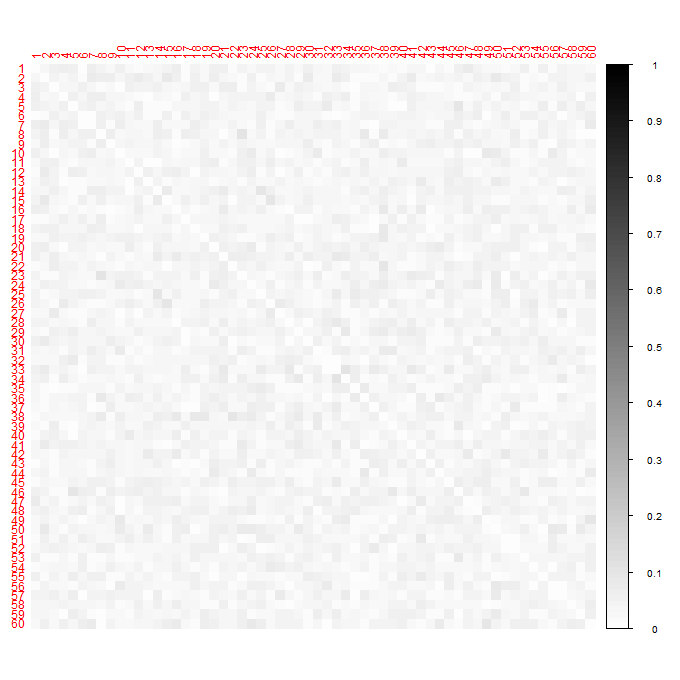}}\hfill
\subfloat[RGlassoWinsor-$\epsilon=0.10$]{\includegraphics[width = 1in]{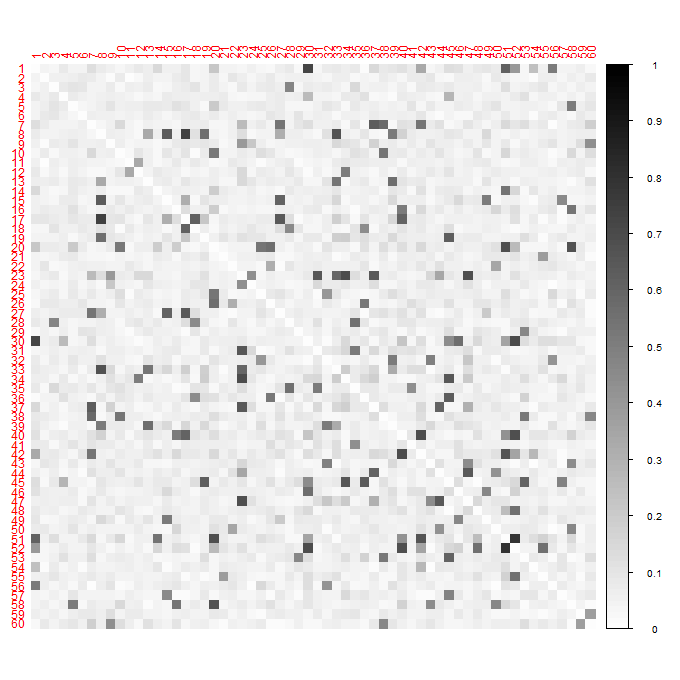}}\hfill
		\caption{Heatmaps for the frequency of adjancency for each pair of nodes over $ N=100$ rreplicates under ICM.  $p=60$ and $n=100$.  The axes  display  the graph  $p$-nodes in a given order. }
	\label{Fig_Glasso_Winsor_Random_p_60}
\end{figure}

\begin{figure}[H]
	\captionsetup[subfloat]{farskip=0.5pt,captionskip=0.5pt}
	\centering
	\subfloat[Glasso-$\epsilon=0$]{\includegraphics[width = 1in]{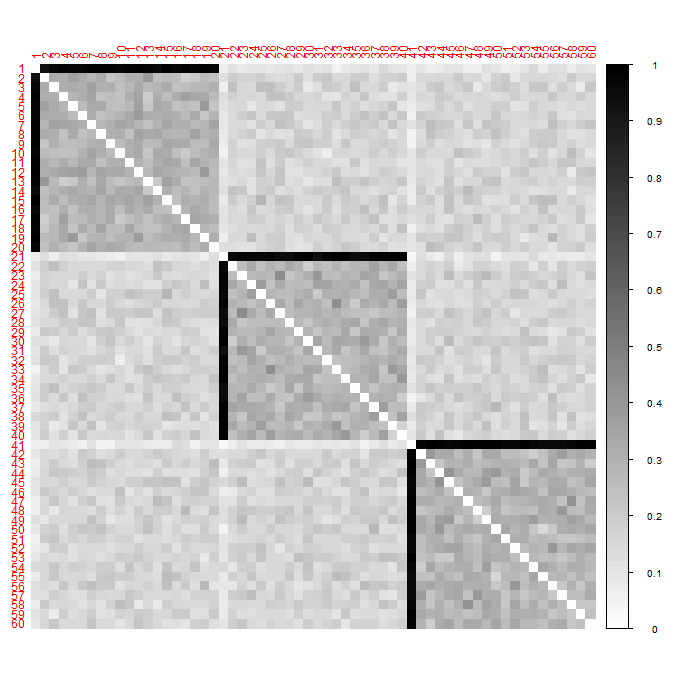}}\hfill
	\subfloat[RGlassoWinsor-$\epsilon=0$]{\includegraphics[width = 1in]{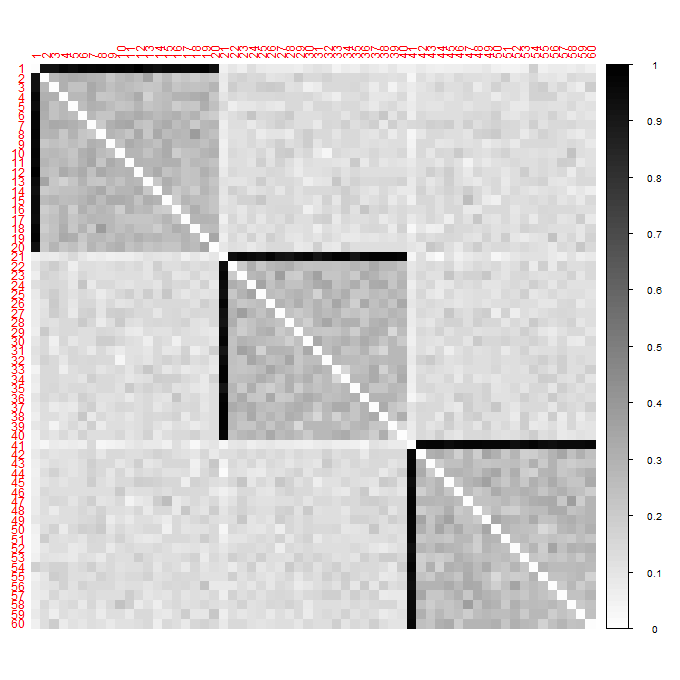}}\hfill
	\subfloat[Glasso-$\epsilon=0.01$]{\includegraphics[width = 1in]{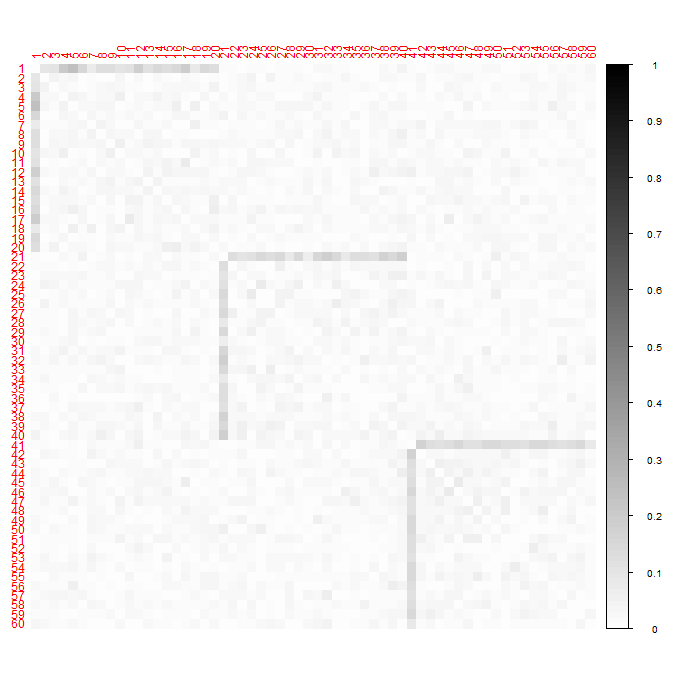}}\hfill	
	\subfloat[RGlassoWinsor-$\epsilon=0.01$]{\includegraphics[width = 1in]{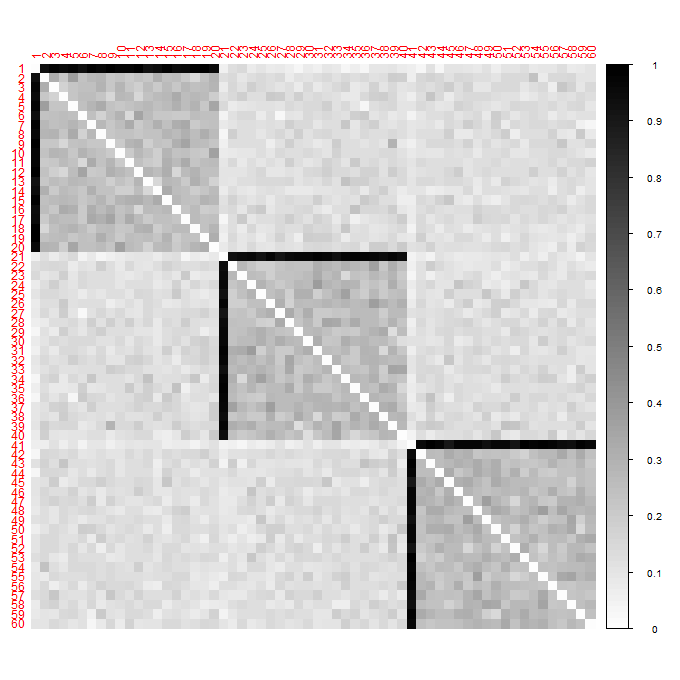}}\hfill	\\
		\subfloat[Glasso-$\epsilon=0.05$]{\includegraphics[width = 1in]{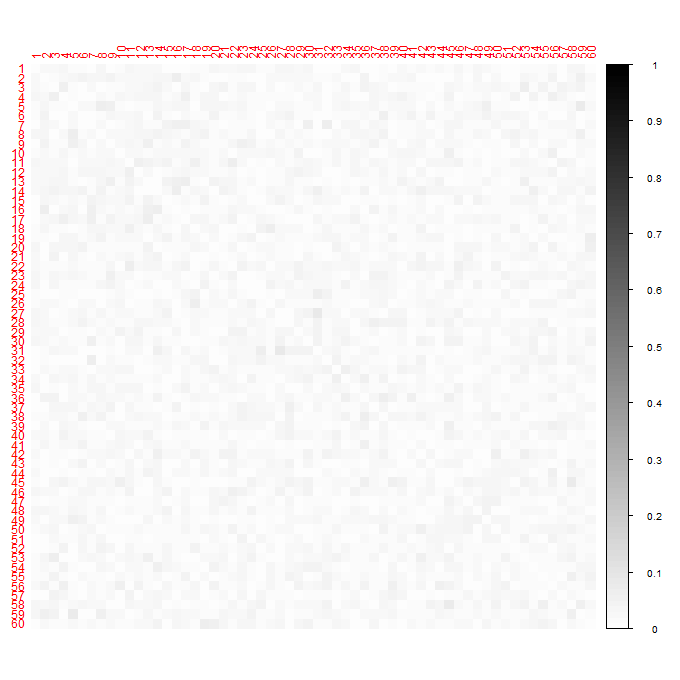}}\hfill	
	\subfloat[RGlassoWinsor-$\epsilon=0.05$]{\includegraphics[width = 1in]{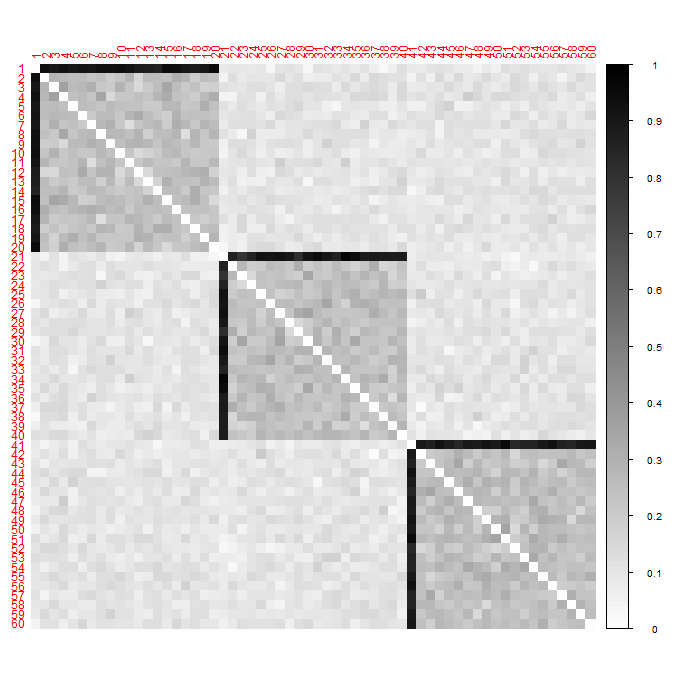}}\hfill	
\subfloat[Glasso-$\epsilon=0.10$]{\includegraphics[width = 1in]{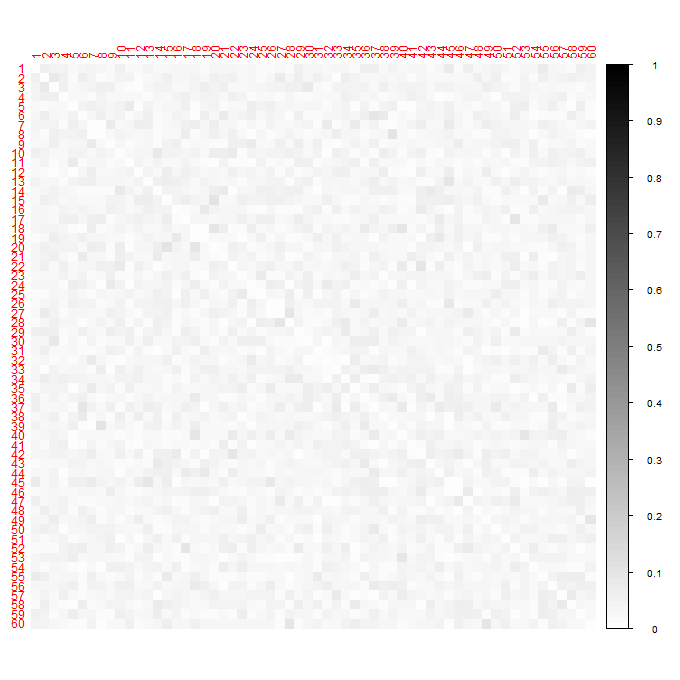}}\hfill	
\subfloat[RGlassoWinsor-$\epsilon=0.10$]{\includegraphics[width = 1in]{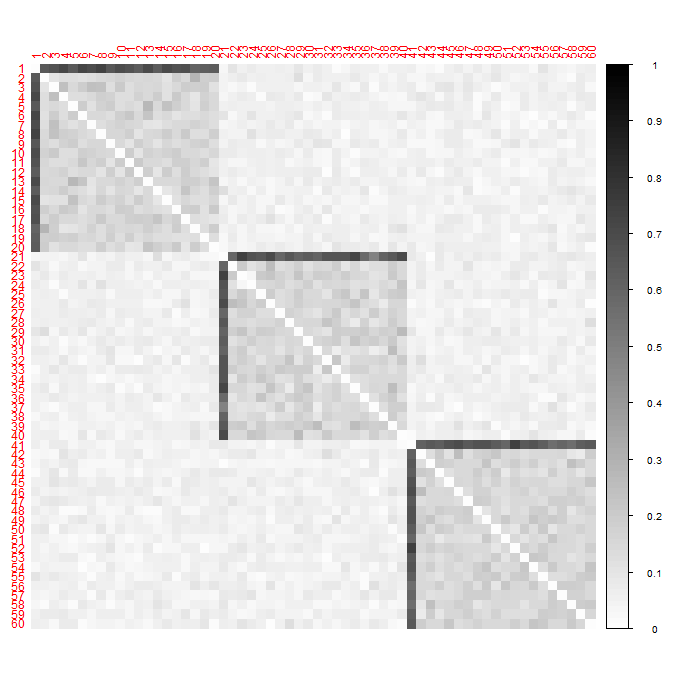}}\hfill	
	\caption{Heatmaps for the frequency of adjancency for each pair of nodes over $ N=100$ rreplicates under ICM.  $p=60$ and $n=100$.  The axes  display  the graph  $p$-nodes in a given order. }
	\label{Fig_Glasso_Winsor_Hub_p_60}
\end{figure}

  Figures \ref{Fig_Glasso_Winsor_NN_p_60}-\ref{Fig_Glasso_Winsor_Hub_p_60}   compare the heatmaps for the true graph and the estimated graphs using Glasso and RGlassoWinsor for clean and 5\% contaminated data, for $p=60$.

\begin{table}[H]
	\mbox{}\hfill
	\begin{minipage}[t]{.48\linewidth}
		\centering{ 
			
			\scalebox{0.46}{
\begin{tabular}{l|rr|rr|rr|rr}
					\hline
					& \multicolumn{2}{c|}{$\epsilon=0$} & \multicolumn{2}{c|}{$\epsilon=0.01$} & \multicolumn{2}{c|}{$\epsilon=0.05$} & \multicolumn{2}{c}{$\epsilon=0.10$} \\
					
					\hline
					& $D_{KL}$ & $m_{F}$ & $D_{KL}$ & $m_{F}$ & $D_{KL}$ & $m_{F}$ & $D_{KL}$ & $m_{F}$  \\
					\hline
					RGlassoWinsor & 10.970 & 9.140 & 11.241 & 9.257 & 12.811 & 9.562 & 16.170 & 9.899\\
					& (0.034) & (0.013) & (0.035) & (0.011) & (0.047) & (0.007) & (0.078) & (0.006) \\
					Glasso & 10.015 & 8.944 & 20.916 & 10.107 & 54.193 & 10.956 & 77.927 & 11.208\\
					& (0.020) & (0.013) & (0.183) & (0.009) & (0.177) & (0.003) & (0.168 & (0.001)\\
					RGlasso$Q_n$ & 21.840 & 10.238 & 26.432 & 10.405 & 47.999 & 10.885 & 79.564 & 11.226\\
					& (0.152) & (0.006) & (0.201) & (0.007) & (0.333) & (0.005) & (0.449) & (0.003)\\
					RGlassoTau & 15.348 & 9.847 & 18.715 & 10.071 & 42.766 & 10.793 & 84.973 & 11.260\\
					& (0.098) & (0.009) & (0.142) & (0.008) & (0.330) & (0.006) & (0.494) & (0.003)\\
					RGlassoGauss & 10.412 & 9.109 & 11.084 & 9.303 & 14.113 & 9.735 & 18.470 & 10.052\\
					& (0.027) & (0.013) & (0.036) & (0.010) & (0.058) & (0.007) & (0.081) & (0.005)\\
					RGlassoSpearman & 10.623 & 9.188 & 11.300 & 9.354 & 14.338 & 9.752 & 18.866 & 10.072\\
					& (0.027) & (0.010) & (0.041) & (0.011) & (0.066) & (0.008) & (0.091) & (0.006)\\
					RGlassoQuad & 13.681 & 9.752 & 14.601 & 9.831 & 18.478 & 10.072 & 24.070 & 10.319\\
					& (0.054) & (0.006) & (0.062) & (0.006) & (0.090) & (0.005) & (0.119) & (0.005\\
					\hline
			\end{tabular}}
		}
		
		\caption{   Model BG under ICM. Comparison  of  means and standard deviations (in brackets) of  $\text{m}_F$  and $D_{KL}$ over $N=100$ replicates. $p=60$, $n=100$.} 
		
		\label{num_performance_se_Block_p_60}%

	\end{minipage}\hfill
	\begin{minipage}[t]{.48\linewidth}
		\centering
		\scalebox{0.46}{
		\begin{tabular}{l|rr|rr|rr|rr}
			\hline
			& \multicolumn{2}{c|}{$\epsilon=0$} & \multicolumn{2}{c|}{$\epsilon=0.01$} & \multicolumn{2}{c|}{$\epsilon=0.05$} & \multicolumn{2}{c}{$\epsilon=0.10$} \\
			
			\hline
			& $D_{KL}$ & $m_{F}$ & $D_{KL}$ & $m_{F}$ & $D_{KL}$ & $m_{F}$ & $D_{KL}$ & $m_{F}$  \\
			\hline
			RGlassoWinsor & 39.840 & 15.542 & 40.583 & 15.624 & 46.022 & 16.101 & 57.991 & 16.698\\
			& (0.096) & (0.015) & (0.086) & (0.012) & (0.118) & (0.009) & (0.196) & (0.008)\\
			Glasso & 36.423 & 15.182 & 73.743 & 17.065 & 188.553 & 18.729 & 269.598 & 19.220\\
			& (0.066) & (0.019) & (0.327) & (0.011) & (0.404) & (0.003) & (0.361) & (0.002)\\
			RGlasso$Q_n$ & 109.766 & 17.928 & 129.704 & 18.199 & 216.252 & 18.954 & 334.403 & 19.462\\
			& (0.507) & (0.008) & (0.552) & (0.007) & (0.819) & (0.005) & (1.058) & (0.003)\\
			RGlassoTau & 68.377 & 17.039 & 86.398 & 17.491 & 193.466 & 18.795 & 357.429 & 19.520\\
			& (0.291) & (0.010) & (0.361) & (0.008) & (0.920) & (0.007) & (1.302) & (0.003)\\
			RGlassoGauss & 37.458 & 15.355 & 39.919 & 15.639 & 51.074 & 16.422 & 65.584 & 16.975\\
			& (0.077) & (0.017) & (0.102) & (0.016) & (0.159) & (0.010) & (0.191) & (0.007)\\
			RGlassoSpearman & 38.929 & 15.553 & 41.225 & 15.767 & 52.450 & 16.470 & 67.646 & 17.032\\
			& (0.100) & (0.017) & (0.120) & (0.016) & (0.179) & (0.011) & (0.199) & (0.007)\\
			RGlassoQuad & 56.280 & 16.676 & 60.054 & 16.811 & 74.871 & 17.235 & 94.063 & 17.650\\
			& (0.196) & (0.010) & (0.214) & (0.009) & (0.264) & (0.007) & (0.310) & (0.006)\\
			\hline
		\end{tabular}}
		
		\caption{   Model BG under ICM. Comparison  of  means and standard deviations (in brackets) of  $\text{m}_F$  and $D_{KL}$ over $N=100$ replicates.  $p=200$, $n=100$.} 
		
		\label{num_performance_se_Block_p_200}%
		
	\end{minipage}\hfill
	\mbox{}
	
\end{table}
 
 \begin{table}[H]
 	\mbox{}\hfill
 	\begin{minipage}[t]{.48\linewidth}
 		\centering{ 
 			
 			\scalebox{0.5}{
 				\begin{tabular}{l|rr|rr|rr|rr}
 					\hline
 					& \multicolumn{2}{c|}{$\epsilon=0$} & \multicolumn{2}{c|}{$\epsilon=0.01$} & \multicolumn{2}{c|}{$\epsilon=0.05$} & \multicolumn{2}{c}{$\epsilon=0.10$}  \\
 					
 					\hline
 					& TPR & TNR & TPR & TNR & TPR & TNR & TPR & TNR     \\
 					\hline
 					RGlassoWinsor & 0.498 & 0.896 & 0.433 & 0.918 & 0.268 & 0.949 & 0.081 & 0.983\\
 					& (0.008)& (0.004)& (0.008)& (0.003)& (0.008)& (0.002)& (0.004)& (0.001)\\
 					Glasso & 0.637 & 0.840 & 0.103 & 0.973 & 0.021 & 0.977 & 0.035 & 0.965\\
 					& (0.007)& (0.004)& (0.005)& (0.001)& (0.002)& (0.001)& (0.002)& (0.002)\\
 					RGlasso$Q_n$ & 0.006 & 1.000 & 0.002 & 1.000 & 0.000 & 1.000 & 0.000 & 1.000\\
 					& (0.001)& (0.000)& (0.000)& (0.000)& (0.000)& (0.000)& (0.000)& (0.000)\\
 					RGlassoTau & 0.141 & 0.972 & 0.078 & 0.987 & 0.002 & 1.000 & 0.000 & 0.999\\
 					& (0.005)& (0.002)& (0.004)& (0.001)& (0.000)& (0.000)& (0.000)& (0.000)\\
 					RGlassoGauss & 0.544 & 0.882 & 0.444 & 0.904 & 0.231 & 0.945 & 0.112 & 0.966\\
 					& (0.008)& (0.004)& (0.008)& (0.003)& (0.007)& (0.002)& (0.005)& (0.002)\\
 					RGlassoSpearman & 0.486 & 0.898 & 0.400 & 0.918 & 0.236 & 0.953 & 0.115 & 0.972\\
 					& (0.007)& (0.003)& (0.009)& (0.003)& (0.007)& (0.002)& (0.006)& (0.002)\\
 					RGlassoQuad & 0.095 & 0.978 & 0.076 & 0.982 & 0.057 & 0.986 & 0.028 & 0.992\\
 					& (0.004)& (0.001)& (0.004)& (0.001)& (0.003)& (0.001)& (0.002)& (0.001)\\
 					\hline
 			\end{tabular}}
 		}
 		
 		\caption{Model BG under ICM.  Comparison  of  means and standard deviations (in brackets) of  TPR and TNR  over $N=100$ replicates. $p=60, n=100$. } 
 		\label{TPTN_bLock_p_60}%
 	\end{minipage}\hfill
 	\begin{minipage}[t]{.48\linewidth}
 		\centering
 		\scalebox{0.5}{		
 				\begin{tabular}{l|r|r|r|r}
 				\hline
 				& \multicolumn{1}{c|}{$\epsilon=0$} & \multicolumn{1}{c|}{$\epsilon=0.01$} & \multicolumn{1}{c|}{$\epsilon=0.05$} & \multicolumn{1}{c}{$\epsilon=0.10$}   \\
 				
 				\hline
 				RGlassoWinsor & 0.323 & 0.315 & 0.241 & 0.118\\
 				& (0.002)& (0.003)& (0.003)& (0.005)\\
 				Glasso & 0.336 & 0.112 & -0.005 & 0.000\\
 				& 0.002)& (0.005)& (0.002)& (0.002)\\
 				RGlasso$Q_n$ & 0.044 & 0.016 & 0.003 & 0.001\\
 				& 0.005)& (0.004)& (0.001)& (0.001)\\
 				RGlassoTau & 0.168 & 0.136 & 0.012 & -0.007\\
 				& 0.003)& (0.004)& (0.003)& (0.001)\\
 				RGlassoGauss & 0.332 & 0.297 & 0.196 & 0.112\\
 				& (0.002)& (0.003)& (0.003)& (0.003\\
 				RGlassoSpearman & 0.316 & 0.291 & 0.220 & 0.131\\
 				& (0.003)& (0.003)& (0.003)& (0.004)\\
 				RGlassoQuad & 0.124 & 0.107 & 0.092 & 0.059\\
 				& (0.003)& (0.004)& (0.004)& (0.004)\\
 				\hline
 		\end{tabular}}
 		\vspace{0.15cm}
 		\centering 
 		\caption{   Model BG under ICM. Comparison  of  means and standard deviations (in brackets) of  MCC over $N=100$ replicates. $p=60$, $n=100$.} 
 	\label{MCC_se_block_p_60}%
 		
 	\end{minipage}\hfill
 	\mbox{}
 	
 \end{table}

 \begin{table}[H]
 	\mbox{}\hfill
 	\begin{minipage}[t]{.48\linewidth}
 		\centering{ 
 			
 		\scalebox{0.5}{
 			\begin{tabular}{l|rr|rr|rr|rr}
 				\hline
 				& \multicolumn{2}{c|}{$\epsilon=0$} & \multicolumn{2}{c|}{$\epsilon=0.01$} & \multicolumn{2}{c|}{$\epsilon=0.05$} & \multicolumn{2}{c}{$\epsilon=0.10$}  \\
 				
 				\hline
 				& TPR & TNR & TPR & TNR & TPR & TNR & TPR & TNR   \\
 				\hline
 				RGlassoWinsor & 0.392 & 0.966 & 0.367 & 0.969 & 0.201 & 0.985 & 0.036 & 0.996\\
 				& (0.007) & (0.001) & (0.006) & (0.001) & (0.005) & (0.001) & (0.002) & (0.000)\\
 				Glasso & 0.547 & 0.939 & 0.076 & 0.984 & 0.010 & 0.987 & 0.016 & 0.982\\
 				& (0.007) & (0.002) & (0.004) & (0.001) & (0.001) & (0.000) & (0.001) & (0.001)\\
 				RGlasso$Q_n$ & 0.000 & 1.000 & 0.000 & 1.000 & 0.000 & 1.000 & 0.000 & 1.000\\
 				& (0.000) & (0.000) & (0.000) & (0.000) & (0.000) & (0.000) & (0.000) & (0.000)\\
 				RGlassoTau & 0.124 & 0.993 & 0.048 & 0.998 & 0.000 & 1.000 & 0.000 & 1.000\\
 				& (0.004) & (0.000) & (0.002) & (0.000) & (0.000) & (0.000) & (0.000) & (0.000)\\
 				RGlassoGauss & 0.482 & 0.953 & 0.395 & 0.961 & 0.168 & 0.981 & 0.070 & 0.988\\
 				& (0.007) & (0.002) & (0.007) & (0.002) & (0.005) & (0.001) & (0.003) & (0.001)\\
 				RGlassoSpearman & 0.415 & 0.960 & 0.366 & 0.964 & 0.198 & 0.980 & 0.088 & 0.988\\
 				& (0.007) & (0.002) & (0.008) & (0.002) & (0.006) & (0.001) & (0.003) & (0.001)\\
 				RGlassoQuad & 0.113 & 0.983 & 0.102 & 0.985 & 0.079 & 0.987 & 0.048 & 0.991\\
 				& (0.005) & (0.001) & (0.004) & (0.001) & (0.003) & (0.001) & (0.002) & (0.001)\\			
 				\hline
 				
 				\hline
 			\end{tabular}
 			
 			}
 		}
 			\caption{Model BG under ICM.  Comparison  of  means and standard deviations (in brackets) of  TPR and TNR  over $N=100$ replicates. $p=200, n=100$. } 
 		\label{TPTN_bLock_p_200}%
 
 	\end{minipage}\hfill
 	\begin{minipage}[t]{.48\linewidth}
 		\centering
 	\scalebox{0.5}{
 \begin{tabular}{l|r|r|r|r}
 	\hline
 	& \multicolumn{1}{c|}{$\epsilon=0$} & \multicolumn{1}{c|}{$\epsilon=0.01$} & \multicolumn{1}{c|}{$\epsilon=0.05$} & \multicolumn{1}{c}{$\epsilon=0.10$}   \\
 	
 	\hline
 	RGlassoWinsor & 0.258 & 0.253 & 0.195 & 0.066\\
 	& (0.001) & (0.002) & (0.002) & (0.002)\\
 	Glasso & 0.271 & 0.062 & -0.003 & -0.002\\
 	& (0.001) & (0.002) & (0.001) & (0.001)\\
 	RGlasso$Q_n$ & 0.002 & 0.000 & 0.000 & 0.000\\
 	& (0.001) & (0.000) & (0.000) & (0.000)\\
 	RGlassoTau & 0.177 & 0.128 & 0.006 & -0.001\\
 	& (0.002) & (0.003) & (0.002) & (0.000)\\
 	RGlassoGauss & 0.272 & 0.243 & 0.148 & 0.073\\
 	& (0.001) & (0.001) & (0.002) & (0.002)\\
 	RGlassoSpearman & 0.254 & 0.238 & 0.169 & 0.092\\
 	& (0.001) & (0.001) & (0.002) & (0.002)\\
 	RGlassoQuad & 0.103 & 0.100 & 0.079 & 0.055\\
 	& (0.002) & (0.002) & (0.002) & (0.001)\\
 	\hline
 \end{tabular}	
 }
 	
 		\vspace{0.2cm}
 		\centering 
 	\caption{   Model BG under ICM. Comparison  of  means and standard deviations (in brackets) of  MCC over $N=100$ replicates.   $p=200$, $n=100$.} 
 
 \label{MCC_se_block_p_200}%
 		
 	\end{minipage}\hfill
 	\mbox{}
 \end{table}

 \begin{table}[H]
 	\mbox{}\hfill
 	\begin{minipage}[t]{.48\linewidth}
 		\centering{ 
 			
 			\scalebox{0.46}{
 			\begin{tabular}{l|rr|rr|rr|rr}
 				\hline
 				& \multicolumn{2}{c|}{$\epsilon=0$} & \multicolumn{2}{c|}{$\epsilon=0.01$} & \multicolumn{2}{c|}{$\epsilon=0.05$} & \multicolumn{2}{c}{$\epsilon=0.10$} \\
 				
 				\hline
 				& $D_{KL}$ & $m_{F}$ & $D_{KL}$ & $m_{F}$ & $D_{KL}$ & $m_{F}$ & $D_{KL}$ & $m_{F}$  \\
 				\hline
 				RGlassoWinsor & 12.676 & 5.220 & 13.105 & 5.273 & 15.506 & 5.561 & 19.174 & 5.909\\
 				& (0.204) & (0.030) & (0.214) & (0.030) & (0.224) & (0.026) & (0.235) & (0.020)\\
 				Glasso & 12.056 & 5.194 & 25.379 & 6.283 & 62.605 & 7.615 & 88.522 & 8.010\\
 				& (0.127) & (0.018) & (0.229) & (0.016) & (0.213) & (0.005) & (0.193) & (0.002)\\
 				RGlasso$Q_n$ & 29.458 & 6.575 & 33.859 & 6.764 & 50.575 & 7.332 & 73.491 & 7.806\\
 				& (0.610) & (0.031) & (0.819) & (0.035) & (0.991) & (0.026) & (0.862) & (0.012)\\
 				RGlassoTau & 27.851 & 6.493 & 32.020 & 6.681 & 50.077 & 7.312 & 81.897 & 7.920\\
 				& (0.564) & (0.031) & (0.796) & (0.036) & (1.043) & (0.029) & (0.796) & (0.010)\\
 				RGlassoGauss & 12.294 & 5.215 & 13.287 & 5.339 & 17.094 & 5.748 & 21.860 & 6.140\\
 				& (0.149) & (0.021) & (0.181) & (0.024) & (0.210) & (0.021) & (0.229) & (0.017)\\
 				RGlassoSpearman & 12.296 & 5.214 & 13.320 & 5.343 & 17.286 & 5.767 & 22.176 & 6.164\\
 				& (0.152) & (0.022) & (0.187) & (0.025) & (0.223) & (0.022) & (0.239) & (0.017)\\
 				RGlassoQuad & 13.616 & 5.384 & 14.809 & 5.520 & 19.189 & 5.936 & 24.611 & 6.329\\
 				& (0.160) & (0.021) & (0.191) & (0.022) & (0.233) & (0.020) & (0.243) & (0.016)\\
 				\hline
 		\end{tabular}}

 		}
 		
 		\caption{   Model $\text{NN}(2)$. Comparison  of  means and standard deviations (in brackets) of  $\text{m}_F$  and $D_{KL}$ over $N=100$ replicates. $p=60$, $n=100$.} 
 		\label{num_performance_se_NN_p_60}%

 	\end{minipage}\hfill
 	\begin{minipage}[t]{.48\linewidth}
 		\centering
 		\scalebox{0.46}{
 				\begin{tabular}{l|rr|rr|rr|rr}
 				\hline
 				& \multicolumn{2}{c|}{$\epsilon=0$} & \multicolumn{2}{c|}{$\epsilon=0.01$} & \multicolumn{2}{c|}{$\epsilon=0.05$} & \multicolumn{2}{c}{$\epsilon=0.10$} \\
 				
 				\hline
 				& $D_{KL}$ & $m_{F}$ & $D_{KL}$ & $m_{F}$ & $D_{KL}$ & $m_{F}$ & $D_{KL}$ & $m_{F}$  \\
 				\hline
 				RGlassoWinsor & 37.597 & 9.035 & 38.772 & 9.133 & 47.601 & 9.807 & 63.232 & 10.698\\
 				& (0.513) & (0.045) & (0.511) & (0.043) & (0.472) & (0.033) & (0.628) & (0.031)\\
 				Glasso & 33.702 & 8.804 & 81.300 & 11.279 & 203.376 & 13.848 & 289.175 & 14.598\\
 				& (0.359) & (0.034) & (0.470) & (0.020) & (0.395) & (0.005) & (0.366) & (0.003)\\
 				RGlasso$Q_n$ & 89.685 & 11.754 & 103.496 & 12.165 & 163.177 & 13.344 & 248.940 & 14.299\\
 				& (1.216) & (0.041) & (1.083) & (0.030) & (1.547) & (0.025) & (1.718) & (0.014)\\
 				RGlassoTau & 83.190 & 11.538 & 96.377 & 11.955 & 165.162 & 13.371 & 284.673 & 14.544\\
 				& (1.171) & (0.042) & (1.229) & (0.038) & (1.554) & (0.024) & (1.707) & (0.011)\\
 				RGlassoGauss & 34.620 & 8.863 & 37.606 & 9.126 & 52.130 & 10.150 & 70.345 & 11.071\\
 				& (0.411) & (0.038) & (0.389) & (0.033) & (0.443) & (0.027) & (0.502) & (0.022)\\
 				RGlassoSpearman & 35.989 & 8.989 & 38.826 & 9.231 & 53.426 & 10.228 & 72.356 & 11.156\\
 				& (0.411) & (0.037) & (0.387) & (0.032) & (0.452) & (0.027) & (0.519) & (0.022)\\
 				RGlassoQuad & 53.251 & 10.222 & 57.165 & 10.440 & 73.848 & 11.210 & 95.469 & 11.962\\
 				& (0.392) & (0.025) & (0.368) & (0.022) & (0.441) & (0.020) & (0.518) & (0.017)\\
 				\hline
 		\end{tabular}} 		
 		
 		\caption{   Model $\text{NN}(2)$. Comparison  of  means and standard deviations (in brackets) of  $\text{m}_F$  and $D_{KL}$ over $N=100$ replicates. $p=200$, $n=100$.} 
 		\label{num_performance_se_NN_p_200}%
 		
 	\end{minipage}\hfill
 	\mbox{}
 	
 \end{table}

\begin{table}[H]
	\mbox{}\hfill
	\begin{minipage}[t]{.48\linewidth}
		\centering{ 
			
			\scalebox{0.5}{
			\begin{tabular}{l|rr|rr|rr|rr}
			\hline
			& \multicolumn{2}{c|}{$\epsilon=0$} & \multicolumn{2}{c|}{$\epsilon=0.01$} & \multicolumn{2}{c|}{$\epsilon=0.05$} & \multicolumn{2}{c}{$\epsilon=0.10$}  \\
			
			\hline
			& TPR & TNR & TPR & TNR & TPR & TNR & TPR & TNR   \\
			\hline
		RGlassoWinsor & 0.284 & 0.999 & 0.265 & 0.999 & 0.185 & 0.999 & 0.125 & 0.999\\
		& (0.014) & (0.000) & (0.013) & (0.000) & (0.010) & (0.000) & (0.006) & (0.000)\\
		Glasso & 0.298 & 0.999 & 0.176 & 0.991 & 0.099 & 0.979 & 0.087 & 0.963\\
		& (0.012) & (0.000) & (0.008) & (0.000) & (0.003) & (0.001) & (0.004) & (0.002)\\
		RGlasso$Q_n$ & 0.079 & 1.000 & 0.085 & 1.000 & 0.084 & 1.000 & 0.073 & 1.000\\
		& (0.004) & (0.000) & (0.005) & (0.000) & (0.004) & (0.000) & (0.002) & (0.000)\\
		RGlassoTau & 0.079 & 1.000 & 0.083 & 1.000 & 0.088 & 0.999 & 0.083 & 0.999\\
		& (0.004) & (0.000) & (0.004) & (0.000) & (0.005) & (0.000) & (0.003) & (0.000)\\
		RGlassoGauss & 0.297 & 0.999 & 0.265 & 0.999 & 0.198 & 0.998 & 0.139 & 0.998\\
		& (0.013) & (0.000) & (0.012) & (0.000) & (0.011) & (0.000) & (0.008) & (0.000)\\
		RGlassoSpearman & 0.300 & 0.999 & 0.279 & 0.999 & 0.222 & 0.998 & 0.159 & 0.998\\
		& (0.013) & (0.000) & (0.012) & (0.000) & (0.012) & (0.000) & (0.009) & (0.000)\\
		RGlassoQuad & 0.276 & 0.993 & 0.253 & 0.993 & 0.224 & 0.992 & 0.188 & 0.992\\
		& (0.011) & (0.001) & (0.011) & (0.001) & (0.011) & (0.001) & (0.009) & (0.001)\\
		\hline
		\end{tabular}		
			
		}
		}
		
		\caption{ Model $\text{NN}(2)$ under ICM.  Comparison  of  means and standard deviations (in brackets) of  TPR and TNR  over $N=100$ replicates. $p=60, n=100$. } 
		\label{TPTN_NN_p_60}%
	\end{minipage}\hfill
	\begin{minipage}[t]{.48\linewidth}
		\centering
		\scalebox{0.5}{		
		\begin{tabular}{l|r|r|r|r}
			\hline
			& \multicolumn{1}{c|}{$\epsilon=0$} & \multicolumn{1}{c|}{$\epsilon=0.01$} & \multicolumn{1}{c|}{$\epsilon=0.05$} & \multicolumn{1}{c}{$\epsilon=0.10$}   \\
			
			\hline
			RGlassoWinsor & 0.486 & 0.471 & 0.394 & 0.316\\
			& (0.012) & (0.011) & (0.010) & (0.007)\\
			Glasso & 0.511 & 0.231 & 0.088 & 0.043\\
			& 0.010 & 0.007 & 0.003 & 0.003\\
			RGlasso$Q_n$ & 0.264 & 0.273 & 0.265 & 0.246\\
			& (0.004) & (0.006) & (0.005) & (0.003)\\
			RGlassoTau & 0.266 & 0.270 & 0.263 & 0.230\\
			& (0.004) & (0.006) & (0.005) & (0.004)\\
			RGlassoGauss & 0.508 & 0.469 & 0.373 & 0.290\\
			& (0.011) & (0.010) & (0.008) & (0.007)\\
			RGlassoSpearman & 0.507 & 0.480 & 0.402 & 0.319\\
			& (0.010) & (0.010) & (0.009) & (0.008)\\
			RGlassoQuad & 0.375 & 0.348 & 0.312 & 0.260\\
			& (0.007) & (0.006) & (0.006) & (0.006)\\
			\hline
	\end{tabular}}
		\vspace{0.2cm}
		\centering 
		\caption{   Model $\text{NN}(2)$. Comparison  of  means and standard deviations (in brackets) of  MCC over $N=100$ replicates. $p=60$, $n=100$.} 
		\label{MCC_se_NN_p_60} 
		
	\end{minipage}\hfill
	\mbox{}
\end{table}

\begin{table}[H]
	\mbox{}\hfill
	\begin{minipage}[t]{.48\linewidth}
		\centering{ 
			
			\scalebox{0.5}{
		\begin{tabular}{l|rr|rr|rr|rr}
		\hline
		& \multicolumn{2}{c|}{$\epsilon=0$} & \multicolumn{2}{c|}{$\epsilon=0.01$} & \multicolumn{2}{c|}{$\epsilon=0.05$} & \multicolumn{2}{c}{$\epsilon=0.10$}  \\
		
		\hline
		& TPR & TNR & TPR & TNR & TPR & TNR & TPR & TNR   \\
		\hline
	RGlassoWinsor & 0.589 & 0.997 & 0.568 & 0.997 & 0.450 & 0.998 & 0.282 & 0.999\\
	& (0.010) & (0.000) & (0.011) & (0.000) & (0.010) & (0.000) & (0.011) & (0.000)\\
	Glasso & 0.641 & 0.997 & 0.402 & 0.988 & 0.157 & 0.967 & 0.107 & 0.946\\
	& (0.007) & (0.000) & (0.007) & (0.000) & (0.004) & (0.001) & (0.003) & (0.001)\\
	RGlasso$Q_n$ & 0.265 & 0.999 & 0.254 & 0.999 & 0.253 & 0.999 & 0.215 & 0.998\\
	& (0.010) & (0.000) & (0.007) & (0.000) & (0.008) & (0.000) & (0.008) & (0.000)\\
	RGlassoTau & 0.267 & 0.999 & 0.264 & 0.999 & 0.242 & 0.998 & 0.208 & 0.994\\
	& (0.009) & (0.000) & (0.009) & (0.000) & (0.008) & (0.000) & (0.007) & (0.000)\\
	RGlassoGauss & 0.630 & 0.997 & 0.604 & 0.996 & 0.471 & 0.995 & 0.320 & 0.995\\
	& (0.008) & (0.000) & (0.009) & (0.000) & (0.009) & (0.000) & (0.009) & (0.000)\\
	RGlassoSpearman & 0.623 & 0.996 & 0.608 & 0.996 & 0.505 & 0.995 & 0.365 & 0.995\\
	& (0.008) & (0.000) & (0.009) & (0.000) & (0.009) & (0.000) & (0.010) & (0.000)\\
	RGlassoQuad & 0.530 & 0.987 & 0.513 & 0.987 & 0.445 & 0.987 & 0.357 & 0.989\\
	& (0.008) & (0.001) & (0.008) & (0.001) & (0.008) & (0.001) & (0.008) & (0.001)\\
	\hline
	\end{tabular}		
		
	}
		}
		
		\caption{ Model $\text{NN}(2)$ under ICM.  Comparison  of  means and standard deviations (in brackets) of  TPR and TNR  over $N=100$ replicates. $p=200, n=100$. } 
		\label{TPTN_NN_p_200}%
	\end{minipage}\hfill
	\begin{minipage}[t]{.48\linewidth}
		\centering
		\scalebox{0.5}{		
				\begin{tabular}{l|r|r|r|r}
				\hline
				& \multicolumn{1}{c|}{$\epsilon=0$} & \multicolumn{1}{c|}{$\epsilon=0.01$} & \multicolumn{1}{c|}{$\epsilon=0.05$} & \multicolumn{1}{c}{$\epsilon=0.10$}   \\
				
				\hline
			RGlassoWinsor & 0.598 & 0.594 & 0.546 & 0.408\\
			& (0.005) & (0.004) & (0.004) & (0.006)\\
			Glasso & 0.640 & 0.269 & 0.057 & 0.020\\
			& (0.003) & (0.003) & (0.001) & (0.001)\\
			RGlasso$Q_n$ & 0.408 & 0.401 & 0.377 & 0.324\\
			& (0.007) & (0.006) & (0.006) & (0.005)\\
			RGlassoTau & 0.413 & 0.404 & 0.354 & 0.207\\
			& (0.006) & (0.006) & (0.005) & (0.003)\\
			RGlassoGauss & 0.625 & 0.584 & 0.458 & 0.334\\
			& (0.004) & (0.004) & (0.004) & (0.003)\\
			RGlassoSpearman & 0.602 & 0.576 & 0.485 & 0.374\\
			& (0.004) & (0.004) & (0.004) & (0.004)\\
			RGlassoQuad & 0.358 & 0.345 & 0.310 & 0.271\\
			& (0.004) & (0.004) & (0.004) & (0.003)\\
			\hline
		\end{tabular}}
		
		\vspace{0.2cm}
		\centering 
		\caption{   Model $\text{NN}(2)$. Comparison  of  means and standard deviations (in brackets) of  MCC over $N=100$ replicates. $p=200$, $n=100$.} 
		\label{MCC_se_NN_p_200}%
		
	\end{minipage}\hfill
	\mbox{}
\end{table}

 \begin{table}[H]
 	\mbox{}\hfill
 	\begin{minipage}[t]{.48\linewidth}
 		\centering{ 
 			
 			\scalebox{0.46}{
 					\begin{tabular}{l|rr|rr|rr|rr}
 					\hline
 					& \multicolumn{2}{c|}{$\epsilon=0$} & \multicolumn{2}{c|}{$\epsilon=0.01$} & \multicolumn{2}{c|}{$\epsilon=0.05$} & \multicolumn{2}{c}{$\epsilon=0.10$} \\
 					
 					\hline
 					& $D_{KL}$ & $m_{F}$ & $D_{KL}$ & $m_{F}$ & $D_{KL}$ & $m_{F}$ & $D_{KL}$ & $m_{F}$  \\
 					\hline
 					RGlassoWinsor & 5.840 & 4.792 & 6.233 & 4.965 & 8.998 & 6.112 & 13.667 & 7.233\\
 					& (0.057) & (0.026) & (0.074) & (0.035) & (0.089) & (0.028) & (0.101) & (0.020)\\
 					Glasso & 4.663 & 4.382 & 26.898 & 8.630 & 75.752 & 10.720 & 102.968 & 11.132\\
 					& (0.028) & (0.019) & (0.217) & (0.020) & (0.271) & (0.006) & (0.195) & (0.002)\\
 					RGlasso$Q_n$ & 8.823 & 6.133 & 6.750 & 5.127 & 30.508 & 9.153 & 56.545 & 10.307\\
 					& (0.080) & (0.024) & (0.055) & (0.028) & (0.295) & (0.021) & (0.334) & (0.010)\\
 					RGlassoTau & 6.240 & 5.087 & 6.369 & 4.946 & 25.478 & 8.725 & 70.107 & 10.635\\
 					& (0.054) & (0.024) & (0.052) & (0.028) & (0.414) & (0.037) & (0.527) & (0.012)\\
 					RGlassoGauss & 5.114 & 4.663 & 5.718 & 4.797 & 10.627 & 6.613 & 16.331 & 7.721\\
 					& (0.035) & (0.020) & (0.038) & (0.020) & (0.082) & (0.021) & (0.090) & (0.014)\\
 					RGlassoSpearman & 5.408 & 4.787 & 5.737 & 4.746 & 10.462 & 6.564 & 16.219 & 7.690\\
 					& (0.037) & (0.020) & (0.038) & (0.020) & (0.085) & (0.023) & (0.087) & (0.015)\\
 					RGlassoQuad & 10.731 & 6.690 & 8.962 & 5.546 & 16.392 & 7.763 & 22.210 & 8.464\\
 					& (0.079) & (0.020) & (0.045) & (0.016) & (0.107) & (0.017) & (0.135) & (0.015)\\
 					\hline
 			\end{tabular}}

 		}
 		
 		\caption{   Model $\text{Rand}$ under ICM. Comparison  of  means and standard deviations (in brackets) of  $\text{m}_F$  and $D_{KL}$ over $N=100$ replicates. $p=60$, $n=100$.} 
 		\label{num_performance_se_Random_p_60}%

 	\end{minipage}\hfill
 	\begin{minipage}[t]{.48\linewidth}
 		\centering
 		\scalebox{0.46}{
 		\begin{tabular}{l|rr|rr|rr|rr}
 			\hline
 			& \multicolumn{2}{c|}{$\epsilon=0$} & \multicolumn{2}{c|}{$\epsilon=0.01$} & \multicolumn{2}{c|}{$\epsilon=0.05$} & \multicolumn{2}{c}{$\epsilon=0.10$} \\
 			
 			\hline
 			& $D_{KL}$ & $m_{F}$ & $D_{KL}$ & $m_{F}$ & $D_{KL}$ & $m_{F}$ & $D_{KL}$ & $m_{F}$  \\
 			\hline
 			RGlassoWinsor & 20.043 & 7.549 & 20.427 & 7.827 & 25.615 & 9.209 & 37.412 & 10.860\\
 			& (0.083) & (0.021) & (0.083) & (0.027) & (0.125) & (0.028) & (0.207) & (0.025)\\
 			Glasso & 21.130 & 7.103 & 81.438 & 13.433 & 230.321 & 17.038 & 326.535 & 17.917\\
 			& (0.194) & (0.016) & (0.474) & (0.025) & (0.493) & (0.007) & (0.365) & (0.003)\\
 			RGlasso$Q_n$ & 23.106 & 8.413 & 27.929 & 9.358 & 63.926 & 12.670 & 130.203 & 15.252\\
 			& (0.072) & (0.026) & (0.174) & (0.039) & (0.468) & (0.033) & (0.672) & (0.021)\\
 			RGlassoTau & 21.434 & 7.897 & 25.409 & 8.926 & 69.364 & 13.042 & 179.583 & 16.333\\
 			& (0.084) & (0.022) & (0.141) & (0.039) & (0.543) & (0.033) & (1.392) & (0.025)\\
 			RGlassoGauss & 20.944 & 7.180 & 21.802 & 7.610 & 29.334 & 9.430 & 42.066 & 11.056\\
 			& (0.176) & (0.013) & (0.144) & (0.012) & (0.075) & (0.024) & (0.139) & (0.021)\\
 			RGlassoSpearman & 21.862 & 7.388 & 22.261 & 7.738 & 29.633 & 9.484 & 43.133 & 11.143\\
 			& (0.163) & (0.011) & (0.127) & (0.014) & (0.080) & (0.024) & (0.154) & (0.022)\\
 			RGlassoQuad & 33.513 & 9.454 & 35.316 & 9.858 & 46.618 & 11.278 & 63.532 & 12.579\\
 			& (0.073) & (0.016) & (0.065) & (0.020) & (0.141) & (0.021) & (0.220) & (0.020)\\
 			\hline
 	\end{tabular}}

 	\caption{   Model $\text{Rand}$ under ICM. Comparison  of  means and standard deviations (in brackets) of  $\text{m}_F$  and $D_{KL}$ over $N=100$ replicates. $p=200$, $n=100$.} 
 	\label{num_performance_se_Random_p_200}%
 	
 \end{minipage}\hfill
 \mbox{}
 
\end{table}

\begin{table}[H]
	\mbox{}\hfill
	\begin{minipage}[t]{.48\linewidth}
		\centering{ 
			
			\scalebox{0.5}{
	\begin{tabular}{l|rr|rr|rr|rr}
		\hline
		& \multicolumn{2}{c|}{$\epsilon=0$} & \multicolumn{2}{c|}{$\epsilon=0.01$} & \multicolumn{2}{c|}{$\epsilon=0.05$} & \multicolumn{2}{c}{$\epsilon=0.10$}  \\
		
		\hline
		& TPR & TNR & TPR & TNR & TPR & TNR & TPR & TNR   \\
		\hline
		RGlassoWinsor & 0.928 & 0.843 & 0.922 & 0.836 & 0.823 & 0.887 & 0.519 & 0.940\\
		& (0.003) & (0.003) & (0.004) & (0.005) & (0.006) & (0.003) & (0.012) & (0.003)\\
		Glasso & 0.960 & 0.795 & 0.234 & 0.973 & 0.010 & 0.985 & 0.032 & 0.968\\
		& (0.002) & (0.003) & (0.011) & (0.001) & (0.001) & (0.001) & (0.002) & (0.001)\\
		RGlasso$Q_n$ & 0.675 & 0.952 & 0.887 & 0.801 & 0.034 & 1.000 & 0.002 & 1.000\\
		& (0.008) & (0.002) & (0.004) & (0.004) & (0.005) & (0.000) & (0.000) & (0.000)\\
		RGlassoTau & 0.856 & 0.868 & 0.892 & 0.800 & 0.160 & 0.993 & 0.000 & 1.000\\
		& (0.005) & (0.003) & (0.004) & (0.004) & (0.015) & (0.001) & (0.000) & (0.000)\\
		RGlassoGauss & 0.946 & 0.826 & 0.942 & 0.763 & 0.745 & 0.870 & 0.479 & 0.924\\
		& (0.003) & (0.003) & (0.002) & (0.004) & (0.007) & (0.003) & (0.009) & (0.003)\\
		RGlassoSpearman & 0.930 & 0.832 & 0.941 & 0.758 & 0.771 & 0.879 & 0.536 & 0.925\\
		& (0.003) & (0.003) & (0.003) & (0.004) & (0.007) & (0.003) & (0.009) & (0.003)\\
		RGlassoQuad & 0.538 & 0.932 & 0.811 & 0.683 & 0.351 & 0.962 & 0.229 & 0.976\\
		& (0.009) & (0.002) & (0.004) & (0.004) & (0.012) & (0.002) & (0.010) & (0.002)\\
		\hline
	\end{tabular}		
		
	}
		}
		
		\caption{ Model $\text{Rand}$ under ICM.  Comparison  of  means and standard deviations (in brackets) of  TPR and TNR  over $N=100$ replicates. $p=60, n=100$. } 
		\label{TPTN_Random_p_60}%
	\end{minipage}\hfill
	\begin{minipage}[t]{.48\linewidth}
		\centering
		\scalebox{0.5}{		
				\begin{tabular}{l|r|r|r|r}
				\hline
				& \multicolumn{1}{c|}{$\epsilon=0$} & \multicolumn{1}{c|}{$\epsilon=0.01$} & \multicolumn{1}{c|}{$\epsilon=0.05$} & \multicolumn{1}{c}{$\epsilon=0.10$}   \\
				
				\hline
				RGlassoWinsor & 0.422 & 0.414 & 0.437 & 0.365\\
				& (0.004) & (0.005) & (0.005) & (0.005)\\
				Glasso & 0.380 & 0.227 & -0.010 & -0.001\\
				& (0.003) & (0.008) & (0.002) & (0.002)\\
				RGlasso$Q_n$ & 0.504 & 0.354 & 0.127 & 0.017\\
				& (0.004) & (0.004) & (0.010) & (0.004)\\
				RGlassoTau & 0.418 & 0.355 & 0.253 & -0.003\\
				& (0.003) & (0.004) & (0.013) & (0.000)\\
				RGlassoGauss & 0.408 & 0.342 & 0.365 & 0.299\\
				& (0.004) & (0.003) & (0.003) & (0.003)\\
				RGlassoSpearman & 0.408 & 0.337 & 0.391 & 0.340\\
				& (0.003) & (0.003) & (0.003) & (0.004)\\
				RGlassoQuad & 0.355 & 0.224 & 0.305 & 0.247\\
				& (0.004) & (0.002) & (0.005) & (0.006)\\
				\hline
		\end{tabular}}
		\vspace{0.2cm}
		\centering 
		\caption{   Model $\text{Rand}$ under ICM. Comparison  of  means and standard deviations (in brackets) of  MCC over $N=100$ replicates. $p=60$, $n=100$.} 
		\label{MCC_se_Random_p_60}%
		
	\end{minipage}\hfill
	\mbox{}
\end{table}

\begin{table}[H]
	\mbox{}\hfill
	\begin{minipage}[t]{.48\linewidth}
		\centering{ 
			
			\scalebox{0.5}{
		\begin{tabular}{l|rr|rr|rr|rr}
			\hline
			& \multicolumn{2}{c|}{$\epsilon=0$} & \multicolumn{2}{c|}{$\epsilon=0.01$} & \multicolumn{2}{c|}{$\epsilon=0.05$} & \multicolumn{2}{c}{$\epsilon=0.10$}  \\
			
			\hline
			& TPR & TNR & TPR & TNR & TPR & TNR & TPR & TNR   \\
			
			\hline
			RGlassoWinsor & 0.903 & 0.869 & 0.888 & 0.882 & 0.764 & 0.921 & 0.479 & 0.951\\
			& (0.003) & (0.003) & (0.004) & (0.003) & (0.006) & (0.002) & (0.010) & (0.002)\\
			Glasso & 0.952 & 0.794 & 0.118 & 0.984 & 0.037 & 0.966 & 0.055 & 0.946\\
			& (0.002) & (0.002) & (0.006) & (0.001) & (0.002) & (0.001) & (0.002) & (0.001)\\
			RGlasso$Q_n$ & 0.839 & 0.866 & 0.799 & 0.895 & 0.636 & 0.946 & 0.508 & 0.961\\
			& (0.004) & (0.003) & (0.006) & (0.003) & (0.008) & (0.002) & (0.008) & (0.001)\\
			RGlassoTau & 0.858 & 0.855 & 0.805 & 0.891 & 0.508 & 0.957 & 0.115 & 0.974\\
			& (0.004) & (0.003) & (0.006) & (0.003) & (0.010) & (0.002) & (0.007) & (0.001)\\
			RGlassoGauss & 0.944 & 0.806 & 0.916 & 0.815 & 0.769 & 0.845 & 0.582 & 0.861\\
			& (0.002) & (0.002) & (0.002) & (0.003) & (0.004) & (0.003) & (0.005) & (0.002)\\
			RGlassoSpearman & 0.932 & 0.803 & 0.912 & 0.817 & 0.800 & 0.851 & 0.635 & 0.866\\
			& (0.002) & (0.002) & (0.002) & (0.003) & (0.004) & (0.003) & (0.005) & (0.002)\\
			RGlassoQuad & 0.736 & 0.784 & 0.714 & 0.801 & 0.621 & 0.835 & 0.525 & 0.852\\
			& (0.003) & (0.002) & (0.003) & (0.002) & (0.004) & (0.002) & (0.005) & (0.002)\\
			\hline
		\end{tabular}		
			
		}
		}
		
		\caption{ Model $\text{Rand}$ under ICM.  Comparison  of  means and standard deviations (in brackets) of  TPR and TNR  over $N=100$ replicates. $p=200, n=100$. } 
		\label{TPTN_Random_p_200}%
	\end{minipage}\hfill
	\begin{minipage}[t]{.48\linewidth}
		\centering
		\scalebox{0.5}{		
				\begin{tabular}{l|r|r|r|r}
				\hline
				& \multicolumn{1}{c|}{$\epsilon=0$} & \multicolumn{1}{c|}{$\epsilon=0.01$} & \multicolumn{1}{c|}{$\epsilon=0.05$} & \multicolumn{1}{c}{$\epsilon=0.10$}   \\
				
				\hline
				RGlassoWinsor & 0.237 & 0.248 & 0.263 & 0.206\\
				& (0.002) & (0.003) & (0.003) & (0.002)\\
				Glasso & 0.192 & 0.078 & 0.002 & 0.000\\
				& (0.001) & (0.004) & (0.001) & (0.001)\\
				RGlasso$Q_n$ & 0.215 & 0.237 & 0.263 & 0.245\\
				& (0.002) & (0.003) & (0.002) & (0.002)\\
				RGlassoTau & 0.210 & 0.233 & 0.235 & 0.054\\
				& (0.002) & (0.003) & (0.002) & (0.003)\\
				RGlassoGauss & 0.197 & 0.196 & 0.177 & 0.133\\
				& (0.001) & (0.002) & (0.001) & (0.001)\\
				RGlassoSpearman & 0.192 & 0.196 & 0.190 & 0.153\\
				& (0.001) & (0.002) & (0.001) & (0.001)\\
				RGlassoQuad & 0.132 & 0.134 & 0.128 & 0.111\\
				& (0.001) & (0.001) & (0.001) & (0.001)\\
				\hline
		\end{tabular}}
	\vspace{0.2cm}
\centering 
\caption{   Model $\text{Rand}$ under ICM. Comparison  of  means and standard deviations (in brackets) of  MCC over $N=100$ replicates. $p=200$, $n=100$.} 
\label{MCC_se_Random_p_200}%
		
	\end{minipage}\hfill
	\mbox{}
\end{table}

  \begin{table}[H]
 	\mbox{}\hfill
 	\begin{minipage}[t]{.48\linewidth}
 		\centering{ 
 			
 			\scalebox{0.46}{
 					\begin{tabular}{l|rr|rr|rr|rr}
 					\hline
 					& \multicolumn{2}{c|}{$\epsilon=0$} & \multicolumn{2}{c|}{$\epsilon=0.01$} & \multicolumn{2}{c|}{$\epsilon=0.05$} & \multicolumn{2}{c}{$\epsilon=0.10$} \\
 					
 					\hline
 					& $D_{KL}$ & $m_{F}$ & $D_{KL}$ & $m_{F}$ & $D_{KL}$ & $m_{F}$ & $D_{KL}$ & $m_{F}$  \\
 					\hline
 					RGlassoWinsor & 5.011 & 5.943 & 5.315 & 6.168 & 7.447 & 7.007 & 12.139 & 8.082\\
 					& (0.046) & (0.023) & (0.049) & (0.024) & (0.066) & (0.023) & (0.103) & (0.019)\\
 					Glasso & 3.973 & 5.664 & 27.864 & 9.499 & 74.258 & 11.171 & 101.902 & 11.560\\
 					& (0.028) & (0.016) & (0.326) & (0.029) & (0.250) & (0.006) & (0.201) & (0.002)\\
 					RGlasso$Q_n$ & 7.637 & 7.119 & 9.617 & 7.600 & 27.930 & 9.684 & 55.192 & 10.811\\
 					& (0.093) & (0.024) & (0.136) & (0.029) & (0.342) & (0.025) & (0.393) & (0.011)\\
 					RGlassoTau & 5.283 & 6.269 & 6.508 & 6.736 & 23.809 & 9.353 & 68.778 & 11.102\\
 					& (0.057) & (0.021) & (0.072) & (0.023) & (0.427) & (0.036) & (0.606) & (0.012)\\
 					RGlassoGauss & 4.316 & 5.835 & 5.205 & 6.288 & 9.277 & 7.524 & 14.755 & 8.478\\
 					& (0.035) & (0.019) & (0.045) & (0.023) & (0.073) & (0.018) & (0.088) & (0.014)\\
 					RGlassoSpearman & 4.616 & 5.989 & 5.347 & 6.332 & 9.108 & 7.445 & 14.529 & 8.415\\
 					& (0.036) & (0.017) & (0.042) & (0.019) & (0.077) & (0.020) & (0.089) & (0.015)\\
 					RGlassoQuad & 9.150 & 7.583 & 10.060 & 7.772 & 14.573 & 8.478 & 20.553 & 9.122\\
 					& (0.080) & (0.018) & (0.075) & (0.016) & (0.099) & (0.014) & (0.123) & (0.013)\\
 					\hline
 			\end{tabular} }

 		}
 		
 		\caption{Model $\text{Hub}$. $p=60, n=100$.  Comparison  of  means and standard deviations (in brackets) of  $\text{m}_F$  and $D_{KL}$ over $N=100$ replicates. $p=60$, $n=100$.} 
 		\label{num_performance_hub_p_60}
 		
 	\end{minipage}\hfill
 	\begin{minipage}[t]{.48\linewidth}
 		\centering
 		\scalebox{0.46}{
 			\begin{tabular}{l|rr|rr|rr|rr}
 			\hline
 			& \multicolumn{2}{c|}{$\epsilon=0$} & \multicolumn{2}{c|}{$\epsilon=0.01$} & \multicolumn{2}{c|}{$\epsilon=0.05$} & \multicolumn{2}{c}{$\epsilon=0.10$} \\
 			
 			\hline
 			& $D_{KL}$ & $m_{F}$ & $D_{KL}$ & $m_{F}$ & $D_{KL}$ & $m_{F}$ & $D_{KL}$ & $m_{F}$  \\
 			\hline
 			RGlassoWinsor & 21.474 & 11.976 & 22.658 & 12.237 & 30.340 & 13.575 & 137.550 & 18.697\\
 			& (0.133) & (0.029) & (0.141) & (0.031) & (0.163) & (0.025) & (3.893) & (0.086)\\
 			Glasso & 17.898 & 11.392 & 94.248 & 17.427 & 251.829 & 20.449 & Inf & 22.052\\
 			& (0.086) & (0.034) & (0.578) & (0.025) & (0.576) & (0.006) & NaN & (0.012)\\
 			RGlasso$Q_n$ & 58.669 & 16.220 & 71.529 & 16.866 & 142.258 & 18.997 & Inf & 21.685\\
 			& (0.313) & (0.019) & (0.398) & (0.019) & (0.739) & (0.015) & NaN & (0.012)\\
 			RGlassoTau & 27.397 & 13.236 & 35.904 & 14.263 & 128.506 & 18.679 & Inf & 21.592\\
 			& (0.185) & (0.027) & (0.234) & (0.026) & (0.817) & (0.019) & NaN & (0.023)\\
 			RGlassoGauss & 19.221 & 11.724 & 22.336 & 12.363 & 36.577 & 14.369 & 181.468 & 19.497\\
 			& (0.102) & (0.028) & (0.129) & (0.029) & (0.174) & (0.022) & (4.486) & (0.070)\\
 			RGlassoSpearman & 20.910 & 12.097 & 23.621 & 12.601 & 37.259 & 14.393 & 178.486 & 19.437\\
 			& (0.118) & (0.030) & (0.133) & (0.027) & (0.194) & (0.024) & (4.266) & (0.073)\\
 			RGlassoQuad & 43.709 & 15.100 & 47.328 & 15.376 & 226.795 & 20.026 & 231.525 & 20.209\\
 			& (0.231) & (0.022) & (0.240) & (0.022) & (6.527) & (0.108) & (4.780) & (0.056)\\
 			\hline
 	\end{tabular}	}

 	\caption{   Model $\text{Hub}$ under ICM. Comparison  of  means and standard deviations (in brackets) of  $\text{m}_F$  and $D_{KL}$ over $N=100$ replicates. $p=200$, $n=100$.} 
 	\label{num_performance_se_hub_p_200}%
 	
 \end{minipage}\hfill
 \mbox{}
 
\end{table}

\begin{table}[H]
	\mbox{}\hfill
	\begin{minipage}[t]{.48\linewidth}
		\centering{ 
			
			\scalebox{0.5}{
			\begin{tabular}{l|rr|rr|rr|rr}
			\hline
			& \multicolumn{2}{c|}{$\epsilon=0$} & \multicolumn{2}{c|}{$\epsilon=0.01$} & \multicolumn{2}{c|}{$\epsilon=0.05$} & \multicolumn{2}{c}{$\epsilon=0.10$}  \\
			
			\hline
			& TPR & TNR & TPR & TNR & TPR & TNR & TPR & TNR   \\
			\hline
			RGlassoWinsor & 0.954 & 0.836 & 0.958 & 0.839 & 0.892 & 0.860 & 0.646 & 0.920\\
			& (0.003) & (0.003) & (0.004) & (0.003) & (0.006) & (0.003) & (0.015) & (0.003)\\
			Glasso & 0.971 & 0.804 & 0.131 & 0.985 & 0.012 & 0.983 & 0.027 & 0.970\\
			& (0.002) & (0.003) & (0.015) & (0.001) & (0.002) & (0.001) & (0.003) & (0.001)\\
			RGlasso$Q_n$ & 0.750 & 0.941 & 0.671 & 0.956 & 0.099 & 0.998 & 0.012 & 1.000\\
			& (0.013) & (0.002) & (0.020) & (0.002) & (0.017) & (0.001) & (0.003) & (0.000)\\
			RGlassoTau & 0.890 & 0.859 & 0.861 & 0.885 & 0.237 & 0.988 & 0.002 & 0.999\\
			& (0.007) & (0.002) & (0.008) & (0.002) & (0.023) & (0.001) & (0.001) & (0.000)\\
			RGlassoGauss & 0.967 & 0.820 & 0.950 & 0.824 & 0.802 & 0.863 & 0.573 & 0.909\\
			& (0.003) & (0.003) & (0.003) & (0.003) & (0.008) & (0.003) & (0.013) & (0.003)\\
			RGlassoSpearman & 0.952 & 0.834 & 0.945 & 0.838 & 0.830 & 0.866 & 0.644 & 0.905\\
			& (0.003) & (0.003) & (0.004) & (0.002) & (0.007) & (0.003) & (0.012) & (0.003)\\
			RGlassoQuad & 0.634 & 0.921 & 0.620 & 0.924 & 0.458 & 0.949 & 0.312 & 0.970\\
			& (0.011) & (0.002) & (0.010) & (0.002) & (0.013) & (0.002) & (0.014) & (0.002)\\
			\hline
		\end{tabular}	
		}
		}
		
		\caption{ Model $\text{Hub}$ under ICM.  Comparison  of  means and standard deviations (in brackets) of  TPR and TNR  over $N=100$ replicates. $p=60, n=100$. } 
		\label{TPTN_Hub_p_60}%
	\end{minipage}\hfill
	\begin{minipage}[t]{.48\linewidth}
		\centering
		\scalebox{0.5}{		
			\begin{tabular}{l|r|r|r|r}
			\hline
			& \multicolumn{1}{c|}{$\epsilon=0$} & \multicolumn{1}{c|}{$\epsilon=0.01$} & \multicolumn{1}{c|}{$\epsilon=0.05$} & \multicolumn{1}{c}{$\epsilon=0.10$}   \\
			
			\hline
			RGlassoWinsor & 0.359 & 0.366 & 0.363 & 0.344\\
			& (0.004) & (0.004) & (0.004) & (0.006)\\
			Glasso & 0.332 & 0.128 & -0.007 & -0.004\\
			& (0.003) & (0.013) & (0.002) & (0.003)\\
			RGlasso$Q_n$ & 0.452 & 0.459 & 0.182 & 0.050\\
			& (0.007) & (0.008) & (0.017) & (0.010)\\
			RGlassoTau & 0.358 & 0.384 & 0.258 & 0.004\\
			& (0.004) & (0.004) & (0.016) & (0.003)\\
			RGlassoGauss & 0.347 & 0.344 & 0.325 & 0.279\\
			& (0.003) & (0.003) & (0.004) & (0.005)\\
			RGlassoSpearman & 0.356 & 0.357 & 0.344 & 0.311\\
			& (0.004) & (0.003) & (0.005) & (0.005)\\
			RGlassoQuad & 0.336 & 0.335 & 0.298 & 0.257\\
			& (0.005) & (0.005) & (0.007) & (0.008)\\
			\hline
		\end{tabular}	
		}
\vspace{0.2cm}
\centering 
\caption{   Model Hub under ICM. Comparison  of  means and standard deviations (in brackets) of  MCC over $N=100$ replicates. $p=60$, $n=100$.} 
\label{MCC_hub_p_60}%
	
\end{minipage}\hfill
\mbox{}
\end{table}

\begin{table}[H]
	\mbox{}\hfill
	\begin{minipage}[t]{.48\linewidth}
		\centering{ 
			
			\scalebox{0.5}{
		\begin{tabular}{l|rr|rr|rr|rr}
			\hline
			& \multicolumn{2}{c|}{$\epsilon=0$} & \multicolumn{2}{c|}{$\epsilon=0.01$} & \multicolumn{2}{c|}{$\epsilon=0.05$} & \multicolumn{2}{c}{$\epsilon=0.10$}  \\
			
			\hline
			& TPR & TNR & TPR & TNR & TPR & TNR & TPR & TNR   \\
			
			\hline
			RGlassoWinsor & 0.925 & 0.934 & 0.909 & 0.938 & 0.810 & 0.955 & 0.188 & 0.995\\
			& (0.003) & (0.001) & (0.004) & (0.002) & (0.005) & (0.001) & (0.020) & (0.001)\\
			Glasso & 0.958 & 0.919 & 0.063 & 0.990 & 0.004 & 0.990 & 0.077 & 0.971\\
			& (0.002) & (0.002) & (0.005) & (0.000) & (0.000) & (0.000) & (0.007) & (0.002)\\
			RGlasso$Q_n$ & 0.028 & 1.000 & 0.012 & 1.000 & 0.001 & 1.000 & 0.012 & 0.999\\
			& (0.003) & (0.000) & (0.001) & (0.000) & (0.000) & (0.000) & (0.005) & (0.000)\\
			RGlassoTau & 0.783 & 0.965 & 0.678 & 0.978 & 0.007 & 1.000 & 0.005 & 1.000\\
			& (0.007) & (0.001) & (0.009) & (0.001) & (0.001) & (0.000) & (0.002) & (0.000)\\
			RGlassoGauss & 0.943 & 0.931 & 0.902 & 0.932 & 0.702 & 0.950 & 0.271 & 0.984\\
			& (0.002) & (0.002) & (0.003) & (0.002) & (0.006) & (0.001) & (0.021) & (0.002)\\
			RGlassoSpearman & 0.924 & 0.933 & 0.896 & 0.936 & 0.749 & 0.950 & 0.350 & 0.984\\
			& (0.003) & (0.002) & (0.004) & (0.002) & (0.006) & (0.001) & (0.025) & (0.002)\\
			RGlassoQuad & 0.556 & 0.965 & 0.518 & 0.967 & 0.132 & 0.995 & 0.123 & 0.996\\
			& (0.008) & (0.001) & (0.009) & (0.001) & (0.018) & (0.001) & (0.017) & (0.001)\\
			\hline
		\end{tabular}		
			
		}
		}
		
		\caption{ Model $\text{Hub}$ under ICM.  Comparison  of  means and standard deviations (in brackets) of  TPR and TNR  over $N=100$ replicates. $p=200, n=100$. } 
		\label{TPTN_Hub_p_200}%
	\end{minipage}\hfill
	\begin{minipage}[t]{.48\linewidth}
		\centering
		\scalebox{0.5}{		
			\begin{tabular}{l|r|r|r|r}
			\hline
			& \multicolumn{1}{c|}{$\epsilon=0$} & \multicolumn{1}{c|}{$\epsilon=0.01$} & \multicolumn{1}{c|}{$\epsilon=0.05$} & \multicolumn{1}{c}{$\epsilon=0.10$}   \\
			
			\hline
			RGlassoWinsor & 0.322 & 0.329 & 0.341 & 0.190\\
			& (0.003) & (0.004) & (0.003) & (0.013)\\
			Glasso & 0.305 & 0.046 & -0.006 & 0.024\\
			& (0.004) & (0.004) & (0.000) & (0.003)\\
			RGlasso$Q_n$ & 0.144 & 0.084 & 0.010 & 0.023\\
			& (0.007) & (0.007) & (0.003) & (0.006)\\
			RGlassoTau & 0.367 & 0.395 & 0.053 & 0.017\\
			& (0.003) & (0.004) & (0.005) & (0.004)\\
			RGlassoGauss & 0.324 & 0.312 & 0.281 & 0.179\\
			& (0.003) & (0.003) & (0.003) & (0.009)\\
			RGlassoSpearman & 0.322 & 0.317 & 0.301 & 0.226\\
			& (0.003) & (0.003) & (0.003) & (0.012)\\
			RGlassoQuad & 0.265 & 0.256 & 0.124 & 0.146\\
			& (0.003) & (0.003) & (0.013) & (0.013)\\
			\hline
		\end{tabular}
		}
	\vspace{0.2cm}
\centering 
\caption{   Model $\text{Hub}$. Comparison  of  means and standard deviations (in brackets) of  MCC over $N=100$ replicates. $p=200$, $n=100$.} 
\label{MCC_se_hub_p_200}%
		
	\end{minipage}\hfill
	\mbox{}
\end{table}

\newpage 

 
\begin{figure}[H]
	\captionsetup[subfloat]{farskip=0.5pt,captionskip=0.5pt}
	\centering
	\subfloat[Glasso-$\epsilon=0$]{\includegraphics[width = 1in]{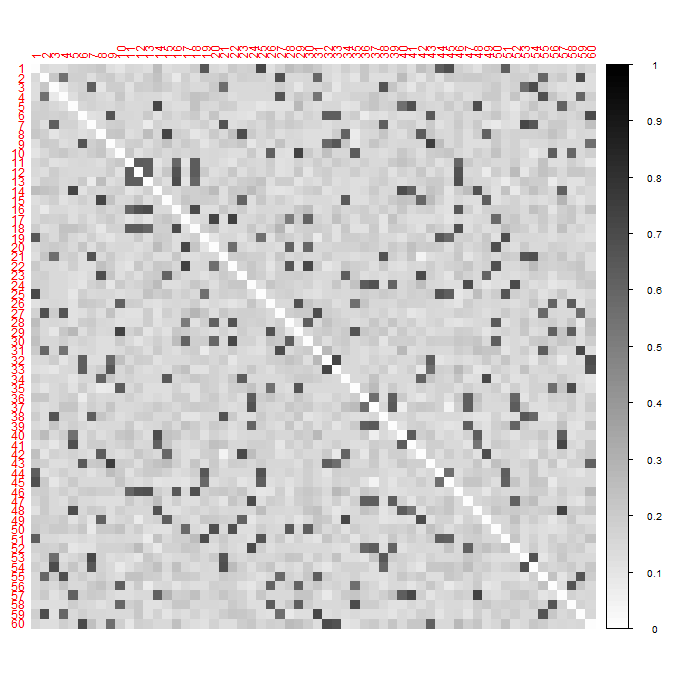}}\hfill
	\subfloat[RGlassoWinsor-$\epsilon=0$]{\includegraphics[width = 1in]{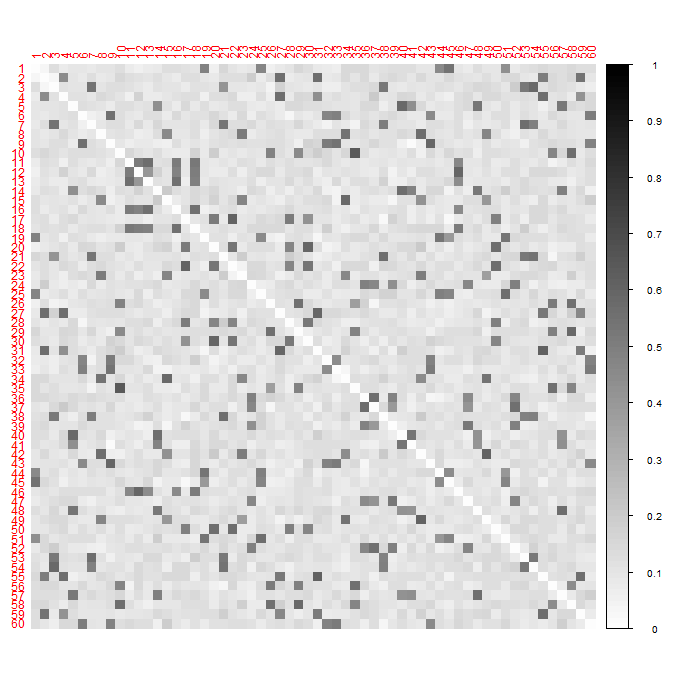}}\hfill
	\subfloat[Glasso-$\epsilon=0.01$]{\includegraphics[width = 1in]{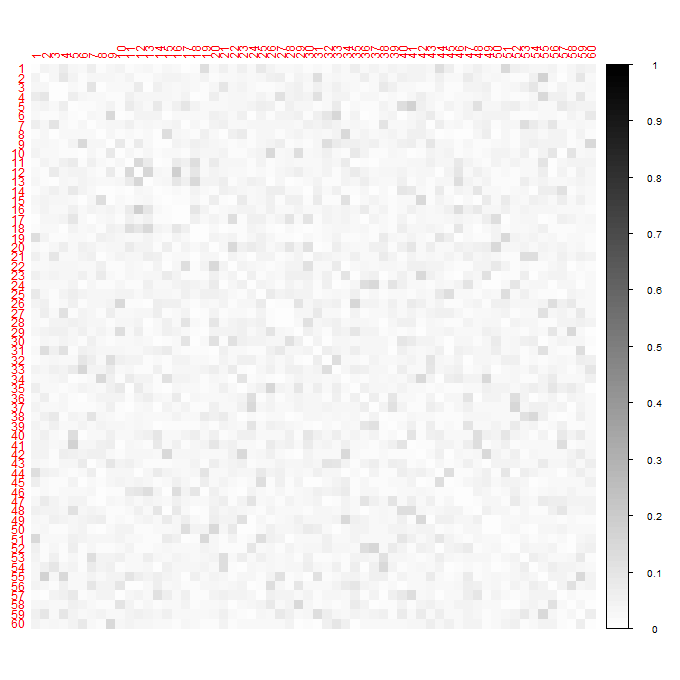}}\hfill
 	\subfloat[RGlassoWinsor-$\epsilon=0.01$]{\includegraphics[width = 1in]{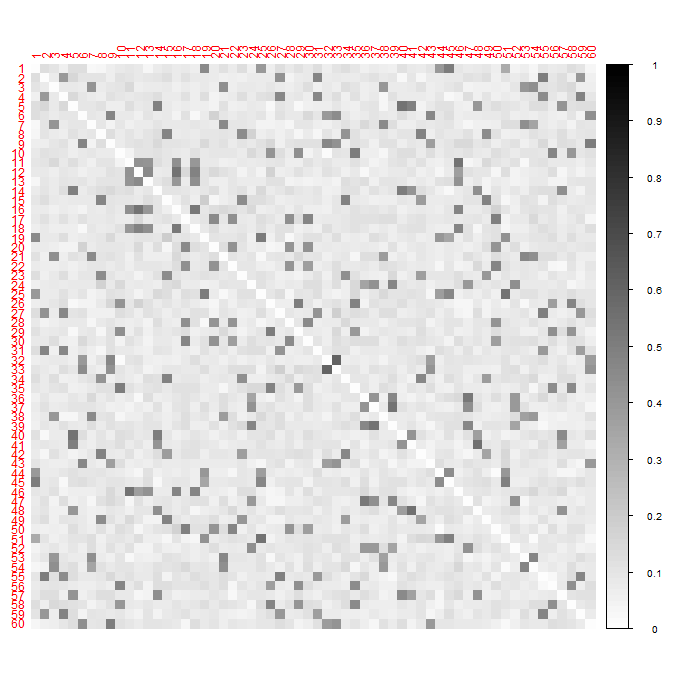}}\hfill \\
	\subfloat[Glasso-$\epsilon=0.05$]{\includegraphics[width = 1in]{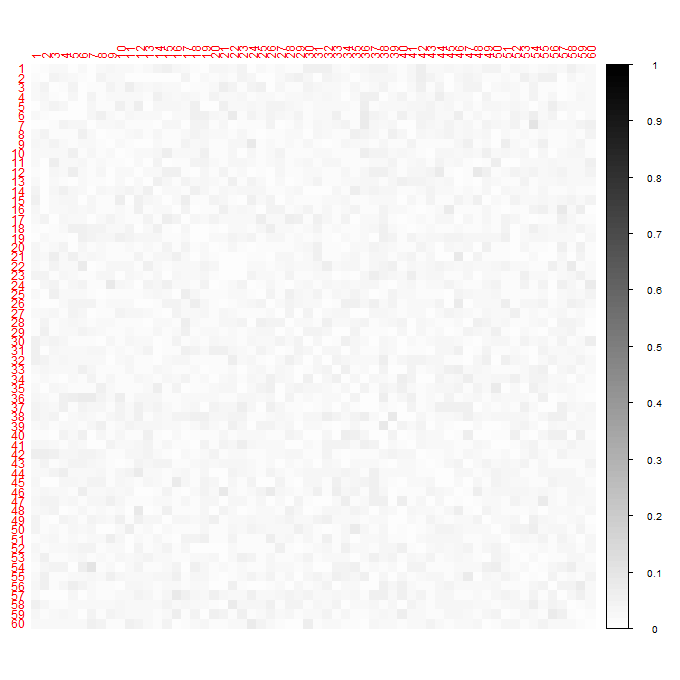}}\hfill
	\subfloat[RGlassoWinsor-$\epsilon=0.05$]{\includegraphics[width = 1in]{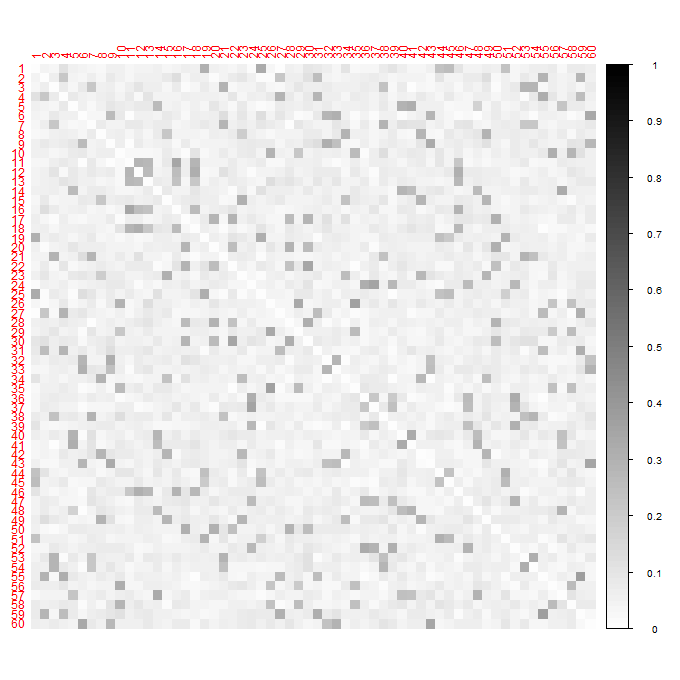}}\hfill
	\subfloat[Glasso-$\epsilon=0.10$]{\includegraphics[width = 1in]{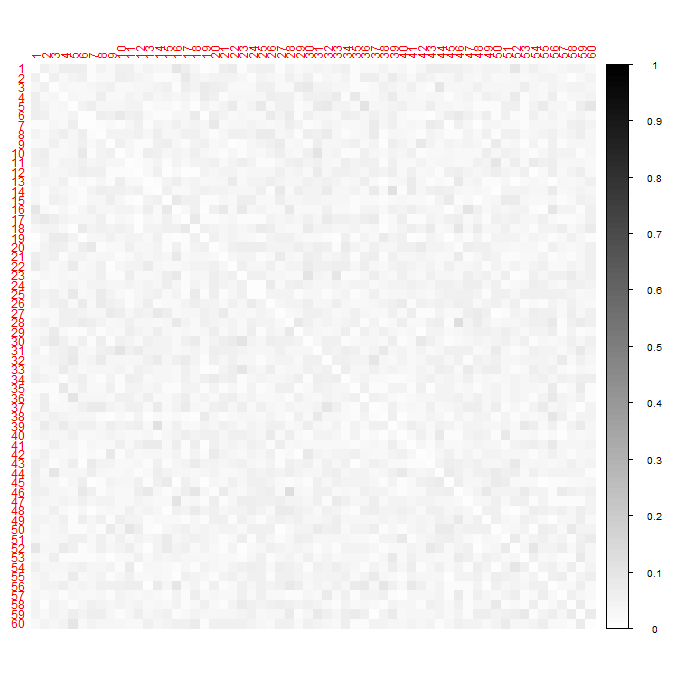}}\hfill
	\subfloat[RGlassoWinsor-$\epsilon=0.10$]{\includegraphics[width = 1in]{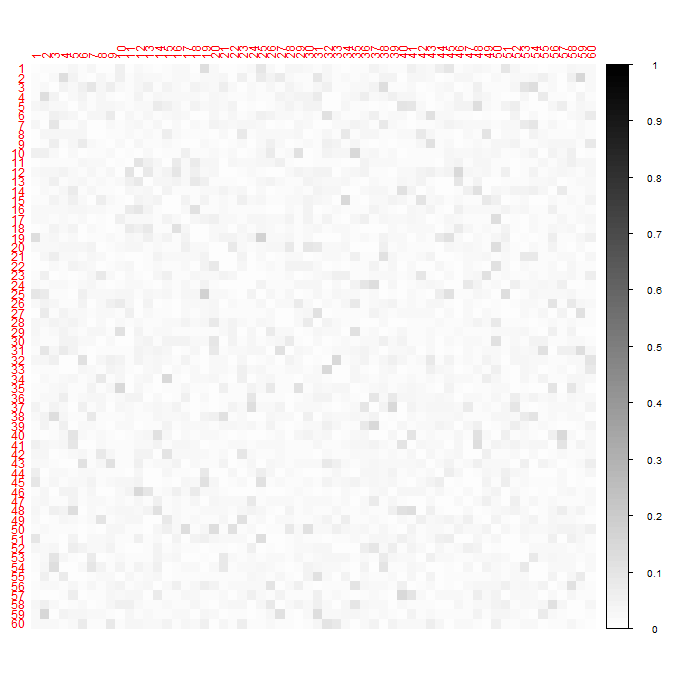}}\hfill

\caption{Heatmaps for the frequency of adjancency for each pair of nodes over  $N=100$   replicates.  ICM. }
	\label{Fig_Glasso_Winsor_Block_p_60}
\end{figure}   
 

  \section*{Appendix B}
  This section contains numerical and  performance classification performance  results for the seven the seven precision matrix estimators  applied to  $\text{AR(1)}$,  $\text{BG}$,  $\text{NN}(2)$, $\text{Rand}$ and $\text{Hub}$ models. We compare two scenarios $p=60$ and $p=200$ with $n=100$ under  THCM.

  \begin{table}[H]
  	\mbox{}\hfill
  	\begin{minipage}[t]{.48\linewidth}
  		\centering{ 
  			
  			\scalebox{0.5}{
\begin{tabular}{l|rr|rr}
  				\hline
  				& \multicolumn{2}{c|}{$\epsilon=0.05$ 
  				} & \multicolumn{2}{c}{$\epsilon=0.10$}   \\
  				\hline
  				& $D_{KL}$ & $m_{F}$ & $D_{KL}$ & $m_{F}$  \\
  				\hline
  				RGlassoWinsor & 8.947 & 5.265 & 8.974 & 5.286\\
  				
  				& 0.153 & 0.039 & 0.154 & 0.039\\
  				
  				Glasso & 63.652 & 10.374 & 63.578 & 10.372\\
  				
  				& 1.355 & 0.040 & 1.388 & 0.040\\
  				
  				RGlassoQn & 21.897 & 8.151 & 36.383 & 9.215\\
  				
  				& 0.995 & 0.107 & 1.674 & 0.118\\
  				
  				RGlassoTau & 13.671 & 6.854 & 31.368 & 8.836\\
  				
  				& 0.740 & 0.110 & 1.512 & 0.112\\
  				
  				RGlassoGauss & 10.525 & 5.776 & 14.832 & 6.696\\
  				
  				& 0.208 & 0.053 & 0.224 & 0.049\\
  				
  				RGlassoSpearman & 9.776 & 5.829 & 14.258 & 6.711\\
  				
  				& 0.264 & 0.057 & 0.284 & 0.059\\
  				
  				RGlassoQuad & 16.301 & 7.666 & 22.257 & 8.356\\
  				
  				& 0.315 & 0.047 & 0.460 & 0.042\\
  				\hline
  		\end{tabular}}
  		}
  		
  	\caption{Model $\text{AR}(1)$ under THCM.    Comparison  of  means and standard deviations (in brackets) of  $\text{m}_F$  and $D_{KL}$ over $N=100$ replicates. $p=60$, $n=100$.} 
  	\label{num_performance_se_ar1_p_60_THCM}%
  		
  	\end{minipage}\hfill
  	\begin{minipage}[t]{.48\linewidth}
  		\centering
  		\scalebox{0.5}{
  		\begin{tabular}{l|rr|rr}
  			\hline
  			& \multicolumn{2}{c|}{$\epsilon=0.05$ 
  			} & \multicolumn{2}{c}{$\epsilon=0.10$}   \\
  			\hline
  			& $D_{KL}$ & $m_{F}$ & $D_{KL}$ & $m_{F}$  \\
  			\hline
  			RGlassoWinsor & 32.558 & 10.495 & 36.889 & 11.059\\
  			
  			& (0.321) & (0.057) & (0.369) &(0.068)\\
  			
  			Glasso & 96.808 & 15.576 & 239.063 & 18.498\\
  			
  			& (7.168) & (0.194) & (14.250) & (0.251)\\
  			
  			RGlassoQn & 118.090 & 17.297 & 176.665 & 18.639\\
  			
  			& (3.229) & (0.096) & (5.117) & (0.097)\\
  			
  			RGlassoTau & 62.238 & 14.253 & 120.482 & 16.651\\
  			
  			& (2.542) & (0.152) & (6.647) & (0.198)\\
  			
  			RGlassoGauss & 32.797 & 10.395 & 39.384 & 11.403\\
  			
  			& (0.324) & (0.068) & (0.576) & (0.102)\\
  			
  			RGlassoSpearman & 33.863 & 10.886 & 42.141 & 11.959\\
  			
  			& (0.377) & (0.054) & (0.680) & (0.093)\\
  			
  			RGlassoQuad & 65.449 & 14.859 & 79.980 & 15.534\\
  			
  			& (0.800) & (0.049) & (1.150) & (0.056)\\
  			\hline
  	\end{tabular}}
  		
  	\caption{Model $\text{AR}(1)$ under THCM.    Comparison  of  means and standard deviations (in brackets) of  $\text{m}_F$  and $D_{KL}$ over $N=100$ replicates. $p=200$, $n=100$.} 
  	  	\label{num_performance_se_ar1_p_200_THCM}%
  	\end{minipage}\hfill
  	\mbox{}
  	  \end{table}
    
    \begin{table}[H]
    	\mbox{}\hfill
    	\begin{minipage}[t]{.48\linewidth}
    		\centering{ 
    			
    			\scalebox{0.5}{
    				\begin{tabular}{l|rr|rr}
    					\hline
    					& \multicolumn{2}{c|}{$\epsilon=0.05$} & \multicolumn{2}{c}{$\epsilon=0.10$}   \\
    					\hline
    					& TPR & TNR & TPR & TNR       \\
    					\hline
    					RGlassoWinsor & 0.962 & 0.709 & 0.959 & 0.712\\
    					
    					& (0.004)& (0.006) & (0.005) & (0.007)\\
    					
    					Glasso & 0.020 & 0.574 & 0.020 & 0.571\\
    					
    					& (0.008) & (0.005) & (0.008) & (0.005)\\
    					
    					RGlassoQn & 0.415 & 0.979 & 0.227 & 0.984\\
    					
    					& (0.035) & (0.003) & (0.032) & (0.003)\\
    					
    					RGlassoTau & 0.799 & 0.905 & 0.452 & 0.928\\
    					
    					& (0.025) & (0.005) & (0.039) & (0.007)\\
    					
    					RGlassoGauss & 0.917 & 0.675 & 0.699 & 0.667\\
    					
    					& (0.009) & (0.005) & (0.023) & (0.002)\\
    					
    					RGlassoSpearman & 0.946 & 0.719 & 0.886 & 0.665\\
    					
    					& (0.010) & (0.008) & (0.012) & (0.003)\\
    					
    					RGlassoQuad & 0.563 & 0.939 & 0.512 & 0.880\\
    					
    					& (0.015) & (0.003) & (0.017) & (0.007) \\
    					\hline
    			\end{tabular}}
    		}
    		
    		\caption{ Model $\text{AR}(1)$ under THCM.  Comparison  of  means and standard deviations (in brackets) of  TPR and TNR  over $N=100$ replicates. $p=60, n=100$. } 
    		\label{TPTN_ar1_p_60_THCM}%
    	\end{minipage}\hfill
    	\begin{minipage}[t]{.48\linewidth}
    		\centering
    		\scalebox{0.5}{		
    		    		\begin{tabular}{l|l|l}
    			\hline
    			& $\epsilon=0.05$ &$\epsilon=0.10$ \\
    			\hline
    			RGlassoWinsor & 0.266 & 0.268\\
    			
    			& 0.005 & 0.005\\
    			
    			Glasso & -0.149 & -0.149\\
    			
    			& 0.003 & 0.003\\
    			
    			RGlassoQn & 0.350 & 0.182\\
    			
    			& 0.023 & 0.021\\
    			
    			RGlassoTau & 0.400 & 0.216\\
    			
    			& 0.011 & 0.016\\
    			
    			RGlassoGauss & 0.226 & 0.137\\
    			
    			& 0.004 & 0.008\\
    			
    			RGlassoSpearman & 0.266 & 0.207\\
    			
    			& 0.006 & 0.004\\
    			
    			RGlassoQuad & 0.351 & 0.220\\
    			
    			& 0.008 & 0.008\\
    			\hline
    		\end{tabular}
    	}
    		\vspace{0.15cm}
    		\centering 
    		\caption{    Model $\text{AR}(1)$ under THCM.    Comparison  of  means and standard deviations (in brackets) of  MCC over $N=100$ replicates. $p=60$, $n=100$.} 
    		\label{MCC_se_ar1_p_60_THCM}%
    	\end{minipage}\hfill
    	\mbox{}
    \end{table}

\begin{table}[H]
	\mbox{}\hfill
	\begin{minipage}[t]{.48\linewidth}
		\centering{ 
			
			\scalebox{0.5}{
				\begin{tabular}{l|rr|rr}
				\hline
				& \multicolumn{2}{c|}{$\epsilon=0.05$} & \multicolumn{2}{c}{$\epsilon=0.10$}   \\
				\hline
					RGlassoWinsor & 0.926 & 0.852 & 0.824 & 0.846\\
					
					& 0.006 & 0.003 & 0.014 & 0.001\\
					
					Glasso & 0.080 & 0.843 & 0.016 & 0.865\\
					
					& 0.019 & 0.004 & 0.008 & 0.010\\
					
					RGlassoQn & 0.004 & 1.000 & 0.035 & 0.993\\
					
					& 0.001 & 0.000 & 0.012 & 0.002\\
					
					RGlassoTau & 0.687 & 0.947 & 0.490 & 0.936\\
					
					& 0.025 & 0.003 & 0.034 & 0.005\\
					
					RGlassoGauss & 0.919 & 0.841 & 0.763 & 0.846\\
					
					& 0.008 & 0.003 & 0.020 & 0.001\\
					
					RGlassoSpearman & 0.949 & 0.843 & 0.886 & 0.839\\
					
					& 0.003 & 0.004 & 0.009 & 0.002\\
					
					RGlassoQuad & 0.570 & 0.935 & 0.588 & 0.883\\
					
					& 0.009 & 0.004 & 0.010 & 0.003\\
					\hline
			\end{tabular}}
		}
		
		\caption{ Model $\text{AR}(1)$ under THCM.  Comparison  of  means and standard deviations (in brackets) of  TPR and TNR  over $N=100$ replicates. $p=200, n=100$. } 
		\label{TPTN_ar1_p_200_THCM}%
	\end{minipage}\hfill
	\begin{minipage}[t]{.48\linewidth}
		\centering
		\scalebox{0.5}{		
			\begin{tabular}{l| l|l}
			\hline
			& $\epsilon=0.05$ &$\epsilon=0.10$ \\
			
			\hline
				RGlassoWinsor & 0.220 & 0.181\\
				
				& 0.004 & 0.004\\
				
				Glasso & -0.019 & -0.030\\
				
				& 0.006 & 0.003\\
				
				RGlassoQn & 0.034 & 0.016\\
				
				& 0.005 & 0.004\\
				
				RGlassoTau & 0.282 & 0.152\\
				
				& 0.012 & 0.011\\
				
				RGlassoGauss & 0.206 & 0.164\\
				
				& 0.003 & 0.005\\
				
				RGlassoSpearman & 0.218 & 0.193\\
				
				& 0.004 & 0.002\\
				
				RGlassoQuad & 0.216 & 0.150\\
				
				& 0.006 & 0.004\\
				\hline
		\end{tabular}}
		\vspace{0.15cm}
		\centering 
		\caption{    Model $\text{AR}(1)$ under THCM.    Comparison  of  means and standard deviations (in brackets) of  MCC over $N=100$ replicates. $p=200$, $n=100$.} 
		\label{MCC_se_ar1_p_200_THCM}%
	\end{minipage}\hfill
	\mbox{}
\end{table}

\begin{table}[H]
	\mbox{}\hfill
	\begin{minipage}[t]{.48\linewidth}
		\centering{ 
			
			\scalebox{0.5}{
				\begin{tabular}{l|rr|rr}
					\hline
					& \multicolumn{2}{c|}{$\epsilon=0.05$ 
					} & \multicolumn{2}{c}{$\epsilon=0.10$}   \\
					\hline
					& $D_{KL}$ & $m_{F}$ & $D_{KL}$ & $m_{F}$  \\
					\hline
					RGlassoWinsor & 12.865 & 9.401 & 12.880 & 9.415\\
				
				& 0.095 & 0.011 & 0.095 & 0.010\\
				
				Glasso & 46.680 & 10.668 & 46.770 & 10.659\\
				
				& 2.142 & 0.041 & 2.126 & 0.043\\
				
				RGlassoQn & 35.388 & 10.630 & 35.011 & 10.619\\
				
				& 0.765 & 0.017 & 0.797 & 0.018\\
				
				RGlassoTau & 25.459 & 10.334 & 25.156 & 10.321\\
				
				& 0.611 & 0.021 & 0.639 & 0.021\\
				
				RGlassoGauss & 13.361 & 9.477 & 13.373 & 9.480\\
				
				& 0.118 & 0.014 & 0.116 & 0.014\\
				
				RGlassoSpearman & 13.137 & 9.530 & 13.115 & 9.524\\
				
				& 0.123 & 0.014 & 0.117 & 0.014\\
				
				RGlassoQuad & 17.083 & 9.989 & 17.151 & 9.993\\
				
				& 0.190 & 0.011 & 0.196 & 0.012\\
				\hline
			\end{tabular} 	}
		}
		
		\caption{Model BG under THCM.    Comparison  of  means and standard deviations (in brackets) of  $\text{m}_F$  and $D_{KL}$ over $N=100$ replicates. $p=60$, $n=100$.} 
		\label{num_performance_se_block_p_60_THCM}%
		
	\end{minipage}\hfill
	\begin{minipage}[t]{.48\linewidth}
		\centering
		\scalebox{0.5}{
			\begin{tabular}{l|rr|rr}
			\hline
			& \multicolumn{2}{c|}{$\epsilon=0.05$ 
			} & \multicolumn{2}{c}{$\epsilon=0.10$}   \\
			\hline
			& $D_{KL}$ & $m_{F}$ & $D_{KL}$ & $m_{F}$  \\
			\hline
			RGlassoWinsor & 44.089 & 15.652 & 44.092 & 15.653\\
				
				& 0.189 & 0.019 & 0.193 & 0.018\\
				
				Glasso & 83.599 & 16.565 & 87.251 & 16.596\\
				
				& 7.690 & 0.119 & 8.238 & 0.124\\
				
				RGlassoQn & 140.510 & 18.311 & 139.861 & 18.303\\
				
				& 1.850 & 0.021 & 1.932 & 0.021\\
				
				RGlassoTau & 93.899 & 17.598 & 92.663 & 17.578\\
				
				& 1.673 & 0.030 & 1.413 & 0.027\\
				
				RGlassoGauss & 43.741 & 15.566 & 43.743 & 15.563\\
				
				& 0.220 & 0.023 & 0.214 & 0.022\\
				
				RGlassoSpearman & 44.324 & 15.751 & 44.320 & 15.747\\
				
				& 0.221 & 0.018 & 0.231 & 0.018\\
				
				RGlassoQuad & 63.239 & 16.883 & 63.405 & 16.893\\
				
				& 0.423 & 0.014 & 0.397 & 0.013\\
				\hline
		\end{tabular}}
		
		\caption{Model BG under THCM.    Comparison  of  means and standard deviations (in brackets) of  $\text{m}_F$  and $D_{KL}$ over $N=100$ replicates. $p=200$, $n=100$.} 
		\label{num_performance_se_block_p_200_THCM}%
	\end{minipage}\hfill
	\mbox{}
\end{table}

\begin{table}[H]
	\mbox{}\hfill
	\begin{minipage}[t]{.48\linewidth}
		\centering{ 
			
			\scalebox{0.5}{
				\begin{tabular}{l|rr|rr}
					\hline
					& \multicolumn{2}{c|}{$\epsilon=0.05$} & \multicolumn{2}{c}{$\epsilon=0.10$}   \\
					\hline
					& TPR & TNR & TPR & TNR       \\
					\hline
					RGlassoWinsor & 0.312 & 0.827 & 0.303 & 0.832\\
				
				& 0.009 & 0.006 & 0.008 & 0.006\\
				
				Glasso & 0.214 & 0.736 & 0.215 & 0.731\\
				
				& 0.013 & 0.009 & 0.011 & 0.009\\
				
				RGlassoQn & 0.001 & 1.000 & 0.001 & 1.000\\
				
				& 0.000 & 0.000 & 0.000 & 0.000\\
				
				RGlassoTau & 0.041 & 0.988 & 0.047 & 0.986\\
				
				& 0.004 & 0.001 & 0.005 & 0.002\\
				
				RGlassoGauss & 0.280 & 0.785 & 0.277 & 0.789\\
				
				& 0.011 & 0.008 & 0.011 & 0.008\\
				
				RGlassoSpearman & 0.303 & 0.839 & 0.307 & 0.834\\
				
				& 0.010 & 0.008 & 0.010 & 0.008\\
				
				RGlassoQuad & 0.062 & 0.977 & 0.059 & 0.978\\
				
				& 0.004 & 0.002 & 0.004 & 0.002\\
				\hline
			\end{tabular}}
		}
		
		\caption{ Model BG under THCM.  Comparison  of  means and standard deviations (in brackets) of  TPR and TNR  over $N=100$ replicates. $p=60, n=100$. } 
		\label{TPTN_block_p_60_THCM}%
	\end{minipage}\hfill
	\begin{minipage}[t]{.48\linewidth}
		\centering
		\scalebox{0.5}{		
			\begin{tabular}{l|l|l}
				\hline
				& $\epsilon=0.05$ &$\epsilon=0.10$ \\
				\hline
			RGlassoWinsor & 0.108 & 0.108\\
		
		& 0.008 & 0.008\\
		
		Glasso & -0.032 & -0.034\\
		
		& 0.006 & 0.006\\
		
		RGlassoQn & 0.005 & 0.006\\
		
		& 0.002 & 0.002\\
		
		RGlassoTau & 0.064 & 0.068\\
		
		& 0.005 & 0.005\\
		
		RGlassoGauss & 0.049 & 0.050\\
		
		& 0.008 & 0.007\\
		
		RGlassoSpearman & 0.115 & 0.116\\
		
		& 0.008 & 0.007\\
		
		RGlassoQuad & 0.070 & 0.067\\
		
		& 0.004 & 0.004\\
		\hline
			\end{tabular}
		}
		\vspace{0.15cm}
		\centering 
		\caption{    Model BG under THCM.    Comparison  of  means and standard deviations (in brackets) of  MCC over $N=100$ replicates. $p=60$, $n=100$.} 
		\label{MCC_se_block_p_60_THCM}%
	\end{minipage}\hfill
	\mbox{}
\end{table}

\begin{table}[H]
	\mbox{}\hfill
	\begin{minipage}[t]{.48\linewidth}
		\centering{ 
			
			\scalebox{0.5}{
					\begin{tabular}{l|rr|rr}
					\hline
					& \multicolumn{2}{c|}{$\epsilon=0.05$} & \multicolumn{2}{c}{$\epsilon=0.10$}   \\
					\hline
					& TPR & TNR & TPR & TNR       \\
				
					\hline
				RGlassoWinsor & 0.293 & 0.890 & 0.298 & 0.888\\
					
					& (0.008) & (0.004) & (0.008) & (0.004)\\
					
					Glasso & 0.009 & 0.004 & 0.009 & 0.004\\
					
				& (0.156) & (0.858) & (0.153) & (0.860)\\
					
					RGlassoQn & 0.000 & 1.000 & 0.000 & 0.999\\
					
					& (0.000) & (0.000) & (0.000) & (0.001)\\
					
					RGlassoTau & 0.053 & 0.977 & 0.052 & 0.979\\
					
				 & (0.006) & (0.003) & (0.006) & (0.003)\\
					
					RGlassoGauss & 0.312 & 0.870 & 0.326 & 0.868\\
					
				& (0.011) & (0.003) & (0.010) & (0.003)\\
					
					RGlassoSpearman & 0.333 & 0.883 & 0.182 & 0.478\\
					
					& (0.011) & (0.005) & (0.019) & (0.044)\\
					
					RGlassoQuad & 0.100 & 0.961 & 0.000 & 0.000\\
					
				& (0.005) & (0.003) & (0.000) & (0.000)\\
					\hline
			\end{tabular}}
		}
		
		\caption{ Model BG under THCM.  Comparison  of  means and standard deviations (in brackets) of  TPR and TNR  over $N=100$ replicates. $p=200, n=100$. } 
		\label{TPTN_block_p_200_THCM}%
	\end{minipage}\hfill
	\begin{minipage}[t]{.48\linewidth}
		\centering
		\scalebox{0.5}{		
			\begin{tabular}{l|l|l}
				\hline
				& $\epsilon=0.05$ &$\epsilon=0.10$ \\
				\hline
			RGlassoWinsor& 0.085 & 0.086 \\
				
				& (0.004) & (0.004)\\
				
				Glasso & 0.007 & 0.006\\
				
				& (0.005) & (0.005)\\
				
				RGlassoQn & 0.000 & 0.000\\
				
				& (0.000) & (0.000)\\
				
				RGlassoTau & 0.039 & 0.040\\
				
				& (0.004) & (0.004)\\
				
				RGlassoGauss & 0.077 & 0.080\\
				
				& 0.005 & 0.005\\
				
				RGlassoSpearman & 0.099 & 0.055\\
				
				& (0.004) & (0.006)\\
				
				RGlassoQuad & 0.050 & 0.000\\
				
				& (0.003) & (0.000)\\
				\hline
			\end{tabular}
		}
		\vspace{0.15cm}
		\centering 
		\caption{    Model BG under THCM.    Comparison  of  means and standard deviations (in brackets) of  MCC over $N=100$ replicates. $p=200$, $n=100$.} 
		\label{MCC_se_block_p_200_THCM}%
	\end{minipage}\hfill
	\mbox{}
\end{table}

\begin{table}[H]
	\mbox{}\hfill
	\begin{minipage}[t]{.48\linewidth}
		\centering{ 
			
			\scalebox{0.5}{
				\begin{tabular}{l|rr|rr}
					\hline
					& \multicolumn{2}{c|}{$\epsilon=0.05$ 
					} & \multicolumn{2}{c}{$\epsilon=0.10$}   \\
					\hline
					& $D_{KL}$ & $m_{F}$ & $D_{KL}$ & $m_{F}$  \\
					\hline
					RGlassoWinsor & 10.859 & 4.900 & 9.784 & 4.633\\
				
				& 0.189 & 0.036 & 0.175 & 0.041\\
				
				Glasso & 53.908 & 6.331 & 76.950 & 6.903\\
				
				& 6.290 & 0.134 & 6.970 & 0.133\\
				
				RGlassoQn & 26.347 & 6.395 & 23.773 & 6.203\\
				
				& 0.607 & 0.037 & 0.733 & 0.048\\
				
				RGlassoTau & 25.291 & 6.335 & 23.284 & 6.178\\
				
				& 0.573 & 0.037 & 0.685 & 0.046\\
				
				RGlassoGauss & 11.170 & 5.029 & 10.318 & 4.848\\
				
				& 0.153 & 0.025 & 0.161 & 0.032\\
				
				RGlassoSpearman & 11.164 & 5.028 & 10.299 & 4.846\\
				
				& 0.156 & 0.026 & 0.167 & 0.033\\
				
				RGlassoQuad & 12.401 & 5.211 & 11.439 & 5.031\\
				
				& 0.162 & 0.024 & 0.176 & 0.031\\
				\hline
			\end{tabular}}
		}
		
		\caption{Model $\text{NN}(2)$ under THCM.    Comparison  of  means and standard deviations (in brackets) of  $\text{m}_F$  and $D_{KL}$ over $N=100$ replicates. $p=60$, $n=100$.} 
		\label{num_performance_se_NN_p_60_THCM}%
		
	\end{minipage}\hfill
	\begin{minipage}[t]{.48\linewidth}
		\centering
		\scalebox{0.5}{
			\begin{tabular}{l|rr|rr}
				\hline
				& \multicolumn{2}{c|}{$\epsilon=0.05$ 
				} & \multicolumn{2}{c}{$\epsilon=0.10$}   \\
				\hline
				& $D_{KL}$ & $m_{F}$ & $D_{KL}$ & $m_{F}$  \\
				\hline
			
			RGlassoWinsor & 32.349 & 8.413 & 28.157 & 7.617\\
			
			& 0.495 & 0.057 & 0.377 & 0.062\\
			
			Glasso & 54.202 & 8.898 & 55.741 & 8.606\\
			
			& 9.939 & 0.152 & 10.879 & 0.174\\
			
			RGlassoQn & 78.592 & 11.365 & 63.849 & 10.749\\
			
			& 1.326 & 0.048 & 0.989 & 0.045\\
			
			RGlassoTau & 73.443 & 11.165 & 59.061 & 10.486\\
			
			& 1.252 & 0.051 & 1.105 & 0.058\\
			
			RGlassoGauss & 30.828 & 8.408 & 26.300 & 7.664\\
			
			& 0.445 & 0.050 & 0.322 & 0.054\\
			
			RGlassoSpearman & 32.069 & 8.546 & 27.294 & 7.810\\
			
			& 0.449 & 0.048 & 0.330 & 0.053\\
			
			RGlassoQuad & 47.485 & 9.836 & 40.033 & 9.208\\
			
			& 0.492 & 0.035 & 0.456 & 0.043\\
			\hline
		\end{tabular}}
		
		\caption{Model $\text{NN}(2)$ under THCM.    Comparison  of  means and standard deviations (in brackets) of  $\text{m}_F$  and $D_{KL}$ over $N=100$ replicates. $p=200$, $n=100$.} 
		\label{num_performance_se_NN_p_200_THCM}%
	\end{minipage}\hfill
	\mbox{}
\end{table}

\begin{table}[H]
	\mbox{}\hfill
	\begin{minipage}[t]{.48\linewidth}
		\centering{ 
			
			\scalebox{0.5}{
				\begin{tabular}{l|rr|rr}
					\hline
					& \multicolumn{2}{c|}{$\epsilon=0.05$} & \multicolumn{2}{c}{$\epsilon=0.10$}   \\
					\hline
					& TPR & TNR & TPR & TNR       \\
					\hline
					RGlassoWinsor & 0.290 & 0.998 & 0.287 & 0.997\\
					
					& 0.015 & 0.001 & 0.016 & 0.000\\
					
					Glasso & 0.202 & 0.929 & 0.167 & 0.888\\
					
					& 0.013 & 0.006 & 0.010 & 0.006\\
					
					RGlassoQn & 0.087 & 1.000 & 0.092 & 0.999\\
					
					& 0.005 & 0.000 & 0.005 & 0.000\\
					
					RGlassoTau & 0.086 & 1.000 & 0.091 & 0.999\\
					
					& 0.005 & 0.000 & 0.005 & 0.000\\
					
					RGlassoGauss & 0.283 & 0.999 & 0.275 & 0.997\\
					
					& 0.012 & 0.000 & 0.013 & 0.000\\
					
					RGlassoSpearman & 0.295 & 0.999 & 0.292 & 0.998\\
					
					& 0.013 & 0.000 & 0.013 & 0.000\\
					
					RGlassoQuad & 0.276 & 0.990 & 0.279 & 0.985\\
					
					& 0.012 & 0.001 & 0.012 & 0.001\\
					\hline
			\end{tabular}}
		}
		
		\caption{ Model $\text{NN}(2)$ under THCM.  Comparison  of  means and standard deviations (in brackets) of  TPR and TNR  over $N=100$ replicates. $p=60, n=100$. } 
		\label{TPTN_NN_p_60_THCM}%
	\end{minipage}\hfill
	\begin{minipage}[t]{.48\linewidth}
		\centering
		\scalebox{0.5}{		
			\begin{tabular}{l|l|l}
				\hline
				& $\epsilon=0.05$ &$\epsilon=0.10$ \\
			\hline
			RGlassoWinsor & 0.476 & 0.451\\
			
			& 0.011 & 0.012\\
			
			Glasso & 0.184 & 0.054\\
			
			& 0.023 & 0.013\\
			
			RGlassoQn & 0.269 & 0.268\\
			
			& 0.006 & 0.005\\
			
			RGlassoTau & 0.270 & 0.273\\
			
			& 0.005 & 0.005\\
			
			RGlassoGauss & 0.481 & 0.445\\
			
			& 0.010 & 0.011\\
			
			RGlassoSpearman & 0.490 & 0.471\\
			
			& 0.010 & 0.010\\
			
			RGlassoQuad & 0.340 & 0.307\\
			
			& 0.007 & 0.007\\
			\hline
			\end{tabular}
		}
		\vspace{0.15cm}
		\centering 
		\caption{    Model $\text{NN}(2)$ under THCM.    Comparison  of  means and standard deviations (in brackets) of  MCC over $N=100$ replicates. $p=60$, $n=100$.} 
		\label{MCC_se_NN_p_60_THCM}%
	\end{minipage}\hfill
	\mbox{}
\end{table}

\begin{table}[H]
	\mbox{}\hfill
	\begin{minipage}[t]{.48\linewidth}
		\centering{ 
			
			\scalebox{0.5}{
				\begin{tabular}{l|rr|rr}
					\hline
					& \multicolumn{2}{c|}{$\epsilon=0.05$} & \multicolumn{2}{c}{$\epsilon=0.10$}   \\
					\hline
					& TPR & TNR & TPR & TNR       \\
						\hline
					RGlassoWinsor & 0.564 & 0.996 & 0.582 & 0.995\\
					
					& 0.011 & 0.000 & 0.011 & 0.000\\
					
					Glasso & 0.595 & 0.985 & 0.600 & 0.982\\
					
					& 0.013 & 0.005 & 0.014 & 0.006\\
					
					RGlassoQn & 0.237 & 0.999 & 0.219 & 0.999\\
					
					& 0.008 & 0.000 & 0.006 & 0.000\\
					
					RGlassoTau & 0.247 & 0.999 & 0.260 & 0.998\\
					
					& 0.008 & 0.000 & 0.009 & 0.000\\
					
					RGlassoGauss & 0.601 & 0.997 & 0.620 & 0.996\\
					
					& 0.009 & 0.000 & 0.008 & 0.000\\
					
					RGlassoSpearman & 0.593 & 0.996 & 0.609 & 0.995\\
					
					& 0.009 & 0.000 & 0.008 & 0.000\\
					
					RGlassoQuad & 0.499 & 0.985 & 0.519 & 0.981\\
					
					& 0.008 & 0.001 & 0.008 & 0.001\\
					\hline
			\end{tabular}}
		}
		
		\caption{ Model $\text{NN}(2)$ under THCM.  Comparison  of  means and standard deviations (in brackets) of  TPR and TNR  over $N=100$ replicates. $p=200, n=100$. } 
		\label{TPTN_NN_p_200_THCM}%
	\end{minipage}\hfill
	\begin{minipage}[t]{.48\linewidth}
		\centering
		\scalebox{0.5}{		
			\begin{tabular}{l|l|l}
				\hline
				& $\epsilon=0.05$ &$\epsilon=0.10$ \\
					\hline
				RGlassoWinsor & 0.566 & 0.535\\
				
				& 0.005 & 0.005\\
				
				Glasso & 0.592 & 0.567\\
				
				& 0.014 & 0.015\\
				
				RGlassoQn & 0.379 & 0.358\\
				
				& 0.006 & 0.006\\
				
				RGlassoTau & 0.390 & 0.377\\
				
				& 0.006 & 0.006\\
				
				RGlassoGauss & 0.603 & 0.580\\
				
				& 0.003 & 0.004\\
				
				RGlassoSpearman & 0.577 & 0.547\\
				
				& 0.004 & 0.004\\
				
				RGlassoQuad & 0.325 & 0.300\\
				
				& 0.004 & 0.003\\
				\hline
			\end{tabular}
		}
		\vspace{0.15cm}
		\centering 
		\caption{    Model $\text{NN}(2)$ under THCM.    Comparison  of  means and standard deviations (in brackets) of  MCC over $N=100$ replicates. $p=200$, $n=100$.} 
		\label{MCC_se_NN_p_200_THCM}%
	\end{minipage}\hfill
	\mbox{}
\end{table}

\begin{table}[H]
	\mbox{}\hfill
	\begin{minipage}[t]{.48\linewidth}
		\centering{ 
			
			\scalebox{0.5}{
				\begin{tabular}{l|rr|rr}
					\hline
					& \multicolumn{2}{c|}{$\epsilon=0.05$ 
					} & \multicolumn{2}{c}{$\epsilon=0.10$}   \\
					\hline
					& $D_{KL}$ & $m_{F}$ & $D_{KL}$ & $m_{F}$  \\
						\hline
					RGlassoWinsor & 5.934 & 4.577 & 7.187 & 4.497\\
					
					& 0.054 & 0.025 & 0.144 & 0.026\\
					
					Glasso & 61.668 & 7.162 & 79.630 & 8.502\\
					
					& 8.077 & 0.334 & 8.035 & 0.314\\
					
					RGlassoQn & 8.777 & 6.034 & 9.406 & 6.020\\
					
					& 0.118 & 0.036 & 0.113 & 0.038\\
					
					RGlassoTau & 6.358 & 4.995 & 7.071 & 4.922\\
					
					& 0.063 & 0.028 & 0.082 & 0.023\\
					
					RGlassoGauss & 5.146 & 4.496 & 5.755 & 4.320\\
					
					& 0.050 & 0.029 & 0.082 & 0.025\\
					
					RGlassoSpearman & 5.455 & 4.679 & 5.974 & 4.466\\
					
					& 0.048 & 0.027 & 0.076 & 0.022\\
					
					RGlassoQuad & 10.254 & 6.538 & 9.857 & 6.284\\
					
					& 0.077 & 0.023 & 0.060 & 0.026\\
					\hline
			\end{tabular}}
		}
		
		\caption{Model Rand under THCM.    Comparison  of  means and standard deviations (in brackets) of  $\text{m}_F$  and $D_{KL}$ over $N=100$ replicates. $p=60$, $n=100$.} 
		\label{num_performance_se_Random_p_60_THCM}%
		
	\end{minipage}\hfill
	\begin{minipage}[t]{.48\linewidth}
		\centering
		\scalebox{0.5}{
			\begin{tabular}{l|rr|rr}
				\hline
				& \multicolumn{2}{c|}{$\epsilon=0.05$ 
				} & \multicolumn{2}{c}{$\epsilon=0.10$}   \\
				\hline
				& $D_{KL}$ & $m_{F}$ & $D_{KL}$ & $m_{F}$  \\
				\hline
		
			RGlassoWinsor & 44.700 & 12.540 & 44.848 & 12.569\\
			
			& 0.259 & 0.043 & 0.249 & 0.043\\
			
			Glasso & 98.156 & 15.942 & 103.344 & 15.908\\
			
			& 8.016 & 0.222 & 9.142 & 0.244\\
			
			RGlassoQn & 116.645 & 17.964 & 116.103 & 17.934\\
			
			& 2.547 & 0.084 & 2.850 & 0.084\\
			
			RGlassoTau & 70.709 & 15.585 & 72.787 & 15.674\\
			
			& 1.731 & 0.112 & 2.152 & 0.125\\
			
			RGlassoGauss & 44.252 & 12.414 & 44.609 & 12.494\\
			
			& 0.284 & 0.046 & 0.320 & 0.057\\
			
			RGlassoSpearman & 45.823 & 12.948 & 45.451 & 12.867\\
			
			& 0.366 & 0.053 & 0.344 & 0.048\\
			
			RGlassoQuad & 72.537 & 15.902 & 72.120 & 15.868\\
			
			& 0.660 & 0.040 & 0.609 & 0.037\\
			\hline
		\end{tabular}}
		
		\caption{Model Rand under THCM.    Comparison  of  means and standard deviations (in brackets) of  $\text{m}_F$  and $D_{KL}$ over $N=100$ replicates. $p=200$, $n=100$.} 
		\label{num_performance_se_Random_p_200_THCM}%
	\end{minipage}\hfill
	\mbox{}
\end{table}

\begin{table}[H]
	\mbox{}\hfill
	\begin{minipage}[t]{.48\linewidth}
		\centering{ 
			
			\scalebox{0.5}{
				\begin{tabular}{l|rr|rr}
					\hline
					& \multicolumn{2}{c|}{$\epsilon=0.05$} & \multicolumn{2}{c}{$\epsilon=0.10$}   \\
					\hline
					& TPR & TNR & TPR & TNR       \\
					\hline
					RGlassoWinsor & 0.909 & 0.852 & 0.870 & 0.861\\
					
					& 0.003 & 0.003 & 0.006 & 0.003\\
					
					Glasso & 0.571 & 0.837 & 0.358 & 0.856\\
					
					& 0.044 & 0.005 & 0.041 & 0.005\\
					
					RGlassoQn & 0.580 & 0.958 & 0.366 & 0.978\\
					
					& 0.015 & 0.002 & 0.019 & 0.002\\
					
					RGlassoTau & 0.817 & 0.869 & 0.742 & 0.881\\
					
					& 0.006 & 0.003 & 0.008 & 0.003\\
					
					RGlassoGauss & 0.927 & 0.825 & 0.895 & 0.836\\
					
					& 0.003 & 0.003 & 0.005 & 0.003\\
					
					RGlassoSpearman & 0.909 & 0.845 & 0.867 & 0.856\\
					
					& 0.003 & 0.003 & 0.005 & 0.003\\
					
					RGlassoQuad & 0.481 & 0.939 & 0.355 & 0.949\\
					
					& 0.010 & 0.002 & 0.013 & 0.003\\
					\hline
			\end{tabular}}
		}
		
		\caption{ Model Rand under THCM.  Comparison  of  means and standard deviations (in brackets) of  TPR and TNR  over $N=100$ replicates. $p=60, n=100$. } 
		\label{TPTN_Random_p_60_THCM}%
	\end{minipage}\hfill
	\begin{minipage}[t]{.48\linewidth}
		\centering
		\scalebox{0.5}{		
			\begin{tabular}{l|l|l}
				\hline
				& $\epsilon=0.05$ &$\epsilon=0.10$ \\
				\hline
			RGlassoWinsor & 0.424 & 0.417\\
			
			& 0.004 & 0.004\\
			
			Glasso & 0.204 & 0.104\\
			
			& 0.020 & 0.019\\
			
			RGlassoQn & 0.468 & 0.382\\
			
			& 0.005 & 0.010\\
			
			RGlassoTau & 0.400 & 0.379\\
			
			& 0.003 & 0.004\\
			
			RGlassoGauss & 0.398 & 0.394\\
			
			& 0.003 & 0.003\\
			
			RGlassoSpearman & 0.414 & 0.408\\
			
			& 0.004 & 0.004\\
			
			RGlassoQuad & 0.335 & 0.264\\
			
			& 0.004 & 0.006\\
			\hline
			\end{tabular}
		}
		\vspace{0.15cm}
		\centering 
		\caption{    Model Rand under THCM.    Comparison  of  means and standard deviations (in brackets) of  MCC over $N=100$ replicates. $p=60$, $n=100$.} 
		\label{MCC_se_Random_p_60_THCM}%
	\end{minipage}\hfill
	\mbox{}
\end{table}

\begin{table}[H]
	\mbox{}\hfill
	\begin{minipage}[t]{.48\linewidth}
		\centering{ 
			
			\scalebox{0.5}{
				%
}
		}
		
		\caption{ Model Rand under THCM.  Comparison  of  means and standard deviations (in brackets) of  TPR and TNR  over $N=100$ replicates. $p=200, n=100$. } 
		\label{TPTN_Random_p_200_THCM}%
	\end{minipage}\hfill
	\begin{minipage}[t]{.48\linewidth}
		\centering
		\scalebox{0.5}{		
			%
		}
		\vspace{0.15cm}
		\centering 
		\caption{    Model Rand under THCM.    Comparison  of  means and standard deviations (in brackets) of  MCC over $N=100$ replicates. $p=200$, $n=100$.} 
		\label{MCC_se_Random_p_200_THCM}%
	\end{minipage}\hfill
	\mbox{}
\end{table}

\begin{table}[H]
	\mbox{}\hfill
	\begin{minipage}[t]{.48\linewidth}
		\centering{ 
			
			\scalebox{0.5}{
				%
}
		}
		
		\caption{Model Hub under THCM.    Comparison  of  means and standard deviations (in brackets) of  $\text{m}_F$  and $D_{KL}$ over $N=100$ replicates. $p=60$, $n=100$.} 
		\label{num_performance_se_Hub_p_60_THCM}%
		
	\end{minipage}\hfill
	\begin{minipage}[t]{.48\linewidth}
		\centering
		\scalebox{0.5}{
			%
}		
		\caption{Model Hub under THCM.    Comparison  of  means and standard deviations (in brackets) of  $\text{m}_F$  and $D_{KL}$ over $N=100$ replicates. $p=200$, $n=100$.} 
		\label{num_performance_se_Hub_p_200_THCM}%
	\end{minipage}\hfill
	\mbox{}
\end{table}

\begin{table}[H]
	\mbox{}\hfill
	\begin{minipage}[t]{.48\linewidth}
		\centering{ 
			
			\scalebox{0.5}{
				%
}
		}
		
		\caption{ Model Hub under THCM.  Comparison  of  means and standard deviations (in brackets) of  TPR and TNR  over $N=100$ replicates. $p=60, n=100$. } 
		\label{TPTN_Hub_p_60_THCM}%
	\end{minipage}\hfill
	\begin{minipage}[t]{.48\linewidth}
		\centering
		\scalebox{0.5}{		
			%
		}
		\vspace{0.15cm}
		\centering 
		\caption{    Model Hub under THCM.    Comparison  of  means and standard deviations (in brackets) of  MCC over $N=100$ replicates. $p=60$, $n=100$.} 
		\label{MCC_se__p_60_THCM}%
	\end{minipage}\hfill
	\mbox{}
\end{table}

\begin{table}[H]
	\mbox{}\hfill
	\begin{minipage}[t]{.48\linewidth}
		\centering{ 
			
			\scalebox{0.5}{
				%
}
		}
		
		\caption{ Model Hub under THCM.  Comparison  of  means and standard deviations (in brackets) of  TPR and TNR  over $N=100$ replicates. $p=200, n=100$. } 
		\label{TPTN_Hub_p_200_THCM}%
	\end{minipage}\hfill
	\begin{minipage}[t]{.48\linewidth}
		\centering
		\scalebox{0.5}{		
			%
		}
		\vspace{0.15cm}
		\centering 
		\caption{    Model Hub under THCM.    Comparison  of  means and standard deviations (in brackets) of  MCC over $N=100$ replicates. $p=200$, $n=100$.} 
		\label{MCC_se__p_200_THCM}%
	\end{minipage}\hfill
	\mbox{}
\end{table}

 	\bibliography{bibliography}
	\bibliographystyle{Chicago}
\end{document}